\documentclass{aa}  
\usepackage{natbib}
\usepackage{xcolor}
\bibpunct{(}{)}{;}{a}{}{,} 
\usepackage{graphicx}
\usepackage{txfonts}
\usepackage{lscape}
\begin{document} 

   \title{Learning mid-IR emission spectra of polycyclic aromatic hydrocarbon populations from observations }

   \author{S. Foschino,
          \inst{1}
          O. Bern\'e, \inst{1}
          \and
          C. Joblin \inst{1}}

   \institute{Institut de Recherche en Astrophysique et Planetologie, Universit\'e de Toulouse, CNRS, CNES, UPS, Toulouse, France,\\
              9 Av. du colonel Roche, 31028 Toulouse Cedex 04, France\\
              \email{\textbf{olivier.berne@irap.omp.eu}}
             }

   \date{Received ??; accepted ??}

 
  \abstract
   {The {\it James Webb Space Telescope} (JWST) will deliver an unprecedented quantity of high-quality spectral data over the 0.6-28\,$\mu$m range. It will combine sensitivity, spectral resolution, and spatial resolution. Specific tools are required to provide efficient scientific analysis of such large data sets.}
  {Our  aim is to  illustrate the potential of unsupervised learning methods to get insights into chemical variations in the populations that carry the aromatic infrared bands (AIBs), more specifically polycyclic aromatic hydrocarbon (PAH) species and carbonaceous very small  grains (VSGs).}
   {We  present a method based on linear fitting and blind signal separation for extracting representative spectra 
   for a spectral data set. The method is fast and robust, which ensures its applicability to JWST 
   spectral cubes. We tested this method on a sample of ISO-SWS data, which resemble  most closely the JWST 
  spectra in terms of spectral resolution and coverage. }
   {Four representative spectra were extracted. Their main characteristics appear consistent with previous studies with populations dominated by cationic PAHs, neutral PAHs, evaporating VSGs, and large ionized PAHs, known as the  PAH$^x$ population. In addition, the 3\,$\mu$m range, which is considered here for the first time in a blind signal separation (BSS)   method, reveals the presence of aliphatics connected to neutral PAHs. Each representative spectrum is found to carry second-order  spectral signatures (e.g., small bands),  which are connected with the underlying chemical diversity of populations. However, the precise attribution of theses signatures remains  limited by the combined small size and heterogeneity of the sample of astronomical spectra available in this study.}
    {The upcoming JWST data will allow us to overcome this limitation. The large {data sets of hyperspectral images} provided by JWST  analysed  with the proposed method, which is fast and robust, will open  promising perspectives for our understanding of the chemical evolution of the AIB carriers.}

      \titlerunning{Learning mid-IR emission spectra of PAHs populations from observations}    

     \authorrunning{Foschino et al.}
   \maketitle
%

\section{Introduction}
\label{sect_intro}
The aromatic infrared bands (AIBs) have held the attention of a large community of astronomers since their discovery in the 1970s \citep{gillett_8-13_1973}. 
At first named unidentified infrared bands (UIBs), they obtained the current denomination of AIBs after  \citet{Leger_puget} and \citet{allamandola_polycyclic_1985} proposed that they can be emitted by large carbonaceous molecules, namely polycyclic aromatic hydrocarbons (PAHs). 
Their limited size (20 to more than 100 carbon atoms) implies that they are easily heated to high temperatures by the absorption of individual UV photons in photo-dissociation regions (PDRs) where the AIBs dominate the mid-IR spectrum. 
Once the molecule absorbs a UV photon, it can dissociate, be ionized or relax through infrared emission \citep[e.g.,][]{montillaud_2013}. The strongest AIBs, at 3.30, 6.20, 7.70, 8.60, 11.30, and 12.70\,$\mu$m correspond to the cooling of the molecule through vibrational stretching and bending modes of its C-H and C-C bonds. 
To study this emission the community   has counted on several space observatories: the Infrared Astronomical Satellite \citep[IRAS,][]{neugebauer_IRAS_1984}, the Infrared Space Observatory \citep[ISO,][]{Kessler1996}, the Spitzer Space Telescope \citep[{\it Spitzer},][]{Werner2004}, and AKARI \citep{Murakami2007}. 
The upcoming James Webb Space Telescope (JWST, \citealt{gardner_JWST2006}) 
has inspired a lot of hope as its capabilities will surpass all former space observatories.

One of the main findings of observational studies is the discovery that the AIB spectrum 
shows significant variability, depending on the astrophysical sources
\citep{peeters_rich_2002}. Variations within individual objects have also been observed 
\citep[e.g.,][]{Joblin96_RN_33and34variation, Sloan_orion_33_97, Giard1992, Cesarsky2000, rapacioli_spectroscopy_2005}. 
These spectral variations can consist of changes in the band positions and widths or in band relative intensities. 
To analyze and interpret these variations, several approaches have been developed over the years. 
The first  is a decomposition method, which dates back to the 1990s and consists in fitting multiple Lorentzian profiles to the AIBs \citep{Boulanger1998}. This was extended in the {\it Spitzer}
era by \citet{SmithPAHfit} who proposed a fitting tool, PAHfit, that  automatically decomposes 
 the AIB spectrum into several individual bands (Gaussian or Drude profiles) 
in the {\it Spitzer} range (5-35\,$\mu$m). These fitting methods have been used widely to extract
band intensities and relate them to star-formation rates in galaxies \citep{Galliano2018_review}, for example.
They have also been used to study the variations of the AIB spectrum, for instance 
by \citet{galliano_variations_2008} who were able to show that the 6.2\,\,$\mu$m over 11.3\,$\mu$m 
band ratio, expected to trace ionization degree of PAHs, 
correlates with the ionization balance parameter $\frac{G_0\sqrt{T_{gas}}}{n_e}$. \citet{peeters_pah_2017} used Gaussian profiles to 
fit individual bands from spectral maps of NGC\,2023 PDRs (north and south) and discussed 
in detail the possible assignments of these bands. 

Another well-known approach for AIB analysis is the classification of \citet{peeters_rich_2002}.  Using continuum-subtracted ISO-SWS spectra in the 6-9\,$\mu$m range, the authors defined the A, B, and C classes based 
on the peak positions of the 6.2, 7.7, and 8.6\,$\mu$m bands. 
Some objects show spectra that belong only to one class (e.g., HII regions are always class
A),  while other objects can show mid-IR spectra belonging to different classes (e.g., post-AGB stars can be found in Class A, B, and C). \citet{hony_ch_2001} proposed a classification for the 10-15 \,$\mu$m range.
This empirical method has the advantage of being simple, but is difficult to 
apply automatically to large samples of observations. In addition, the strategies
of \citet{peeters_rich_2002} and \citet{hony_ch_2001} rely on subregions of the spectrum, and do not consider the full AIB spectrum.
 
 A more detailed analysis of the AIB spectrum consists in using IR absorption spectra of PAHs that have been either measured or computed, as available in the NASA Ames PAH IR Spectroscopic Database \citep[PAHdb,][]{bauschlicher_nasa_2010,boersma_nasa_2014, bauschlicher_nasa_2018} or the theoretical spectral database of PAHs \citep{malloci_-line_2007}. Using this approach provides further insights into the chemical diversity of the interstellar PAH population
\citep[e.g.,][]{hudgins05, boersma09, simon10, bauschlicher_2009,candian15}. The limitation of this approach concerns the challenge of modeling,  in a precise way, the emission spectra of PAHs in astrophysical environments \citep{mulas06,boersma_properties_2013}, especially when band profiles need to be simulated \citep{pech02, mulas06b}.

The last type of analysis approach involves blind signal separation (BSS) methods.
\citet{rapacioli_spectroscopy_2005} analyzed ISOCAM observations of NGC\,7023 and 
$\rho$-Ophiucus by using singular value decomposition (SVD) coupled 
with a Monte Carlo search algorithm and a positivity criterion. Three elementary 
spectra were extracted and attributed to cationic PAHs, neutral PAHs, 
and carbonaceous very small grains (PAH+, PAH0, and VSGs, respectively). 
The PAH$^+$ spectrum exhibits mainly the 6.2, 7.7, 8.6, 11.3, and 
12.7\,$\mu$m bands.  It is dominated by the 7.7\,$\mu$m bands (hereafter the 7.7\,$\mu$m complex).
The PAH$^0$ spectrum 
is  similar, but with a 11.3\,$\mu$m much stronger than that of the PAH$^+$ spectrum.
The VSG population shows a spectrum with a strong contribution of the continuum at wavelengths longer than 10\,$\mu$m. Its 7.7\,$\mu$m complex is redshifted to 
7.8\,$\mu$m with respect to the first two spectra. The 8.6\,$\mu$m band is blended with the 7.8\,$\mu$m band.

Subsequently, \citet{berne_analysis_2007}  used non-negative matrix factorization (NMF) on the
{\it Spitzer}-IRS spectro-imagery observation of NGC\,7023, $\rho$-Ophiucus, and Ced\,201. 
They extracted three representative spectra consistent with those obtained 
by \citet{rapacioli_spectroscopy_2005}, but over a wider wavelength range and with 
higher spectral resolution. 
\citet{rosenberg_coupled_2011} conducted a coupled NMF--database fitting analysis
of NGC\,7023~NW {\it Spitzer} IRS-SH spectra, from 10 to 15\,$\mu$m where they also identified these three populations, and assigned the 11.0, 11.2, and 11.3\,$\mu$m bands to 
PAH$^+$, PAH$^0$, and VSGs, respectively.
\citet{joblin_carriers_2008} averaged the representative spectra extracted by \citet{rapacioli_spectroscopy_2005}
and \citet{berne_analysis_2007} to build template spectra for each of the three mean populations, i.e., 
PAH$^0$, PAH$^+$, and VSGs. Then they used them to fit the spectra of the  HII regions and planetary nebulae (PNe). They found that the three template spectra were not sufficient to fit PNe spectra in which the 7.7\,$\mu$m complex is redshifted to $\sim$ 7.8\,$\mu$m. The authors then invoked  a fourth template spectrum to better fit this spectrum type. Guided by the  PAH$^+$ spectrum and calculated IR absorption spectra available in the databases, they constructed the PAH$^x$ spectrum and attributed it to large ionized PAHs. 
Later, these four templates were incorporated into the PAH Toulouse Astronomical Templates (PAHTAT) tool\footnote{The IDL code PAHTAT is available for download at \url{http://userpages.irap.omp.eu/~cjoblin/PAHTAT}} \citep{pilleri_evaporating_2012}, which can be used to fit {\it Spitzer}-IRS spectra with these four template spectra. So far, much of the success of the BSS methods in providing a scenario of evolution of the AIB populations with UV processing resides in the study of NGC\,7023, a nebula that is sufficiently extended to observe chemical variations even at the spatial scale of a few arcseconds sampled by the previous IR missions. 

The JWST will provide hyperspectral images with unprecedented details combining middle spectral ($R\sim 3000$) and high spatial (up to 0.1") resolutions over the full 0.6 to 28 \,$\mu$m wavelength range.
The high sensitivity of the instruments will increase the number of observable astrophysical objects (including fainter regions of PDRs toward molecular clouds, the diffuse interstellar
medium, and distant galaxies). These improvements will increase
the quantity of the collected hyperspectral image information, with respect  to previous missions, by several orders of magnitude. Dedicated methods for  analyzing AIB emission in these large data sets are thus required. 

The present study aims to provide a fast and robust method based on BSS for the analysis of these data sets. In Sect.~\ref{sect_approach} we propose such a method. In Sect.~\ref{sect_results} we present our test
 of this method on ISO-SWS data and  the obtained set of representative spectra.
 In Sect.~\ref{sect_discussion} we discuss how this method can help  interpret the observations in terms of chemical populations.  We  conclude in Sect.~\ref{conclusion}.

\section{Description of the Approach}
 \label{sect_approach}

\subsection{Challenges of large JWST data sets}
\label{subsect_JWST}
 The next generation of space infrared spectroscopic observations
will be provided by the JWST, whose  launch is planned in 2021. It will
deliver mid-IR spectral cubes of unprecedented quality over 
the 0.6 to 28 \,$\mu$m range once the data from the near infrared spectrometer (NIRSpec) andmid-IR imager (MIRI) instruments are combined. 
 A summary of its performance is available in Table~\ref{Table_JWSTPerf}. 
Overall, the MIRI and NIRSpec integral field unit (IFU) mode characteristics combine and surpass the advantages of former space observatories as ISO, {\it Spitzer}, and {\it Akari}. The achieved medium spectral resolution will be well adapted to the study of the AIBs. 
MIRI cubes will have different sizes depending on the channel, with typical values between 17\,$\times$\,19\,$\times$\,11000 and 26\,$\times$\,29\,$\times$\,5000, whereas NIRSpec IFU cubes will typically be 30\,$\times$\,30\,$\times$\,12000. The memory needed for one data cube will reach several hundred megabytes. Efficient algorithms will thus be required to analyze these large data sets.

In this paper we propose an approach based on linear methods allowing a reduction in  the dimensionality of data to that of a few {\it \emph{representative}} spectra
using a blind signal separation (BSS) approach.
The first step of the method (which is also an improvement over  earlier BSS applications to mid-IR spectra) is to extract the AIB emission from the observations (i.e., without continuum emission, and gas lines). 
This is done with the  linear fitting method described in Sect.~\ref{subsect_aibextraction}. 
The BSS method used to perform dimensionality reduction on the AIB extracted spectra is described in Sect.~\ref{subsect_BSS}. The initialization process of NMF algorithms is described in Sect.~\ref{subsect_initialization}. The method is summarized in Sect.~\ref{subsect_summary}.

\subsection{AIB extraction}

In this section we present a method for extracting the AIB emission from observations using a linear model to maintain a low computational time.
We note that the model presented here is not physically motivated, and  only  extracts the AIB emission and reduces the impact of noise. Thus, the degeneracy (i.e.,  that many solutions can provide a good fit) induced by the use of a large catalog is of limited concern, as the objective is mainly to have a good fit to extract a clean AIB spectrum and not to use fit parameters for further analysis. The proposed algorithm uses predominantly relatively localized spectral structures like Gaussians to model bands, lines, and  plateaus, in particular in the 6-9\,$\mu$m region \citep{peeters_rich_2002}, whereas much broader components like blackbodies are used for large-scale spectral structures like the continuum.

\label{subsect_aibextraction}
\subsubsection{Model}
\label{subsubsect_model}

Figure \ref{fig_exemplespec} shows typical ISO-SWS spectra. They contain broad AIBs, gas lines,  continuum, and instrumental noise. Our aim was to  extract the AIB contribution. We can model the observation vector $s_{obs}$ by $s_{mod}$ as 
\begin{equation}
 s_{mod}= s_{AIB}+ s_{cont} + s_{gas}+s_{inst}
 \label{eq_model}
,\end{equation}
where $s_{AIB}$,  $s_{cont}$, $s_{gas}$, and $s_{inst}$ correspond respectively to the model of AIBs, continuum, gas lines, and instrumental noise. Each component (but not  the instrumental  noise) is composed of a list of spectra. Section~\ref{subsubsect_catalogs} describes how the lists are constructed.

\subsubsection{Model components}
\label{subsubsect_catalogs}
The objective of the present model is only to represent the data mathematically in a way that allows us to extract easily an estimation of an AIB spectrum from an observation. \\
The first component of Eq.~\ref{eq_model}, $s_{AIB}$, corresponds to the AIB emission expressed as
\begin{equation}
s_{AIB}= g\times A,
\end{equation}where  $g$ is a vector of coefficients and {\bf A} a 
matrix whose lines are Gaussians.

To build matrix {\bf A}, we first define the  intervals of wavelength over which AIBs can exist. For each interval we define a number of Gaussians with a given width $\sigma$ (Table \ref{tab_bandcatalog}). We note that for a given interval $\sigma$ can take several discrete values. Each Gaussian position is spread out evenly over the interval of wavelengths.\medskip

The continuum emission model, $s_{cont}$, is expressed as
\begin{equation}
s_{cont}= k\times B,
\end{equation}where $k$ is a vector of coefficients and  {\bf B} a matrix whose lines are defined as follows. Each continuum $B_{\lambda,i}$ is defined by

\begin{equation}
B_{\lambda,i}(\lambda,T)= \frac{1}{\lambda}B_{\lambda}(\lambda,T)\times exp^{-C_{ext}N_H}.
\end{equation}
We use discrete values of T in an interval of temperature between T$_{min}$ and T$_{max}$ with a temperature step of $\Delta$T. 
Likewise, several discrete values of N$_H$ can be taken into account. The value of  
C$_{ext}$ is taken from \citet{WeingartnerDraine2001} for the Milky Way extinction curve.\medskip

The third component of the model is the gas emission, $s_{gas}$, described as
\begin{equation}
s_{gas}= p\times L,
\end{equation}where $p$ is a vector of coefficients and {\bf L} a matrix whose lines are Gaussians. The same principle as for the AIB emission is applied. However, the lines are much narrower. Each set of Gaussians is distributed over an interval $\Delta \lambda$ around the central wavelength $\mu$ of the line. 
It must be noted that all matrices {\bf A}, {\bf B}, and {\bf L} are pre-computed. Thus, in the fitting optimization only the coefficients in vectors $g$, $k$, and $p$ have to be adjusted. This implies that the model is fully linear since equations 1, 2, 3, 4, and 5 only use additive and multiplicative operators. 
\subsubsection{Optimization}
\label{subsubsect_optimization}
To estimate $s_{mod}$ we minimized the L2 norm between $s_{obs}$ and $s_{mod}$:
\begin{equation}
        \hat{s}_{mod}=  \underset{s_{mod}}{min}{|| s_{obs}-s_{mod}||^2.}
                \label{eq_optimization}
\end{equation}To perform this minimization, we used the non- negative least square (nnls) function of the scipy package implemented in Python. 

This is a critical aspect of the method. The fully linear model described in Sect.~\ref{subsubsect_catalogs} allows us to use the nnls function, which is much faster than the  Levenberg-Marquardt method   used   in PAHfit. As an illustration, we compared the computational cost between a Python version of PAHfit and our method (also in Python) when applied to a single Spitzer-IRS spectrum of the NGC 7023 NW photo-dissociation region, from 5 to 14\,$\mu$m. 
The fitting part in PAHfit (algorithm setup and the fit itself) runs for 194 seconds, while our code lasts 0.47 second. Compared to PAHfit, the main advantage of our method is that it  provides a good AIB extraction 400x faster. This suggests that PAHfit in its current form would be hardly  applicable on a 0.6-28 $\mu$m IFU cube from JWST (containing 900 spectra of 40k spectral points), while the linear method would still run over a reasonable timescale.

This method can be applied to each of the observed spectra, yielding for each source an estimate of $s_{mod}$ (i.e., $\hat{s}_{mod}$).  For each model, this allows us to isolate an estimate of $s_{AIB}$ (i.e., $\hat{s}_{AIB}$) to then construct {\bf X$_{aib}$}, the matrix containing the modeled contribution of AIB for each object. In other words, a line of {\bf X$_{aib}$} is the $\hat{s}_{AIB}$ vector obtained for a given object.

\subsection{NMF blind signal separation method}
\label{subsect_BSS}
\subsubsection{Choice of BSS method}
The BSS methods are based on the assumption that the signal measured
(mid-IR spectra in our case) is a mixture of elementary sources (spectra). The goal of 
these methods is to extract each of these sources from the data.  They have been used in 
several fields, such as acoustics \citep{abrard_timefrequency_2005} and  IR astronomical
spectroscopy (\citet{berne_analysis_2007}, \citet{rosenberg_coupled_2011}). 
A common assumption on the data mixture is the linear instantaneous
model, which implies that all the n spectra of the data set are a 
linear combination of r elementary sources. This is expressed as
\begin{equation}
X = M\times S,
\end{equation}where {\bf X} is the $n\times m$ data matrix whose lines are observed spectra, {\bf M} the $n\times r$ weight matrix, and {\bf S} the $r \times m$ source matrix.  
There are four main classes of methods used to perform BSS:  independent component analysis (ICA), sparse component analysis (SCA), non-negative matrix factorization (NMF),  and Bayesian methods. Here we used NMF since it requires  that the elementary sources are non-negative and the data satisfy the linear instantaneous mixture with non-negative coefficients. These conditions are verified in principle in the present case  except for linearity when extinction becomes important (see \citet{berne_analysis_2007}). 
NMF provides an estimation of {\bf X} by the product of the non-negatives matrices of weights {\bf W} and of sources {\bf H} as 
\begin{equation}
X \approx W\times H,
\end{equation}where {\bf W} is an estimate of {\bf M} and {\bf H} of {\bf S}.
The lines of {\bf H} contain an approximation of the elementary spectra
which we call "representative spectra".
An algorithm for  obtaining {\bf W} and {\bf H} was introduced by \citet{lee_learning_1999,lee_algorithms_2001} (see details of the algorithm in \citet{berne_analysis_2007}). \citet{lin_projected_2007} proposed an algorithm for performing NMF that converges faster using the projected gradient method instead of multiplicative updates. We used the latter method, which minimize the Euclidean distance between matrix {\bf X$_{aib}$} and the matrix product $W\times H$, implemented in MATLAB. \medskip

\subsubsection{Choice of r}
\label{subsubsect_choicer}
The choice of r (i.e., the number of representative spectra) is a crucial and non-trivial problem in BSS. As it is unknown, a criterion is needed to fix it. Several methods were proposed in the literature.
\citet{rapacioli_spectroscopy_2005} and \citet{berne_blind_2012} used a method based on the standard deviation $\sigma_r$ of the reconstruction residual matrix with a varying number r. The number r is increased until $\sigma_r$ reaches the noise level of the observed data. 
To apply this method, a correct measure of the data sample noise level is needed (e.g., an empty pixel). 

Other methods are  based on the dimension of the vector space spanned by the data set. The method of  singular value decomposition  (SVD, \citet{luo_using_2000}) applied to a set of spectra can help us find the dimension of the  vector space in which all spectra can be expressed.
\citet{boulais_geometrical_2015} used this method on astrophysical data.  A steep transition in the eigenvalues of the covariance matrix of {\bf X} indicate the dimension of its vector space. 

Finally, \citet{berne_analysis_2007} proposed a user-based method to identify r, which consists of increasing r until artifacts or physically non-relevant spectra are extracted.

\subsection{Initialization of NMF with MASS}
\label{subsect_initialization}
\subsubsection{Variability of the NMF solutions}
 
Both the  \citet{lee_algorithms_2001} and \citet{lin_projected_2007} algorithms use iterative updates of matrices W and H. The most common method used to initialize {\bf W}
and {\bf H} is to set their values randomly. Earlier studies \citep{berne_analysis_2007,rosenberg_coupled_2011} show that solutions of NMF applied to mid-IR astronomical with random initialization show variability in the results.  
To overcome this issue, \citet{rosenberg_coupled_2011} used a Monte Carlo (MC) analysis to extract an average solution and error bars. This method is  costly  in terms of time,
however, as it multiplies the total time cost by the number of MC-NMF runs. This is an issue because the convergence time increases with the number of points, and JWST data matrices will contain a hundred times more points.

The robustness of NMF representative spectra extraction relies on how these mathematical variabilities can be reduced. In the following we present a method that allows us to obtain a unique solution for a given data set. Using a different  data set, i.e., a different X matrix, implies a change in the solution, i.e., a change in {\bf W} and {\bf H} matrices.

\subsubsection{Initialization of {\bf H} by MASS}
To circumvent the issues associated with random initialization, \citet{boulais_methodes_2017} developed an innovative way to initialize {\bf W} or {\bf H} using a geometrical method, the Maximum Angle Signal Separation (MASS) algorithm.
The MASS algorithm extracts from the data the observed spectra that are the farthest from an angular point of view in the space vector generated by the data set. 
Practically, MASS computes the scalar product between each pair of spectra of the data set and, in a first step, isolates the pair that minimize it and creates a subspace vector with them. 
Then each remaining spectrum is projected on this subspace and the one that maximizes the angle between itself and its orthogonal projection on the first subspace is isolated. 
This spectrum is added to the first two spectra and the operation starts again until the algorithm isolates r spectra. The output is an estimation of the matrix {\bf H} called {\bf S$_{mass}$} based on the observed spectra. 

Based on tests performed on synthetic spectra, \citet{boulais_methodes_2017} showed that initializing {\bf H} by {\bf S$_{mass}$} significantly reduces  the variability of the solution of NMF, although {\bf W} is still initialized randomly. 
It was also shown using these synthetic data that this form of initialization provides a solution that is more reliable compared to what can be obtained using a MC approach. For example, when {\bf H} is initialized by {\bf S$_{mass}$}, the normalized root mean square error measured is 7\% against 18\% with MC-NMF alone.

Overall, since the dispersion of the solution is significantly reduced, only one run of MASS-NMF allows us to obtain a more accurate result than MC-NMF and a significant gain of time.
 A detailed comparison of the MC method results and initialization with MASS will be provided in a subsequent paper \citep{boulais2019}, following \citet{boulais_methodes_2017}.
We  therefore included  this initialization scheme in the algorithm presented in this study.

\subsection{Summary of the MASS-NMF method}
\label{subsect_summary}
In summary, the proposed method for extracting representative spectra from the JWST data is the following:

\begin{itemize}
\item {Spectral resolution and sampling are homogenized over  the whole data sample;}

\item From the results of the fit, a data matrix, {\bf X$_{aib}$}, whose lines are the AIB spectra extracted from observations is created (Sect. \ref{subsubsect_optimization});
\item The number of  representative spectra r is defined by the user (Sect \ref{subsubsect_choicer});
\item The MASS algorithm is applied to {\bf X$_{aib}$}, yielding {\bf S$_{mass}$}, a first gross estimate of
the results (Sect.~\ref{subsect_initialization}); 
\item NMF, with {\bf H} initialized with the values of {\bf S$_{mass}$}, is applied to {\bf X$_{aib}$} until convergence is reached (Sect.~\ref{sect_results}).
\end{itemize}

 \begin{table}
 \centering
\begin{tabular}{|c|c|c|}
\hline
Sources &type&scan speed\\
\hline
 BD +303639&PN&3 \\
 \hline
 IRAS 07027 - 7934 &PN&2\\
\hline
 He 2 113&PN&2 \\
\hline
 IRAS 17047 - 5650&PN&1\\
\hline
IRAS 18317 - 0757&CHII&2 \\
\hline
IRAS 23030 + 5958 &CHII&2\\
\hline
 IRAS 19442 + 2427&CHII&2\\
\hline
 IRAS 22308 + 5812&CHII&2\\
\hline
 IRAS 18032 - 2032&CHII&2\\
\hline
 IRAS 23133 + 6050&CHII&2\\
\hline
 IRAS17279 - 3350&CHII&2\\
\hline
IRAS 18502 + 0051 &CHII&2\\
\hline
W3A0 2219  + 6125 &CHII&2\\
\hline
IRAS 12063 - 6259 &CHII&2\\
\hline
 Orion Bar (1) BRGA&HII&2\\
\hline
 Orion Bar (2)  D5&HII&2\\
\hline
 Orion Bar (3) D5&HII& 2\\
\hline
 Orion Bar (4) H2S1&HII&4\\
\hline
S106 IRS4&HII&2\\
\hline
 G327.3 - 0.5&HII &1\\
\hline
M17 (8) IRAM POS 8$^a$&HII&2\\
\hline
 M17 (6) IRAM POS 7$^a$&HII&2\\
\hline
M17 (5) IRAM POS 6$^a$ &HII&4\\
\hline
 M17 (2) IRAM POS 4$^a$ &HII&2\\
\hline
 M17 (3) IRAM POS 5$^a$&HII& 2\\
\hline
IRAS 15384-5348&HII&4\\
\hline
NGC7023&RN&2\\
\hline
IRAS 03260 + 3111&RN&3\\
 \hline
 IRAS 16555 - 4237&RN&2\\
\hline
  Red Rectangle &Post AGB&4\\
\hline
 M82&Galaxy&4\\
\hline
\end{tabular}
 \caption[]{List of sources used in our sample. All the data come from the \citet{sloan_uniform_2003} catalog and are taken in  SWS01 mode. The last column gives the scan speed used for the corresponding spectrum. $^a$For precise details  about the position of the M17 spectra, see \citet{Verstraete1996_M17} }
         \label{tab_sources}
   \end{table}

\section{Test on real data}
 \label{sect_results}
\subsection{Data set and test parameters}
In this section, we describe the ISO-SWS data on which the present method is tested in Sect.~\ref{subsubsect_isodata}. The model parameters used are described in Sect.~\ref{subsubsect_AIBparam}. The choice of the number on representative spectra is described in Sect.~\ref{subsubsect_datachoiceofr}.

\subsubsection{Data}
\label{subsubsect_isodata}
The ISO-SWS observations cover the 2.4 to 45 \,$\mu$m range, which overlaps that of the  JWST-NIRSpec--MIRI combined data (0.6 to 28~$\mu m$, \citet{gardner_JWST2006}).
We limit our study to the $2.6-15\mu$m range over which the major PAH bands are located. 
Since there is no spatial information, we gathered the spectra of different astrophysical objects from the \citet{sloan_uniform_2003} catalog of highly processed data product spectra.
More specifically, we selected spectra showing clear AIB emission and no noticeable extinction (see Sect.~\ref{subsect_BSS}).
The compiled list of objects and their types is given in Table~\ref{tab_sources}. In total, $n=31$ spectra were used coming from reflection nebulae (RN), PNe, HII, and Compact HII (CHII) regions, as well as one post-AGB star and one starburst galaxy (i.e., Red Rectangle and M82, respectively). 
It appears that some spectra have a negative part due to calibration issues. For example, the IRAS 19442+2427 spectrum has negative values over the wavelength region from 2.6 to $\sim 6.1$\,$\mu$m. This is circumvented by adding a constant to the spectrum corresponding to the absolute value of the minimum intensity of this spectral region. It has no impact on the analysis since the aim of the present paper is to study  the AIB emission that will be extracted using a non-negative least-squares algorithm (see Sect.~\ref{sect_approach}).

The selected observations came from different proposals and involve various observing modes. An ISO-SWS spectrum is made of the combination of several discrete segments. 
To homogenize the spectra, we selected the spectral resolution of the mode with the lowest spectral resolution present in our data sample, which corresponds to the scan speed 2 of mode 01.  
Then we applied Gaussian convolution using a kernel with width corresponding to the lowest spectral resolution of the same segment in the observing mode 01 in scan speed 2. 
Another necessity is to have  dimensional consistency (i.e., the same spectral sampling for all spectra). A wavelength vector with step size of $\frac{\Delta\lambda}{10}$ (where $\Delta\lambda=\frac{\lambda_{ref}}{R}$) was generated, and the data was interpolated onto this new grid. 

\begin{figure}
   \centering
   \includegraphics{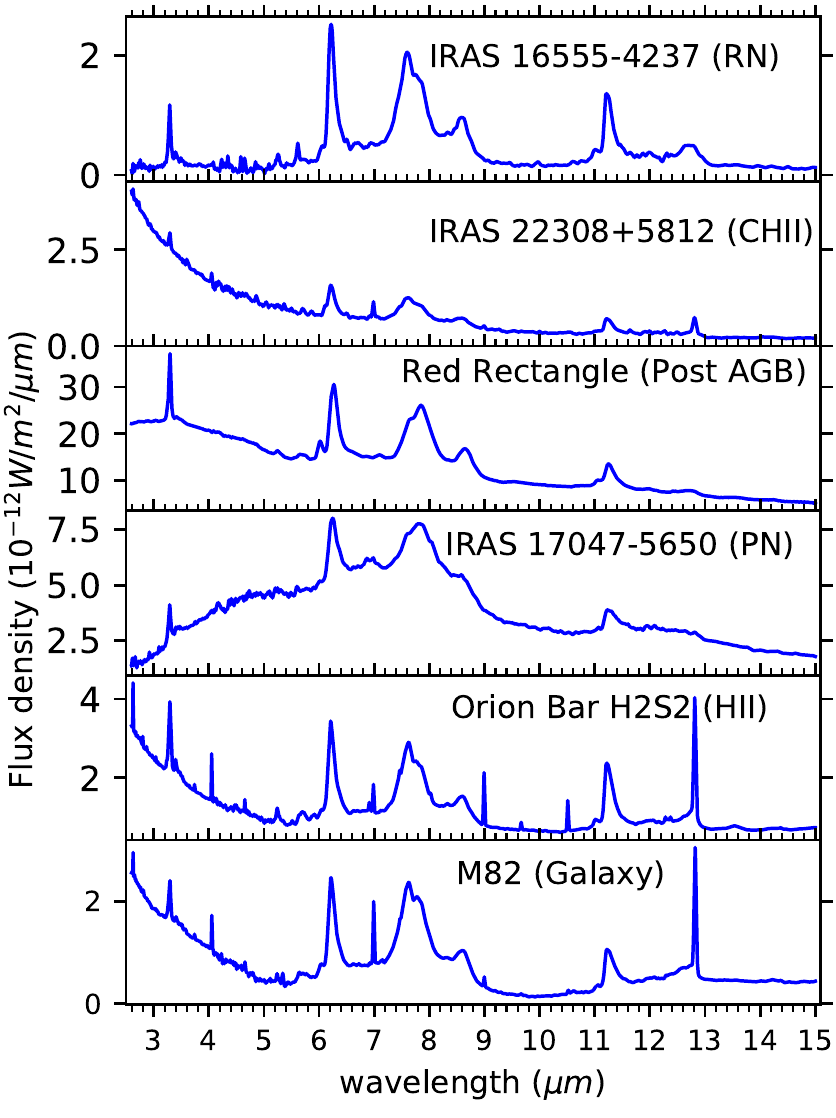}
      \caption{Examples of ISO-SWS spectra used in this study. See Table~\ref{tab_sources} for the complete list of objects.}
         \label{fig_exemplespec}
   \end{figure}

This yields $n=31$ spectra of $m=6799$ points with a spectral resolution constant over each segment, as given in Table~\ref{tab_wavesegments}.
Figure~\ref{fig_exemplespec} shows an example of spectra from each type of object.

\begin{table}
\centering
\begin{tabular}{|c|c|c|c|}
\hline
SWS segment & spectral range & $R$  & $\lambda_{ref}$\\
\hline
1B & 2.6 - 3.02 & 380 & 2.6\\
\hline
1D & 3.02 - 3.52 & 420 & 3.1\\
\hline
1E & 3.52 - 4.08 & 340 & 3.75\\
\hline
2A & 4.08 - 5.30 & 400 & 4.08\\
\hline
2B & 5.30 - 7.00 & 260 & 5.8 \\
\hline
2C & 7.00 - 12.40 & 310 & 7.00\\
\hline 
3A & 12.40 - 16.50 & 320 & 12.9\\
\hline
\end{tabular}
 \caption[]{The wavelength range and lowest resolution in segment are reported for each ISO SWS segment. This is applicable for the observing mode considered in this paper (i.e., SWS 01 scan speed 2).}
         \label{tab_wavesegments}
   \end{table}

\subsubsection{AIB extraction model parameters}
\label{subsubsect_AIBparam}
\begin{figure*}
   \centering
   \includegraphics{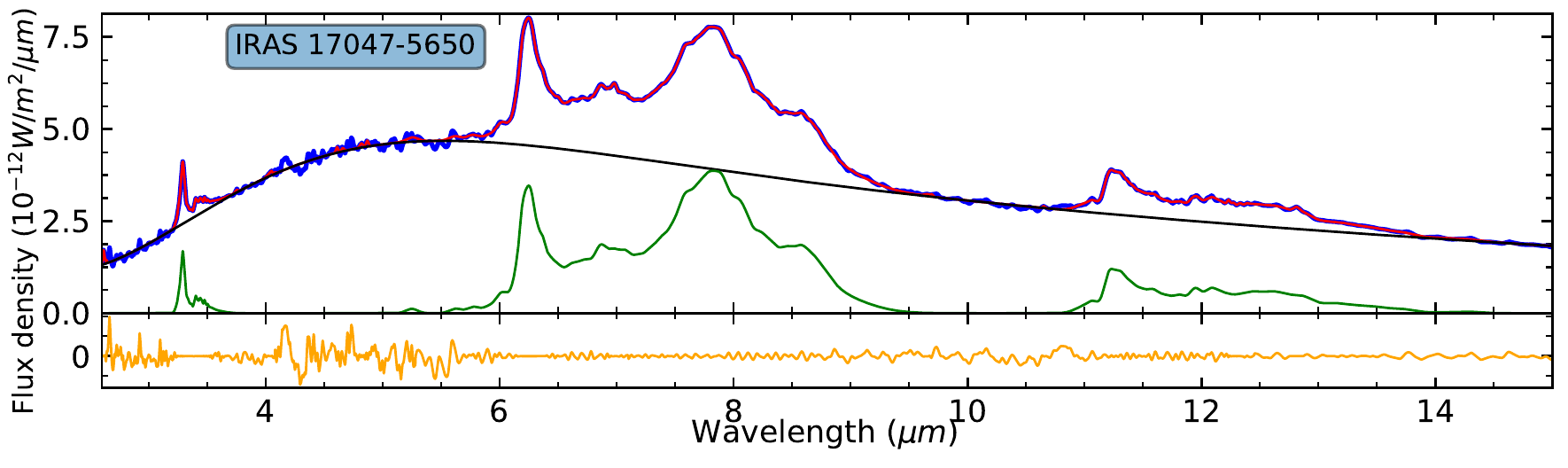}
      \includegraphics{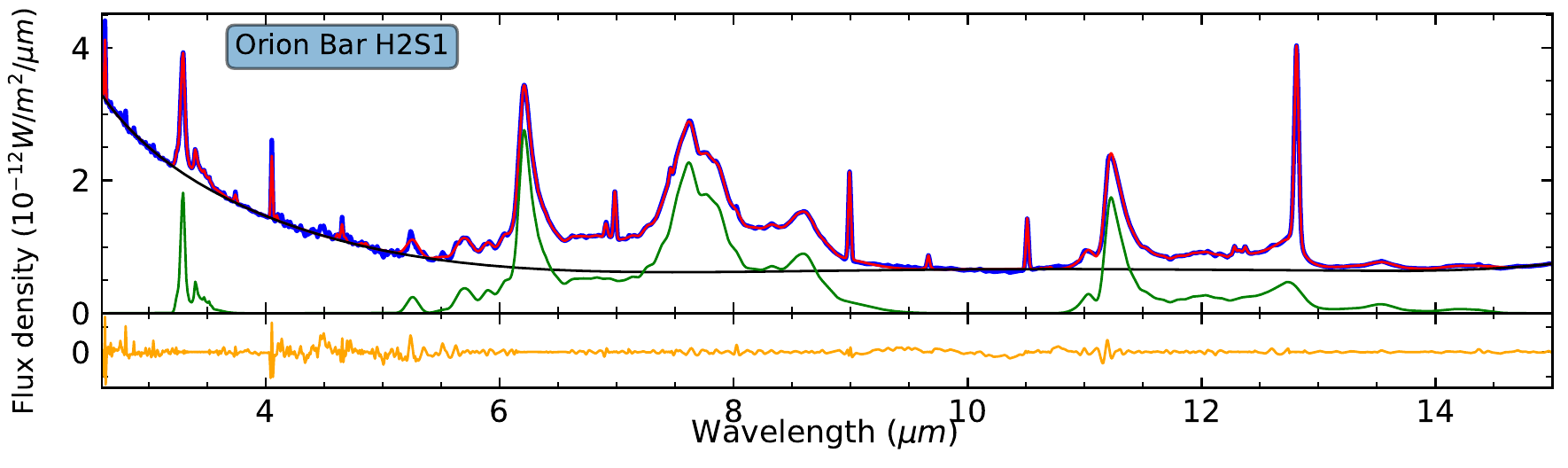}

      \caption{Examples of AIB extraction from mid-IR spectra (green line). The observed spectra are shown in blue, the fit in red, and the continuum in gold. The residual of the fit is shown in orange in the  box below each panel. The rest of the AIB extractions are available in Appendix B.}
         \label{fig_aibextraction}
   \end{figure*}

The parameters of the AIB extraction model discussed in Sect.~\ref{subsect_aibextraction} are presented as follows. Table~\ref{tab_bandcatalog} shows the intervals of wavelengths on which we place the catalog of Gaussians modeling the AIBs. No intervals are considered between 3.55\,$\mu$m and 5.195\,$\mu$m because of the lack of known AIBs in this region. Table~\ref{tab_BBcatalog} shows the characteristics of the modified blackbody catalog. Table~\ref{tab_gascatalog} shows the characteristics of the Gaussian catalog corresponding to the gas lines (i.e., the identification of the transitions and the Gaussian parameters). 
Figure~\ref{fig_aibextraction} shows two examples of extraction of the AIB emission. All our fits are found to be of good quality.
\begin{table}
 \caption[]{List of Gaussian parameters used to fit the AIB spectra.}             
\centering          
\begin{tabular}{c c c c c}     
\hline\hline    
Interval ($\mu$m) & number of Gaussians & $\sigma$ ($\mu$m) \\
\hline
3.25 - 3.55 & 50 & 0.005\\
3.25 - 3.55 & 50 &0.018 \\
5.195 - 5.44 & 50 & 0.095\\
5.195 - 5.44 & 50 &0.05\\
6.15 - 6.4 & 30 & 0.005 \\
5.5 - 6.9 & 40 & 0.05\\
6.91 - 8.8 & 100 & 0.3\\
6.91 - 8.8 & 100 & 0.15\\
6.91 - 8.8 & 100 & 0.08\\
11.0 - 11.7 & 30 & 0.08\\
11.0 - 11.7 & 30 & 0.045\\
11.77 - 12.17 & 20 & 0.1\\
11.77 - 12.17 & 20 & 0.05 \\
12.35 - 12.8 & 9 & 0.1\\
12.85 - 13.15 & 5 & 0.1\\
13.2 - 14.6 & 14 & 0.1\\
\hline            
         \label{tab_bandcatalog}      
\end{tabular}
\end{table}

\begin{table}
 \caption[]{Temperature parameters of  the continuum}             
\centering          
\begin{tabular}{c c c c c }    
\hline\hline       
                      
T$_{min}$ (K)& T$_{max}$ (K)& $\Delta$T (K)& N & total number \\    
\hline                        
   & & & 0 &\\
    40 & 6000 & 20 & 1.6 e22 &897  \\      
   & & & 5e22 &\\
\hline                 
         \label{tab_BBcatalog}
\end{tabular}
\end{table} 

\begin{table}
\caption{List of Gaussian parameters used to fit the gas lines.}             
\centering          
\begin{tabular}{c c c c c}      
\hline\hline       
Identification & $\mu$ ($\mu$m) & $\sigma$ ($\mu$m)  &  $\Delta \lambda$ ($\mu$m) &n \\   
\hline                        
    H I & 2.62 & 0.005 & 0.02 & 15 \\      
   H I & 3.74& 0.009 & 0.02&10  \\
   H I & 4.05 & 0.009 &0.05&20 \\
   H I & 4.65 & 0.009 & 0.04 & 10\\
   H I &4.84 &0.009 &0.04 & 10\\
   H$_2$ S(5) & 6.91 &0.012 &0.02&10\\
\lbrack Ar II \rbrack   &6.97  & 0.009 &0.05 &10\\
    H I  & 7.45 & 0.009 &0.04 &10\\
  \lbrack Ar III\rbrack  &8.99 & 0.009 & 0.06&10\\
 H$_2$ (S3)  & 9.66& 0.009 & 0.04&10\\
 \lbrack S IV\rbrack   & 10.51& 0.009 &0.06 &10\\
 \lbrack Ni II\rbrack   & 10.67 & 0.009 & 0.045&10\\ 
 H$_2$ (S2)  & 12.27 & 0.009 & 0.04&10\\
  H I & 12.36 & 0.009 &0.04 &10\\
   \lbrack Ne II\rbrack   &12.81 & 0.009 & 0.06 &30\\
\hline                  
         \label{tab_gascatalog}
\end{tabular}

\end{table}

\subsubsection{Choice of r}
\label{subsubsect_datachoiceofr}
Several ways to determine  r are discussed in Sect.~\ref{subsubsect_choicer}. 
 The method used by \citet{rapacioli_spectroscopy_2005} and \citet{berne_blind_2012} is not usable here due the impossibility of having 
a correct measure of the noise. 
The  observations were not done with the same instrumental modes, which means the noise values of the spectra are not uniform.

\begin{figure}
  \includegraphics{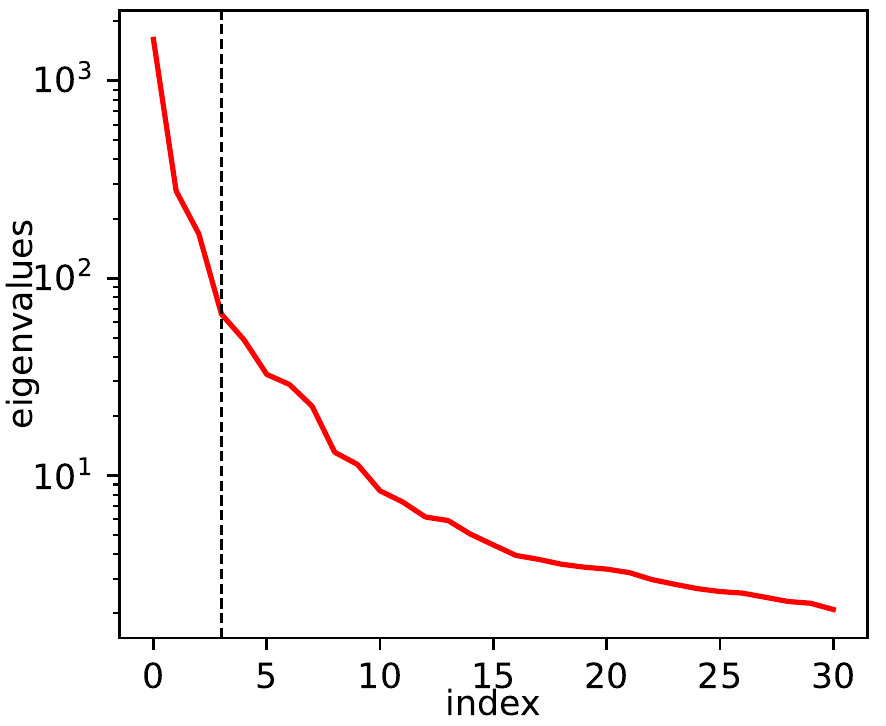}
  \caption{Eigenvalues of the covariance matrix of the AIB spectra (X$_{aib}$). The dashed line corresponds to $r=4$, which is the dimension we  selected. See Sect.~\ref{subsubsect_choicer} for details.}
     \label{fig_svd}
\end{figure}

Figure \ref{fig_svd} shows the eigenvalues of the covariance matrix obtained by applying SVD to the data sample used in the present study. 
No sharp step is observed, which precludes the choice of r by this method. The
\citet{berne_analysis_2007} method is also not successful here, as these artifacts never appear when increasing r. 
Since all of the     methods known to us for  identifying r automatically failed, we relied on previous studies.
\citet{rapacioli_spectroscopy_2005} found $r=4$, with three physical spectra and one spectrum dominated by artifacts. 
\citet{berne_analysis_2007} and \citet{rosenberg_coupled_2011} both found r=3.
However, the following studies show that an additional spectrum was needed (PAH$^x$, see Sect.~\ref{sect_intro}) in particular when PNe are included (which is the case here). 
Thus, $r=4$ is also likely and is used for the following test.

\subsection{Results and performance}
The output of the present method is described in Sect.~\ref{subsect_elementary}. The reconstruction of each spectrum of the data set using the set of repesentative spectra is described in Sect.~\ref{subsect_reconstruction}. In Section~\ref{subsect_total_fraction_mixing}, we discuss the influence of the flux of each spectrum on the output of the method. Finally, we briefly discuss the time cost of this method in the perspective of its application on JWST data sets. 
\subsubsection{Representative spectra}
\label{subsect_elementary}

\begin{figure*}
   \centering
   \includegraphics{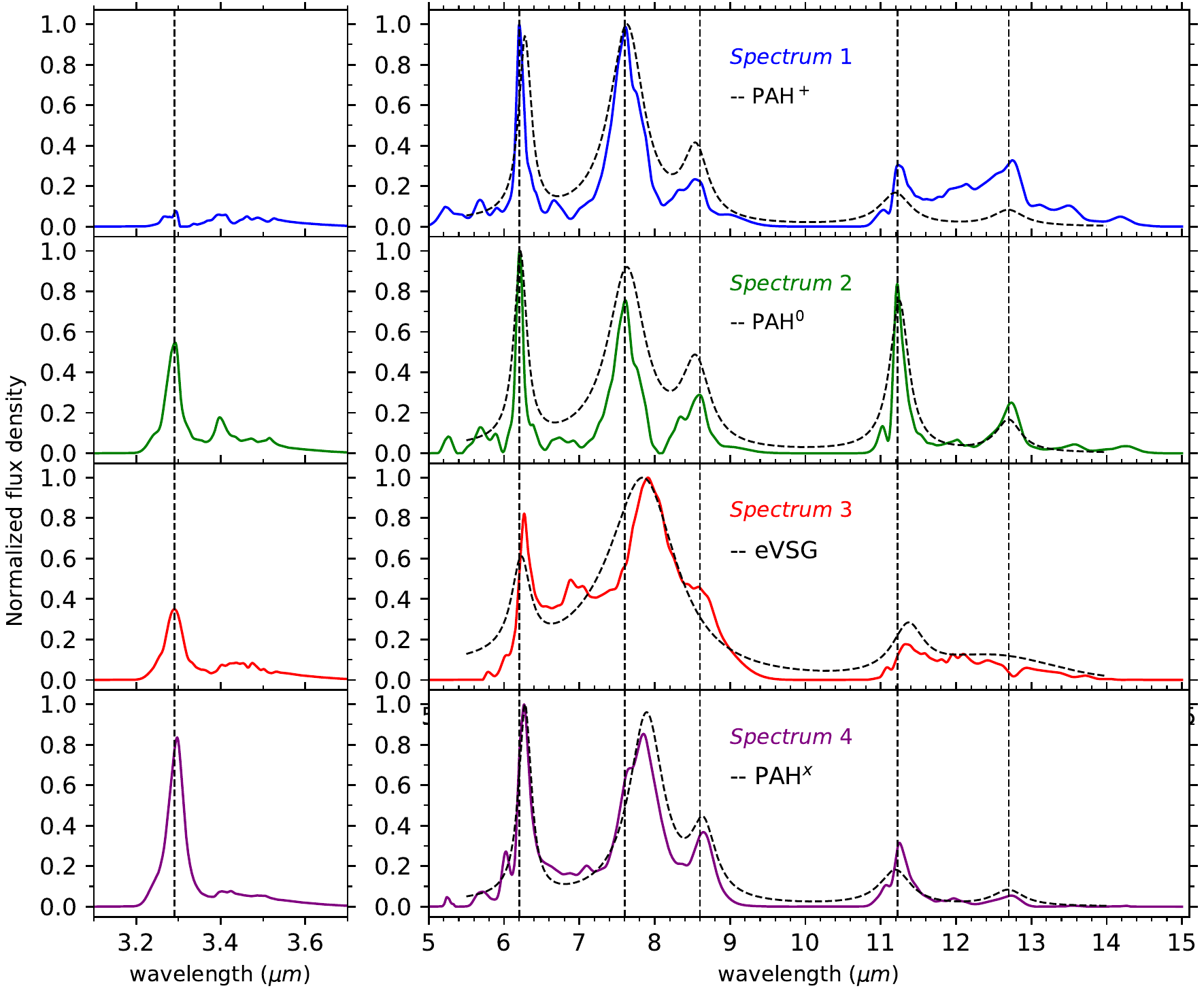}
      \caption{Comparison between PAHTAT templates (dashed lines for  PAH$^+$, PAH$^0$, eVSG, and PAH$^x$; \citet{pilleri_evaporating_2012}) and extracted representative spectra in this work. All spectra are homogeneous to $W/m^2/\mu m$ and normalized to their maximum values. The vertical dashed lines correspond to the positions of the major bands of spectrum 2 (3.29, 6.20, 7.60, 8.60, 11.22, and 12.70\,$\mu$m).}
         \label{fig_templates}
   \end{figure*}

Figure \ref{fig_templates} displays the four representative spectra obtained by applying the method
described in Sect. 3, with $r=4$, on the ISO-SWS data set. The dashed lines in this figure show the template spectra from the PAHTAT fitting tool (\citet{pilleri_evaporating_2012}, see Sect~\ref{sect_intro}).
The first conclusion that can be drawn from this figure is that the extracted spectra using MASS-NMF
applied to the ISO data are qualitatively in good agreement with those of \citet{pilleri_evaporating_2012}, and in particular we recognize 
the four chemical populations presented in this earlier study and ascribed to  cationic PAHs (PAH$^+$), neutral PAHs (PAH$^0$), evaporating very small grains (eVSGs), and PAH$^x$ (i.e., large ionized PAHs). These assignments are discussed in further detail in Sect.~\ref{subsect_assignment}; however, for the sake of simplicity we use this chemical denomination from now on.

\subsubsection{Spectral reconstruction of observations with representative spectra}
\label{subsect_reconstruction}

\begin{figure*}
   \centering
   \includegraphics{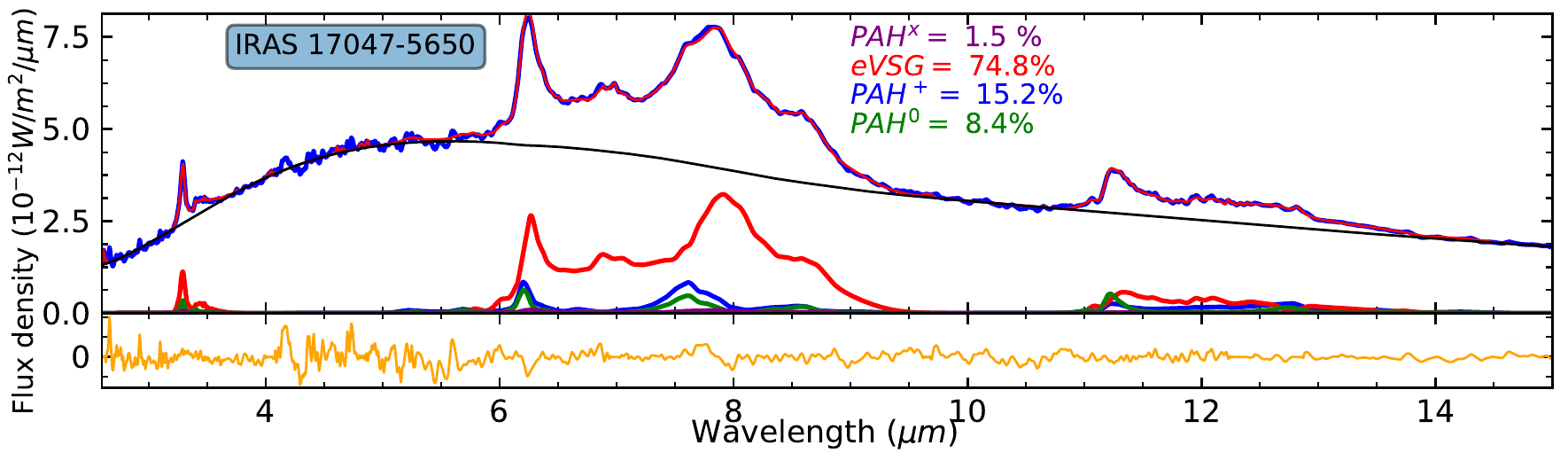}
   \includegraphics{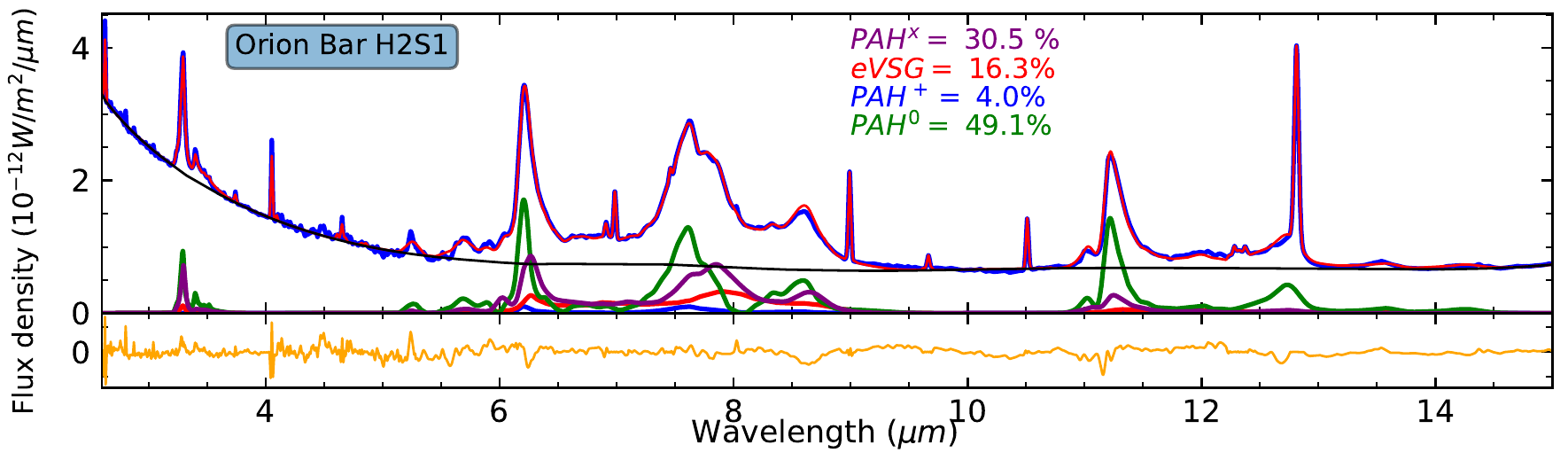}
      \caption{Two examples of spectral reconstruction of IRAS 17047-5650 (upper panel) and Orion Bar H2S1 (bottom panel). The thin black line shows the continuum, blue and red show respectively the data and the model. The four representative spectra for the eVSG, PAH$^+$, PAH$^0$ and PAH$^x$ populations are shown in red, blue, green, and purple for respectively .}
         \label{fig_rec}
   \end{figure*}

One way to assess the quality of the decomposition is to reconstruct the observed spectra using the extracted representative spectra described in the previous section. 
To perform this reconstruction (or fit), we used the algorithm described in Sect.~\ref{subsect_aibextraction} (see Eq.~\ref{eq_optimization}). 
We used the same catalogs of blackbodies and Gaussian functions to model the continuum and gas line emissions; however, the AIBs are now reconstructed using the four representative spectra. 
Figure~\ref{fig_rec} shows two examples of spectral reconstruction. 
The fits for all the objects reported in Table~\ref{tab_sources} are given in Appendix~\ref{appendix_rec}, and the proportions of each spectrum for each reconstruction is given in Table~\ref{tab_proportions}.

\begin{figure}
   \centering
\includegraphics{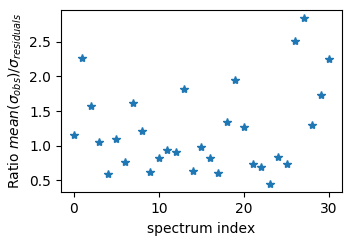}      
\caption{Ratio of the variances  of the residuals to the uncertainties }
         \label{fig_reconstruction_quality}
   \end{figure}

The fits are generally found to be of good quality. 
To estimate this more quantitatively, we compared the variance of the residual to the noise of the observations. A good reconstruction implies that the two quantities are of the same order. However we do not have a reliable measure of the data noise since each spectrum was obtained at a different time and not all of them were recorded with the same setting. 
A good measure of the noise in a hyperspectral image would be from a pixel with no AIB signal, which we do not have here. 
In the absence of an empirical noise estimate, we use the uncertainties provided by the pipeline, as described in \citet{sloan_uniform_2003}. 
These are instrumental uncertainties related to the measurement method, not a measure of the total noise, and therefore likely an underestimate of the total noise. The ratio of the variance of the residuals of the reconstruction to the mean of the instrumental uncertainties is  compared in Fig.~\ref{fig_reconstruction_quality} for each source.
The values are between 0.34 and 2.5, which shows that the variance of the residuals is on the order of the instrumental uncertainty, implying a good quality of the reconstruction.

\citet{joblin_carriers_2008} used the four PAHTAT template spectra reported in Fig.~\ref{fig_templates} 
and two broad features at 8.2 and 12.3\,$\mu$m to fit the 5.5 -- 14\,$\mu$m spectra of a set of PNe and CHII regions after smoothing the observations to the spectral resolution of the templates ($\lambda/\Delta\lambda$=45).
All SWS spectra used in this earlier study are also included in our sample.
We can thus compare the results obtained in each study from the spectral reconstruction 
in terms of the contributions derived for each population (see Table~A.2
of \citet{joblin_carriers_2008} and Table~\ref{tab_proportions}) in Fig.~\ref{fig_relativecommon}. Although differences in detail  can be seen (e.g., a larger contribution of eVSGs in the new study compared to the previous one), an overall agreement is found. This demonstrates that the inclusion of higher resolution data in the decomposition method has preserved the general spectral information in terms of the defined PAH$^0$, PAH$^+$, eVSG, and PAH$^x$ contributions. {However, the representative spectra obtained here contain more spectral information and therefore possibly chemical information, as is discussed in Sect.~\ref{subsect_chemicaldiversity}}.

\subsubsection{Importance of flux in the extraction of the PAH$^x$ spectrum}
\label{subsect_total_fraction_mixing}

The  last line of Table~\ref{tab_proportions} shows the sum of the flux emitted by each population  computed from the reconstruction of the observed spectra presented in Sect~\ref{subsect_reconstruction}. 
The prominent emission of the PAH$^x$ population compared to others is noticeable. This is due to the presence in the sample of the Red Rectangle (RR), whose emission accounts for 55\% of the total PAH$^x$ flux. This source is 12 times stronger in flux than the average of the rest of the sample, and 3 times stronger than the second brightest spectrum (IRAS 07027-7934).
This is important to keep in mind since the version of  NMF used in this paper aims to reduce the Euclidean distance between the data matrix {\bf X} and the estimated {\bf W} and {\bf H} matrices (see Sect.~\ref{subsect_BSS}). 
This implies that the algorithm will provide one vector in the solution that is close 
(in terms of Euclidean distance) to this observed spectrum. This is what happens here 
in the extraction, where the PAH$^x$ spectrum is very similar to the RR  
spectrum. If a normalization of the observed spectra to an integrated flux density 
of one is performed (and the RR  contribution is thus artificially reduced) 
before running NMF, then the population identified as PAH$^x$ in Fig.~\ref{fig_templates}
(spectrum 4) is not extracted by the algorithm. This underlines the fact that the presence 
of the RR in our sample is crucial to the extraction of the PAH$^x$ spectrum. 

\subsubsection{Computation time}
On the ISO-SWS data set of dimensions $(31\times6799)$, the method runs in 10 minutes on a laptop computer. We generated, based on the ISO data and using interpolations, a data set with the dimensions corresponding to an observation of one field of view of NIRSpec and MIRI combined (i.e., $30\times30\times 40 000$) and applied the proposed method to it. We find that the computation time on a server is on the order of one hour.

\section{Towards a chemical interpretation of extracted spectra}
\label{sect_discussion}

In this section we briefly discuss the characteristics of each representative spectrum.
We compare them to spectra obtained in earlier studies to propose a possible chemical interpretation of these spectra.
This is done to illustrate the methodology, thanks to the   global similarities we observed  in the representative spectra in the studies and to the second-order variations due to the specificity of the  data set used. These assignments should not be considered  definitive, in particular given the limited size of the sample and the absence of spatial information. Hopefully, the use of JWST data will allow us to achieve a more secure chemical assignment thanks to larger samples and spatially resolved observations. We also discuss how the classes of 
\citet{peeters_rich_2002} can be connected to these extracted spectra.  In Sect.~\ref{subsect_chemicaldiversity}, we present some considerations on chemical diversity and on the possibilities that a MASS-NMF analysis can open 
when dealing with the  data sets that will be provided by JWST.

\subsection{Chemical assignment of extracted spectra and their connection to standard classes}
\label{subsect_assignment}

Improvements in the extracted spectra can be clearly  seen in Fig.~\ref{fig_templates}  compared to the spectra presented in \citet{pilleri_evaporating_2012}. These concern both the wavelength coverage (i.e., inclusion of the 3\,$\mu$m region) and the spectroscopic details that are present in the new spectra. 
Table \ref{tab_bandscharac} lists the positions of the main detected bands, and of some more minor bands where     we are confident that they are associated with real bands and   not artifacts of the method.  Some bands are well known, such as the satellite bands in the 3.4-3.5 \,$\mu$m range. Other bands are identified because they are observed  in several representative spectra, at close positions.
The positions of the major bands observed in spectrum 2 (3.29, 6.20, 7.61, 8.60, 11.22, and 12.73\,$\mu$m) are used as arbitrary reference positions and are indicated by vertical dashed lines in Fig.\, \ref{fig_templates}. Overall, these results demonstrate that an application of the MASS-NMF method 
to the JWST data sets is promising, and likely to provide a scheme of analysis providing unprecedented detail. We describe here the spectral characteristics of the representative spectra and compare them to earlier studies to interpret their chemical origin.

    {\bf Spectrum 1: PAH cations.} Spectrum 1 (Fig. \ref{fig_templates}) is dominated by two bands centered at 6.20\,$\mu m$ and 7.61\,$\mu m$. The other major bands peak at 8.53, 
    11.24, and 12.75\,$\mu m$.  
    Spectrum 1 shares some characteristics with spectrum 2, but has a higher value of the band intensity ratio of the 7.7\,$\mu$m complex (C-C stretch) to the 11.20\,$\mu$m band (C-H out-of-plane bend). Another major difference concerns the 3\,$\mu$m region where spectrum 1 has no significant bands, contrary to spectrum 2.
    All these characteristics point to PAH cations ($PAH^+$) being the carriers of spectrum 1, in line with experimental and theoretical studies \citep[e.g.,][]{defrees93_cation,Szczepanski_Vala_93_cation, Hudgins_Allamandola_99_cations} and the study by \citet {boersma_properties_2013} using the NASA Ames PAH database.
    We note that the 6.20\,$\mu m$ band is blueshifted compared to the PAHTAT template, 6.20\,$\mu$m relative to 6.27\,$\mu$m. However, the position of this peak varies in the various BSS extractions of the PAH$^+$ spectrum; for instance, it is reported at 6.28\,$\mu$m in \citet{rapacioli_spectroscopy_2005} and 6.24\,$\mu$m in the extraction performed by \citet{berne_analysis_2007}.  However, the shift of 0.08\,$\mu$m in previous templates relative to spectrum 1 is less than the spectral resolution of the ISOCAM data ($\sim$ 0.14\,$\mu$m) that was used in these studies.
    Our results can also be compared to the study of \citet{rosenberg_coupled_2011} on NGC\,7023.
    The authors applied NMF to {\it Spitzer} spectra of NGC\,7023 obtained over the 10-15 $\mu$m range at $R~\sim 600$. They found a PAH cation spectrum dominated by a band at 11.00\,$\mu$m, albeit with relatively large uncertainties (see lower left panel of their Fig. 3). A band at 11.0\,$\mu$m is indeed present in spectrum 1, although the main one is located at 11.24 $\mu$m. This indicates that the representative spectra that can be extracted depend on the considered spectral range and objects
    (see discussion on chemical diversity in Sect.~\ref{subsect_chemicaldiversity}). 
    \medskip

     {\bf Spectrum 2: PAH neutrals}. This spectrum presents clear peaks at 3.29, 6.20, 7.61, 8.59, 11.22, and 12.73\,$\mu$m. In addition to the 3.29\,$\mu$m band, there is a weaker satellite band at 3.40\,$\mu$m over a plateau, which extends from 3.4 to 3.6\,$\mu$m and comprises very weak bands at 3.43, 3.47, and 3.51\,$\mu$m. The 11.22 band is much stronger than in spectrum 1, being close to the 7.7$\,\mu$m complex emission, which points to neutral species ($PAH^0$) being the dominant carriers \citep{compiegne_aromatic_2007,Joblin96_ngc1333}. This is also consistent with the previous decomposition studies \citep{rapacioli_spectroscopy_2005,berne_analysis_2007, pilleri_evaporating_2012}. The position of the 11.2\,$\mu$m C-H band observed here (11.22\,$\mu$m) is in good agreement with that of the neutral PAH spectrum (11.20\,$\mu$m) derived by \citet{rosenberg_coupled_2011}. 
     
    \medskip

    {\bf Spectrum 3: eVSGs}. This spectrum exhibits bands that are redshifted compared to the bands of the $PAH^+$ and $PAH^0$ spectra and are found to be superimposed on a plateau (Fig.~\ref{fig_templates}).  
     Although there are notable differences, in particular in the  7.7\,$\mu$m complex, we associate this spectrum with the population of  eVSGs  described in \citet{pilleri_evaporating_2012}.
    This population spectrum has a dominant contribution in the PNe spectra.
     We note in addition that the position of the 11.20\,$\mu$m band is shifted to $\sim$11.35\,$\mu$m, which is similar to the value of the position observed for this band in the VSG spectrum extracted by \citet{rosenberg_coupled_2011}, i.e., $\sim$ 11.40\,$\mu$m (see their fig. A1).  
     \medskip

 {\bf Spectrum 4: PAH$^x$?} Spectrum 4 shares some similarities with the $PAH^+$ spectrum (spectrum 1),  but with a significant redshift of the main bands corresponding to the C-C stretch and C-H in-plane bend motions. 
 The shift is the strongest for the band at 7.85\,$\mu$m,  compared with a position at 7.61\,$\mu$m in the $PAH^+$ and $PAH^0$ spectra. 
 This is the characteristic put forward by  \citet{joblin_carriers_2008} for the so-called PAH$^x$, which would be large ionized PAHs (cations or anions of more than 100 carbon atoms). 
 We note that the latter authors had constructed the PAH$^x$ spectrum, whereas here this spectrum results from the extraction procedure. 
 We found the presence of a strong 3.3\,$\mu m$ band associated with the PAH$^x$ component, also slightly redshifted compared to the band of $PAH^0$.  
 However, we should  consider  that the extraction of the PAH$^x$ spectrum is only possible due to the presence of the intense RR spectrum in the data set (see Sect.~\ref{subsect_total_fraction_mixing}). Its extraction is promising but still to be confirmed on a sample containing more RR-like objects.

 \medskip

Finally, although the spectrum of every object is a mix of the four representative spectra, we note that the class A objects of \citet{peeters_rich_2002} are dominated by the PAH$^0$ and PAH$^+$ populations, whereas the class B objects are dominated by eVSG and PAH$^x$ populations.

\begin{table*}
\caption{Relative contribution of each representative spectrum in the data sample}             
\centering          
\begin{tabular}{ c c c c c c c c c  } 
\hline\hline       
Object &type &class\tablefootmark{(a)} &Spec. 1 & Spec. 2 & Spec. 3 & Spec. 4 &Total AIB flux&R$_{AIB-C}$\tablefootmark{(b)}\\    
\hline                
   &&& $PAH^+$ & $PAH^0$ & $eVSG$ & $PAH^x$& 10$^{-12}$W/m$^2$&\\
   \hline
M17 2&HII&A&0.41& 0.18& 0.28& 0.12& 2.91& 0.14\\
M17 3&HII&A&0.41 & 0.19& 0.23& 0.17& 2.58&0.18 \\
M17 5&HII&A&0.55& 0.06& 0.25& 0.15& 2.28&0.26 \\
M17 6&HII&A&0.35& 0.20& 0.31& 0.14& 0.93&0.66  \\
M17 8&HII&A&0.30  & 0.28& 0.32& 0.09& 0.69& 0.14\\
Orion Bar H2S1&HII&A&0.04& 0.49& 0.16& 0.31& 3.95&0.36 \\
Orion Bar BRGA&HII&A&0.15  & 0.45 & 0.01& 0.39& 3.22& 0.24\\
Orion Bar D5 (2)&HII&A&0.15& 0.43& 0.09& 0.33& 5.76&0.26 \\
Orion Bar D5 (3)&HII&A&0.13& 0.44& 0.09& 0.34& 5.33&0.43 \\
S106 &HII&A&0.36& 0.22& 0.18& 0.25&3.52&0.16 \\
G 327.3 - 0.5&HII&A&0.62& 0.02& 0.10& 0.26& 4.17& 0.15\\
IRAS 15384-5348&HII&A&0.50& 0.11& 0.19& 0.21& 2.79&0.26  \\
BD +30 3639 &PN&B&0.00& 0.21 & 0.61& 0.18&4.08&0.33 \\
He2 113&PN&B&0.00& 0.09& 0.70& 0.22& 4.34&0.28 \\
IRAS 07027-7934&PN&B&0.00& 0.00& 0.87& 0.13& 1.37&0.16   \\
IRAS 17047-5650&PN&B&0.15& 0.08& 0.75& 0.02& 7.88&0.20 \\
M82&Starburst &A&0.40& 0.11& 0.36& 0.13& 3.76&0.67 \\
NGC7023&RN&A&0.19& 0.38& 0.23& 0.20& 1.05& 0.80\\
IRAS 16555-4237&RN&A&0.26 & 0.33  & 0.18& 0.24&3.06& 1.73\\
IRAS 03260+3111&RN&A&0.23& 0.32& 0.28& 0.17& 2.36&1.04\\
IRAS 18317-0757&CHII&A&0.55 & 0.02& 0.16& 0.27&2.71&0.19\\
IRAS 23030+5958&CHII&A&0.24& 0.38 & 0.00& 0.38& 0.56&0.20 \\
IRAS 19442+2427&CHII&A&0.56& 0.01& 0.36& 0.08 &2.20&0.74 \\
IRAS 22308+5812&CHII&A&0.31& 0.27& 0.19& 0.23& 1.16& 0.13\\
IRAS 18032-2032&CHII&A&0.61& 0.00& 0.22& 0.17& 1.68&0.78 \\
W3A02219+6125&CHII&A&0.51& 0.20& 0.11& 0.18& 3.60& 0.18\\
IRAS 18502+0051&CHII&A&0.60& 0.00& 0.15& 0.25 & 1.09&0.13 \\
IRAS 17279-3350&CHII&A&0.60& 0.00& 0.28& 0.12&1.08&0.12  \\
IRAS 23133+6050&CHII&A&0.27& 0.30& 0.08& 0.35& 2.37&0.62 \\
IRAS 12063-6259&CHII&A&0.39& 0.20 & 0.16& 0.25&1.85&0.47 \\
Red Rectangle&Post AGB&B&0.00& 0.00 & 0.04& 0.96&22.63&0.15 \\
\hline
Mean contribution&&&0.32& 0.19& 0.26&  0.23&\\
\hline              
Sum flux&10$^{-12}$W/m$^2$&& 24.85 & 17.79 & 25.62 & 39.84\\
\hline
         \label{tab_proportions}
    
\end{tabular}
\tablefoot{
\tablefoottext{a}{From \citet{peeters_rich_2002}.}
\tablefoottext{b}{Represents the ratio of the total AIB flux of an astrophysical object to the continuum flux of the same object.}}

\end{table*}

\subsection{Chemical diversity}
\label{subsect_chemicaldiversity}

The improved spectral resolution and larger spectral range of our representative spectra can provide new information on the emitting populations.
Table \ref{tab_bandscharac} reports the positions of the main detected bands and of some minor bands where we are confident that they are associated with real bands and not artifacts of the method. In the following we more specifically focus on the differences that can be noted for a given type of spectral feature in the representative spectra.

We first consider the 3-4\,$\mu$m range, which is studied here for the first time using BSS methods. Figure~\ref{fig_templates} shows that all populations except the carriers of spectrum 1 have a contribution in this spectral range. For PAHs, it is known that ionization induces a collapse of the intensity of the C-H stretch \citep{pauzat_ionisation_1995}. The effect decreases with increasing size. For large species containing 100 carbons or more, the 3.30\,$\mu$m band has a comparable intensity for neutrals and cations \citep{bauschlicher_2009}. On the other hand, the 3.30\,$\mu$m band is in general more intense for anions than for neutrals, as shown by density functional theory calculations, which are gathered in the Theoretical PAH database\footnote{http://astrochemistry.oa-cagliari.inaf.it/database/} and the NASA Ames PAH IR Spectroscopic Database\footnote{www.astrochemistry.org/pahdb/}.
These results strengthen the assignment we made for spectra 1, 2, and 4. Spectrum 1 is dominated by PAH cations, whereas spectrum 2 is dominated by neutrals. The presence of a strong 3.30\,$\mu$m band in spectrum 4 supports the idea that their carriers, some large ionized PAHs, are anions rather than cations, whereas both species were proposed in \citet{joblin_carriers_2008}. Even so, the new data raise a number of questions. The  absence of the 3.30\,$\mu$m band in spectrum 1 would go against the fact that the carriers are large PAH cations, unless they are not sufficiently excited to emit at 3.30\,$\mu$m. The fact that spectrum 3 attributed to eVSGs exhibits a 3.30\,$\mu$m band is also surprising considering that eVSGs are expected to be larger species and to then emit at lower temperatures than PAHs \citep{rapacioli_spectroscopy_2005}. It  should be kept in mind, however,  that the retrieved representative spectra reflect the dominant contributions in the sample of spectra and do not have the ability to separate chemical populations that are present at an intensity that is too low compared to other populations. Spectrum 3 may therefore contain a mix of eVSGs and smaller PAH-like species that could be produced by their evaporation \citep{pilleri_mixed_2015}.

In addition to the 3.30\,$\mu m$ band, spectrum 2 also exhibits the 3.40\,$\mu m$ band, 
as well as minor bands at 3.43, 3.47, and 3.51\,$\mu$m. All these bands have been seen in several sources  \citep[e.g.,][]{pilleri_mixed_2015,Sloan_orion_33_97,ohsawa_akari_2016}. They have been attributed to aliphatic bonds in methyl side groups attached to PAHs \citep{Joblin96_RN_33and34variation} or in superhydrogenated PAHs \citep{bernstein_1996}. These species are expected to have another characteristic band at $\sim$6.9\,$\mu m$ \citep{steglich_2013}, which is indeed present in the PAH$^0$ spectrum. There is also a significant band at 6.9\,$\mu$m in the eVSG spectrum, which might be related to the aliphatic content  of this population \citep{pilleri_mixed_2015}.
For spectra 1, 3, and 4, no band at 3.40\,$\mu$m is detected, but a broad  plateau centered around 3.45\,$\mu m$ is present. \citet{pilleri_mixed_2015} detected this plateau in NGC\,7023 and concluded from a spatial analysis that it is related to the emission from hot bands and combination bands involving aromatic bonds.

We  see in Sect. \ref{subsect_elementary} that the 6.20\,$\mu$m band peak position varies significantly, from 6.20 to 6.27\,$\mu$m, between spectra 1, 2, and 3, 4.  In addition, spectrum 4 has a satellite band at 6.00\,$\mu$m. This feature has been observed in a number of objects and was previously attributed to a C=O stretching mode in oxygenated PAHs \citep{peeters_rich_2002}, C-C mode in complexes of PAHs with Si atoms \citep{joalland_2009} or C=C mode from an olefinic side group \citep{hsia_6_2016}.
The range of the 7.70\,$\mu$m complex indicates some diversity between the populations, although the PAH$^0$ and PAH$^+$ spectra appear relatively close with common bands at 7.61, 7.52/7.53, and 7.76/7.77.

There are significant differences in the 11-15\,$\mu$m range within the PAH spectra. The band at 11.20\,$\mu$m is the dominant one followed by the 12.70\,$\mu$m band. However the 11.20\,$\mu$m band is much more intense in spectrum 2 (PAH$^0$) than in  spectrum 1 (PAH$^+$), where  it is accompanied by other features at $\sim$12\,$\mu$m. This would indicate more chemical diversity in the PAH$^+$ population. 

To conclude, we can see that the representative spectra carry some information on the chemical diversity involved in each population. 
Even so, our sample is limited and rather heterogeneous. 
We therefore have  to be cautious with any detailed band assignment. More robust information will be obtained while analyzing data obtained at high spatial resolution on specific regions in which the chemical evolution of the populations is driven by the local physical conditions such as the UV radiation field. Our four template spectra are consistent with previous studies showing the importance of UV processing.
Among the PAH populations, the PAH$^0$ population is the least processed, which is illustrated by the presence of more fragile aliphatic side groups.
The PAH$^+$ population is expected to be more processed and ionized. 
It is possible that the trends  observed in the 3.30 and 11-13\,$\mu$m ranges (see above) reveal some first steps of dehydrogenation, although not much data are available to support this scenario \citep{pauzat_ionisation_1995}.
The PAH$^x$ population indicates the presence of large PAH anions. 
The spectrum of eVSGs is more difficult to comment and this also reflects our poor knowledge on the nature of this population.

\section{Conclusions}
\label{conclusion}
In this paper, we present a new linear (thus fast) and robust method for the analysis of the AIB spectra in preparation of the large data sets that will be provided by the JWST. 
After an automatic extraction of the AIBs from each observed spectrum (see Sect.~\ref{subsect_aibextraction}), a hybrid blind signal separation method (see Sect~\ref{subsect_BSS}), i.e., NMF initialized using the MASS algorithm (see Sect.~\ref{subsect_initialization}), is applied on the extracted AIBs, 
and provides a set of  representative spectra (here, $r=4$).
The method reduces the dimensionality of the study of a data set to 
that of a reduced number of representative spectra, physically and chemically interpretable in terms of populations 
of PAHs and VSGs.
The speed of the method allows us to analyze large sets of data in a limited time 
compared to non-linear methods or hybrid Monte Carlo--NMF methods. 
As a test, we used ISO-SWS archival data (see Sect.~\ref{subsubsect_isodata}) which most closely  
approached  the spectral range and resolution of JWST data. Four representative  
spectra shown in Fig.~\ref{fig_templates} were extracted, and they demonstrate that the method is
applicable. Based on comparison with previous studies, we propose  the following interpretation:

\begin{itemize}
    \item  Three representative spectra were assigned to $PAH^0$, $PAH^+$, and $eVSG$ spectra, in agreement with previous studies.
    \item We attributed the fourth spectrum to the $PAH^x$ population, initially invoked and constructed by \citet{joblin_carriers_2008} and extracted here automatically for the first time. However, this spectrum should  be confirmed on a larger sample because of its specific behavior.
    \item We found that the spectral classes of \citet{peeters_rich_2002} can be explained by a dominant PAH$^0$ and PAH$^+$ emission for class A, and by eVSG and PAH$^x$ emission for class B.
    \item The 3\,$\mu$m range was included for the first time in such a study, and reveals the presence of aliphatics connected to neutral PAHs, and that the anionic character of the PAH$^x$ population is supported by this new analysis.
    \item Second-order spectral signatures (e.g., small bands) have been identified, showing that more chemical diversity can be investigated thanks to an improved spectral resolution. This list should however be taken   with caution, considering the combined small size and heterogeneity of the sample of astronomical spectra used in this study.
\end{itemize}

The method presented in this study is readily applicable to JWST hyperspectral images that will be provided by MIRI and NIRSpec IFUs, and we expect that it will allow us to refine the scenarios of PAH formation, evolution, and destruction in galaxies.

\begin{acknowledgement}
We thank the referee for his/her comments which helped to significantly improve the clarity of the manuscript. We want to thank also Paolo Pilleri for his help during the beginning of this work, and for his scientific comments and his technical support.
This work was supported by the French Programme Physique et Chimie du Milieu Interstellaire (PCMI) funded by the Conseil  National  de  la  Recherche  Scientifique  (CNRS)  and Centre National d'\'etudes Spatiales (CNES). The research leading to these results has also received funding from the European Research Council under the European Union Seventh Framework Programme (FP/2007-2013) ERC-2013-SyG, Grant Agreement No. 610256 NANOCOSMOS.

\end{acknowledgement}

\bibliography{biblio}
\bibliographystyle{aa}

\begin{appendix}

\section{Summary of MIRI and NIRSpec  characteristics}
Table~\ref{Table_JWSTPerf} summarizes the principal characteristics of the instruments NIRSpec and MIRI on board the JWST.\\
\begin{table}[!h]
\caption{Performance of the JWST.}             
\centering          
\begin{tabular}{ c c c } 
\hline\hline  
&NIRSpec&MIRI\\
\hline
$\lambda (\mu$m) & 0.6-5.0&4.88-28.55\\
\hline
$R=\frac{\Delta\lambda}{\lambda}$&$\sim2700$&$\sim 3000$\\
\hline
Spaxel size & 0.1" & 0.196-0.387"\\
\hline
Sensitivity & $\sim 45~\mu Jy$\tablefootmark{(a)} & $\sim 70~\mu$Jy\tablefootmark{(b)}\\

  \label{Table_JWSTPerf}

\end{tabular}
\tablefoot{Sensitivities are calculated in order to have a 10$\sigma$ detection with a 10k seconds exposure 
for MIRI and 13 
\tablefoottext{a}{At 3.30\,$\mu$m with a  G395H-F290LP  disperser-filter configuration (highest spectral resolution).}
\tablefoottext{b}{At 7.70\,$\mu$m with MRS mode (highest spectral resolution).}

}
\end{table}

\section{Summary of the principal bands in population spectra}
Table~\ref{tab_bandscharac} gives the position and the assignment of the bands extracted from the different representative spectra. They are classified as main peaks, minor bands, and shoulders, which are bands blended with a nearby main band.

\begin{sidewaystable*}
\caption{List of bands observed in the template spectra with their plausible assignments. }
\begin{center}
    
\begin{tabular}{c c c c c c }     
\hline\hline       
Band designation ($\mu$m) & Assignments & PAH$^+$& PAH$^0$ & eVSG & PAH$^x$ \\    
\hline                        
\multicolumn{6}{c}{Main peaks}\\
\hline

3.3 & Aromatic C-H stretch & - & 3.29 & 3.29 & 3.30\\      
6.2 & Aromatic C-C stretch  & 6.20 & 6.20 & 6.27 & 6.27\\
7.7 complex &  Aromatic C-C stretch & 7.61 & 7.61  & 7.91  & 7.85 \\
8.6 & Aromatic C-H in-plane bend & {\it 8.56} (8.52,8.60) & 8.60 & 8.58  & 8.65 \\
11.2 & Aromatic C-H out-of-plane bend & 11.24 & 11.22 & {\it 11.35} (11.32,11.38)& 11.25 \\
12.7 & Aromatic C-H out-of-plane bend & 12.75 & 12.73 & - & 12.73\\
\hline                        
\multicolumn{6}{c}{Shoulders}\\
\hline
3.3 & Aromatic C-H stretch &  & 3.24 &  & 3.24 \\ 
6.2 & Aromatic C-C stretch  &  &  &  & 6.22 \\
7.7 complex & Aromatic C-C stretch  & 7.52,7.76,7.88 & 7.53,7.77 & 7.72,8.06 & 7.65 \\
8.6 & Aromatic C-H in-plane bend &  & 8.53 &  & \\
12.5 & Aromatic C-H out-of-plane bend & 12.53 & 12.50 & - & 12.50 \\
\hline                        
\multicolumn{6}{c}{Minor bands}\\
\hline
3.4 & Aliphatic C-H stretch & - &  3.40, 3.43  & [3.38-3.52] & [3.38-3.52] \\
3.5 & Aliphatic C-H stretch, Combination bands & - & 3.47 & - & -\\
3.5 & Aliphatic C-H stretch, Combination bands & - & 3.51 & - & - \\
5.2 & Combination bands/aromatic & 5.23 & 5.26 & - & 5.24\\
5.7 & Combination bands/aromatic  & 5.68 & 5.69 & 5.79 & 5.70\\
6.0 & - & - & - & 6.02 & 6.02 \\
6.7 & Aromatic C-C stretch? & 6.66 & 6.73 & - & 6.62\\
6.9 & Aliphatic C-H deformation? & - & 6.92 & 6.88 & 6.87\\
7.0 & ?& -&-&7.04& 7.09\\
8.3 & Aromatic C-H in-plane bend? & 8.32 & 8.34  & 8.45 & 8.35\\
11.0 & Aromatic C-H out-of-plane bend & 11.03 & 11.02 & 11.09 & 11.07\\
11.5 & Aromatic C-H out-of-plane bend & - & - & 11.58,11.81 & -\\
12.0 & Aromatic C-H out-of-plane bend & 11.92 & - &  11.96 & 11.97 \\
12.0 & Aromatic C-H out-of-plane bend & 12.13 & - & 12.09 & - \\
12.4 & Aromatic C-H out-of-plane bend & - & - & 12.42 & - \\
13.5 & Aromatic C-H out-of-plane bend & 13.50 & 13.56 & 13.33,13.71 & 13.56\\
14.2 & Aromatic C-H out-of-plane bend & 14.18 & 14.25 & - & 14.25\\

\hline                        
\multicolumn{6}{c}{Plateaus}\\
\hline
3.20-3.60&&-&med&med& weak\\
6-9&&-&-&strong&med\\
11-13.50&&med&-&med& weak\\

\hline    
   \label{tab_bandscharac}
         
\end{tabular}
\end{center}

\tablefoot{Values in $italics$ correspond to mean value of complexes of two bands whose estimated peak positions are shown in parentheses. The sign ``-'' is used when no significant band could be extracted.}
\end{sidewaystable*}

\section{Comparison with Joblin et al. (2008)}
\label{appendix_comp_Joblin_thisstudy}
We present in this section the figure comparing the results from \citet{joblin_carriers_2008} and those from this work for common objects in both studies; see the text in Sect.~\ref{subsect_reconstruction} for details.

      \begin{figure*}
   \centering
\includegraphics[scale=0.8]{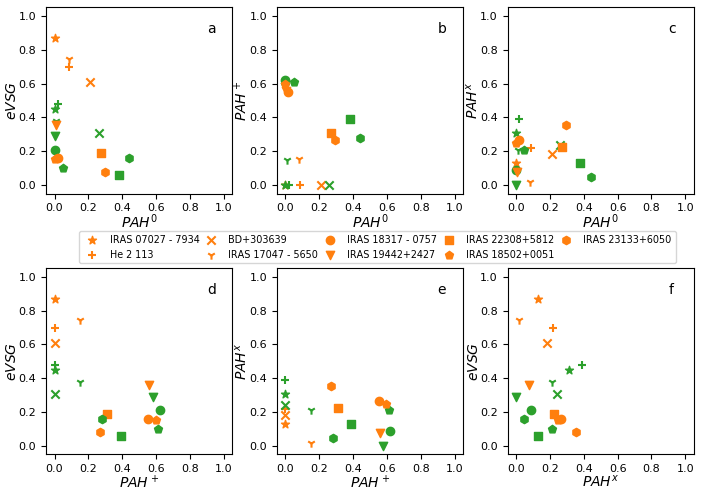}      
\caption{Comparison of the relative populations in the astrophysical objects in common between the study by \citet{joblin_carriers_2008} (in green) and this paper (in orange). 
The middle box shows the symbol attribution for each object.
The contributions from \citet{joblin_carriers_2008} have been normalized to the sum of the PAHs and eVSG contributions.}
         \label{fig_relativecommon}
   \end{figure*}

\section{Extraction of AIB emission from all observed spectra}
\label{appendix_rec}
All figures in this section show the extraction of AIB emission from all the observed spectra using the method described in Sect.~\ref{subsect_aibextraction}.

\begin{figure*}
\centering
\includegraphics{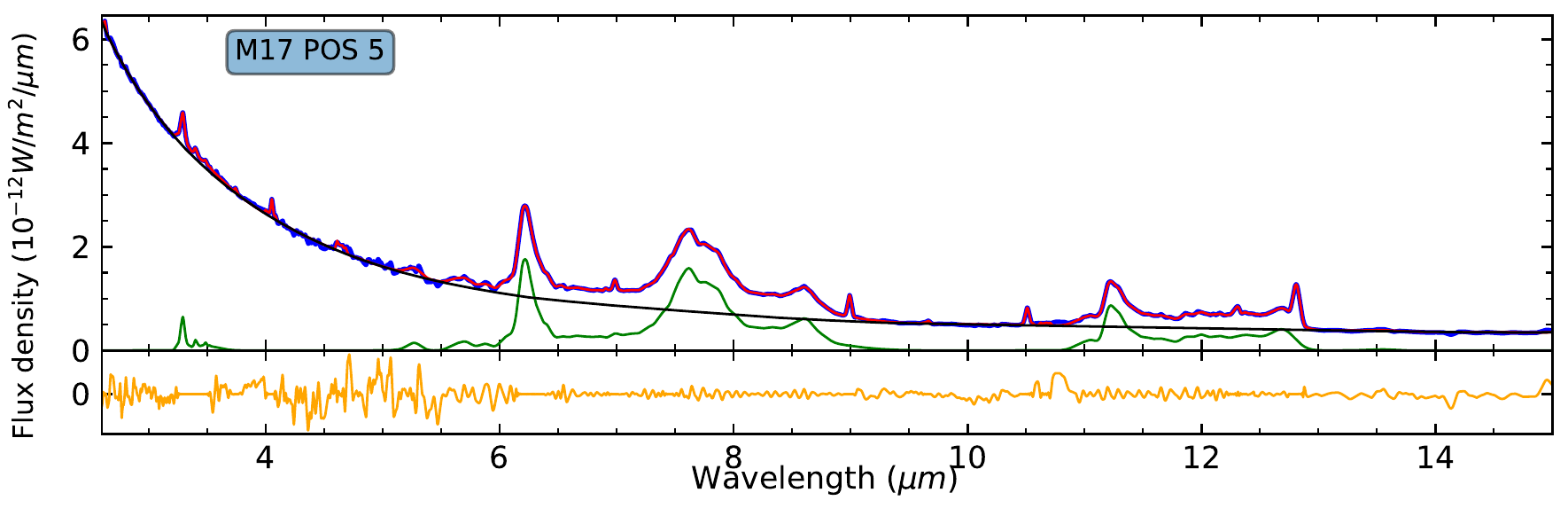}      
\includegraphics{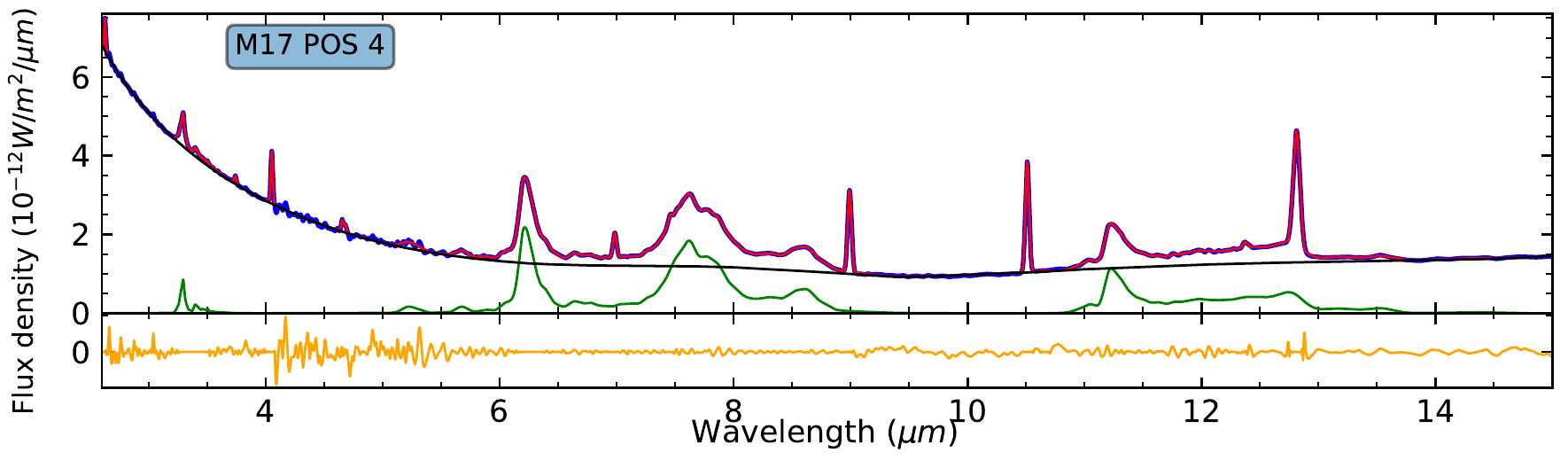}      
\includegraphics{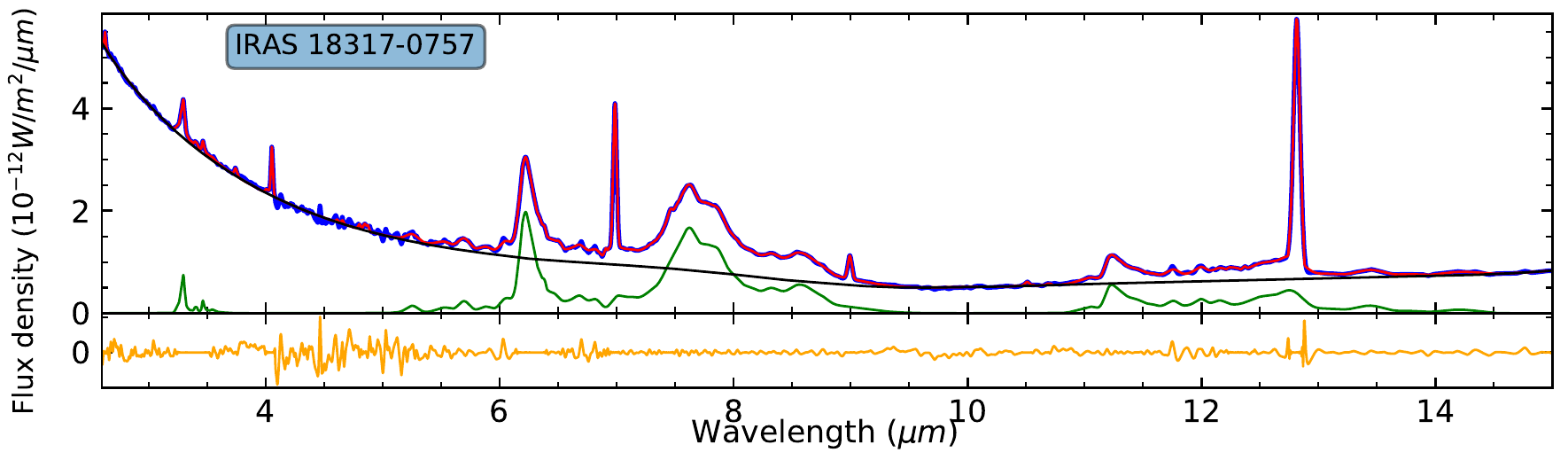}      
\includegraphics{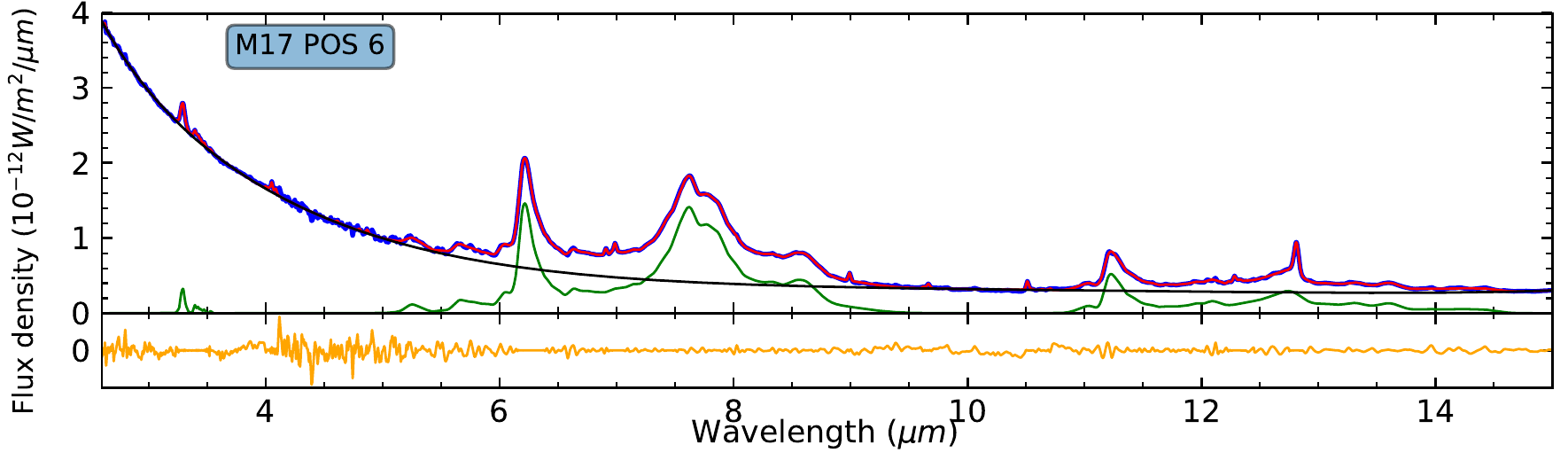}      
\caption{Fit (red) to the data (blue) to extract the AIB contribution $\hat{s}_{AIB}$ (green) using the method described in Sect.\ref{subsect_aibextraction}. The black line shows the estimated continuum $\hat{s}_{cont}$ and orange the estimated contribution from noise $\hat{s}_{inst}$. The name of the corresponding astrophysical object is given in the blue box inside the plot.}
\end{figure*}

\begin{figure*}
\centering
\includegraphics{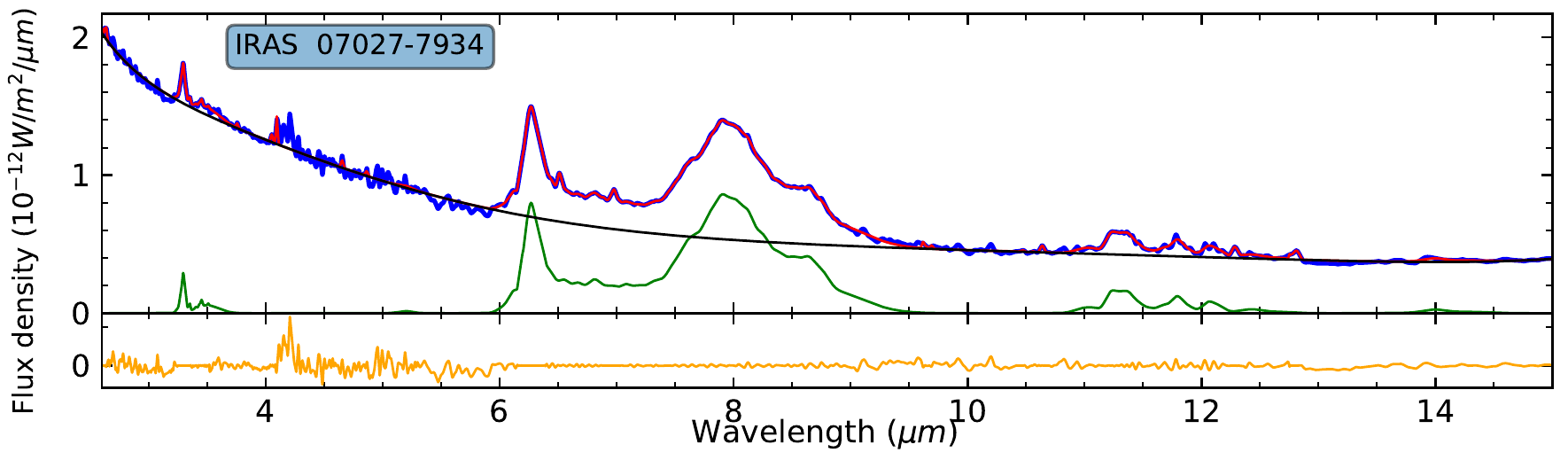}      
\includegraphics{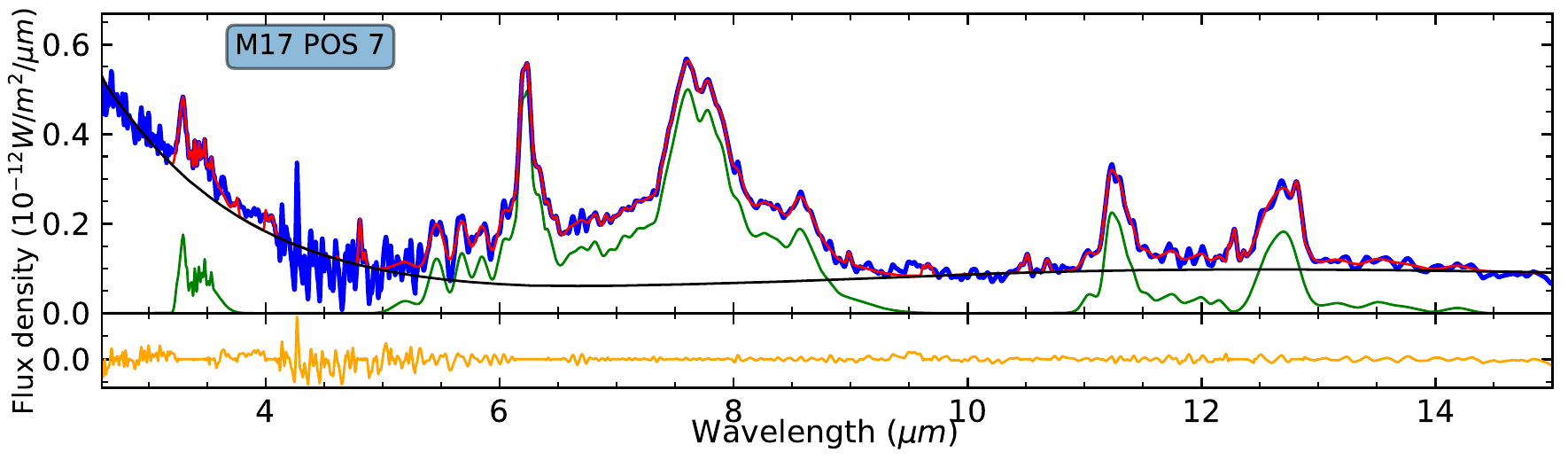}      
\includegraphics{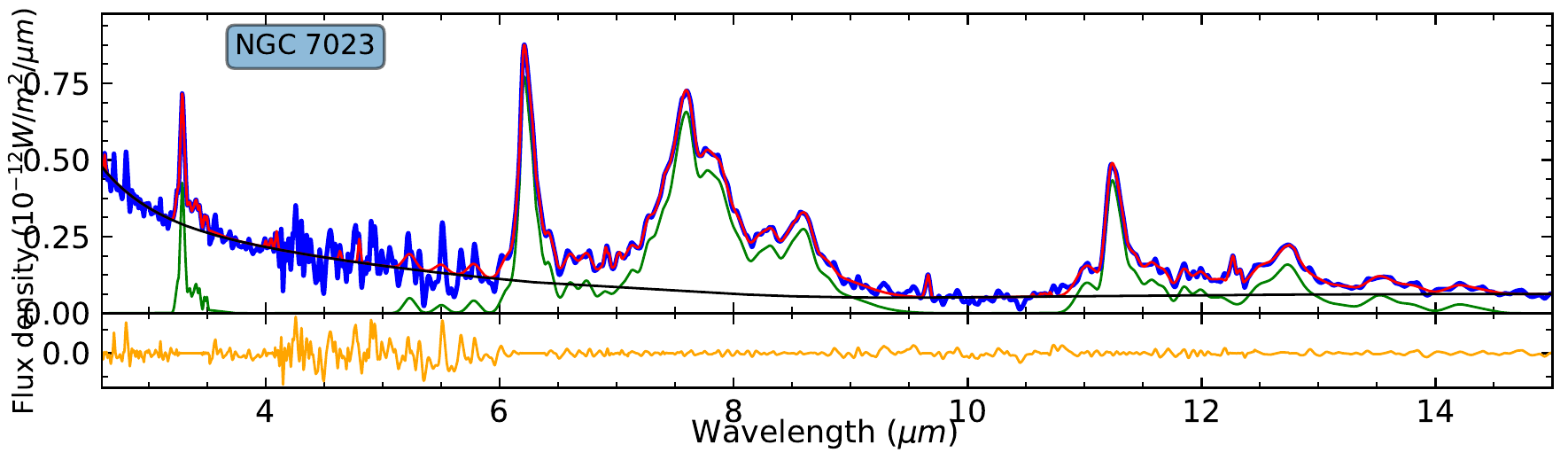}      
\includegraphics{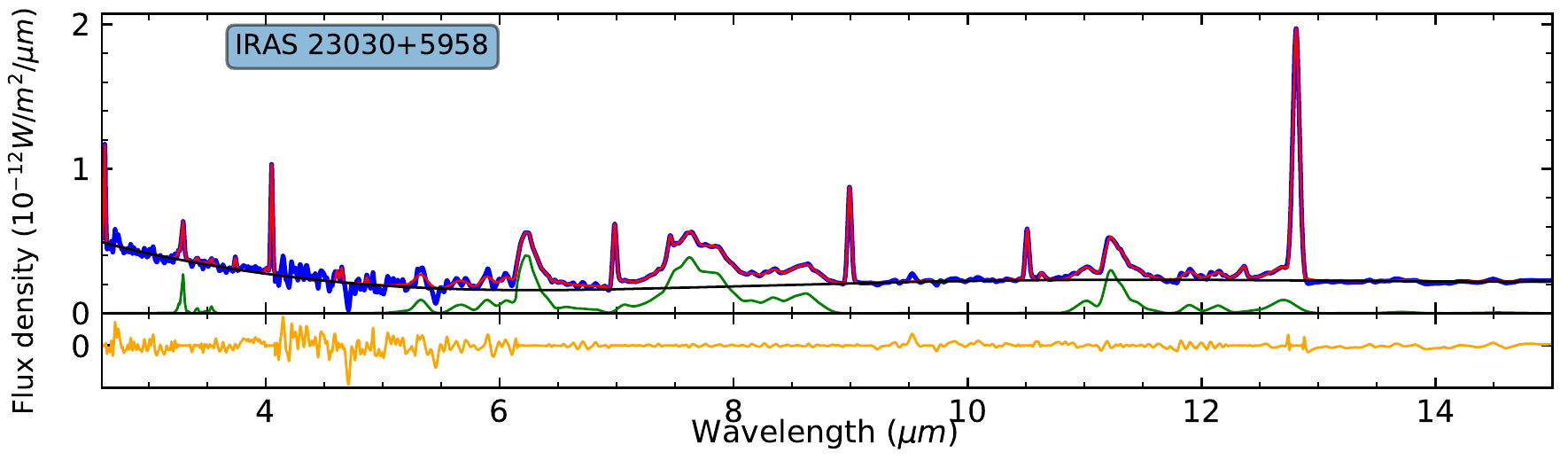}      
\caption{Continued.}
\end{figure*}

\begin{figure*}
\centering
\includegraphics{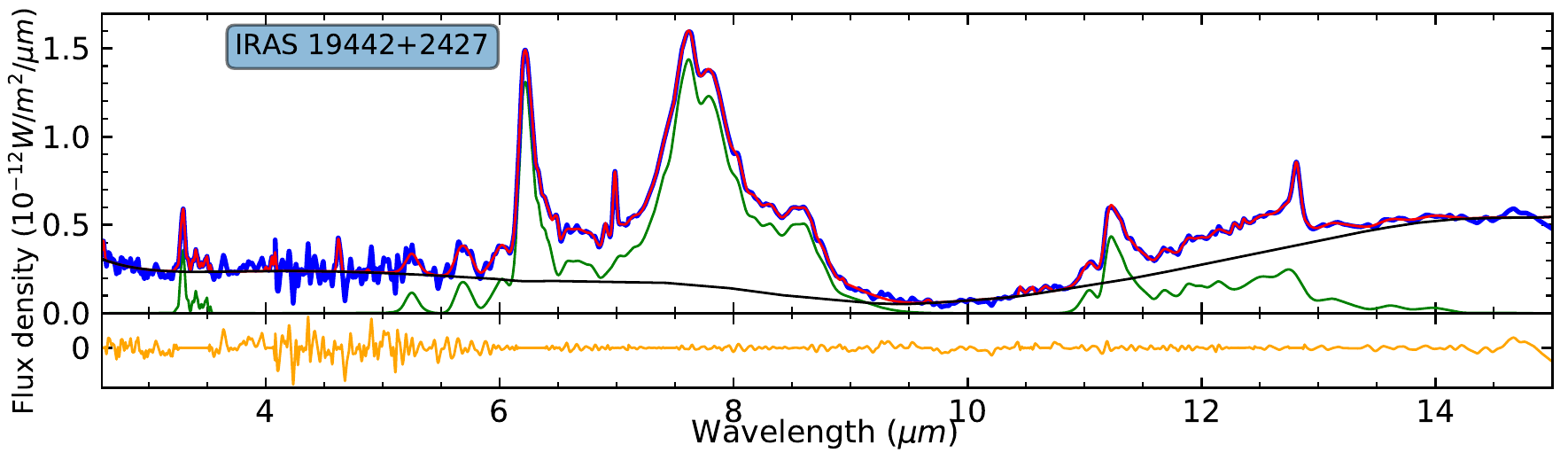}      
\includegraphics{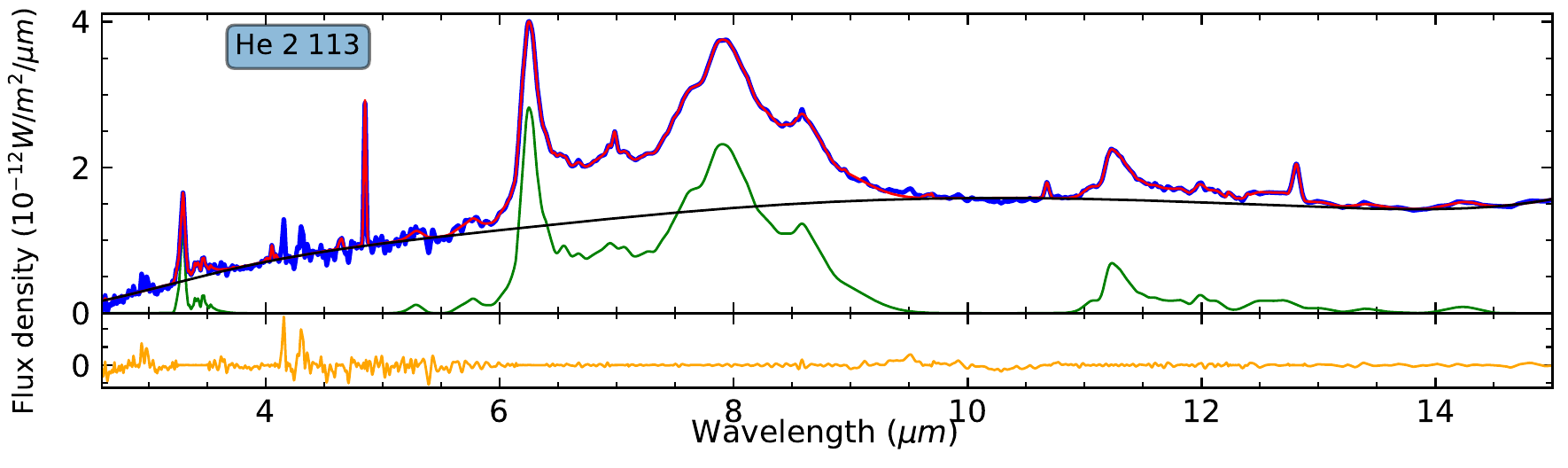}      
\includegraphics{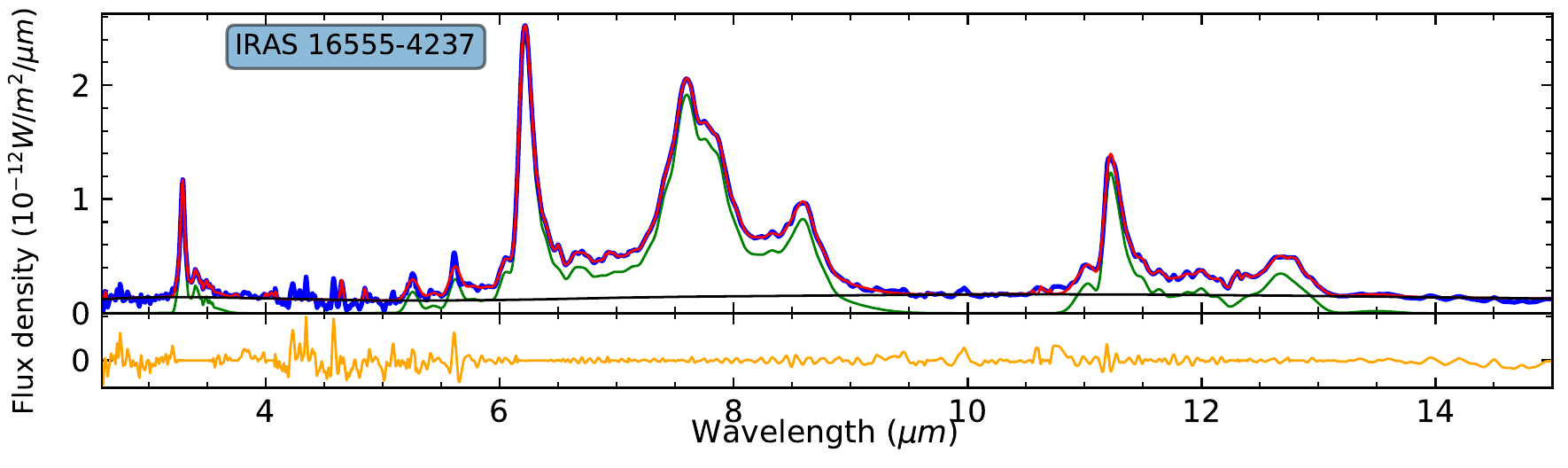}      
\includegraphics{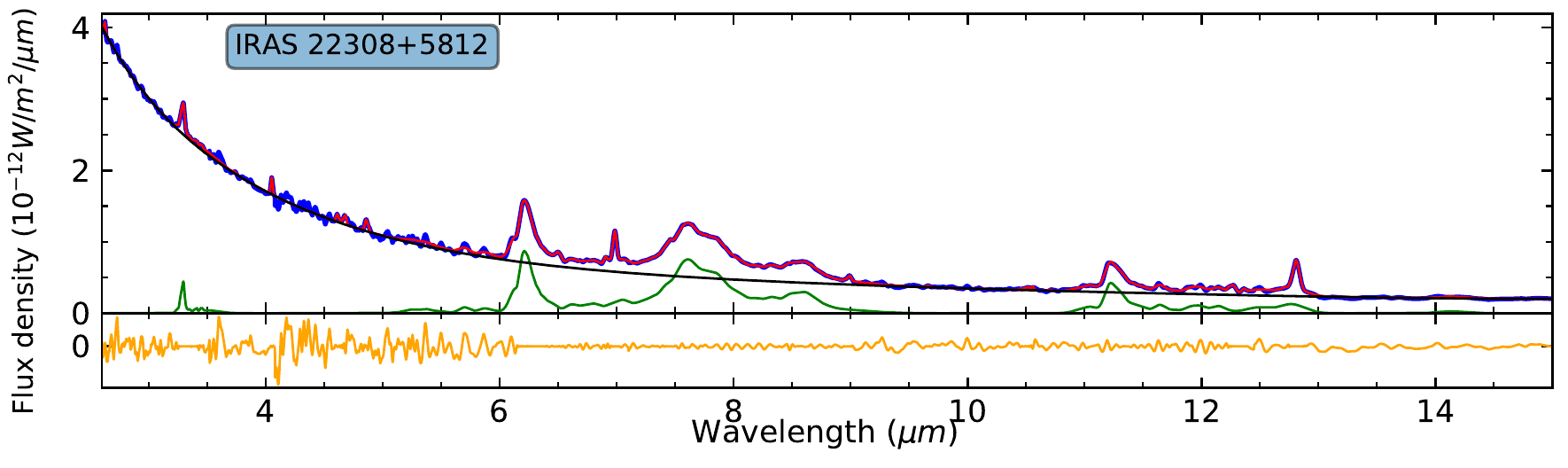}      
\caption{Continued.}
\end{figure*}

\begin{figure*}
\centering
\includegraphics{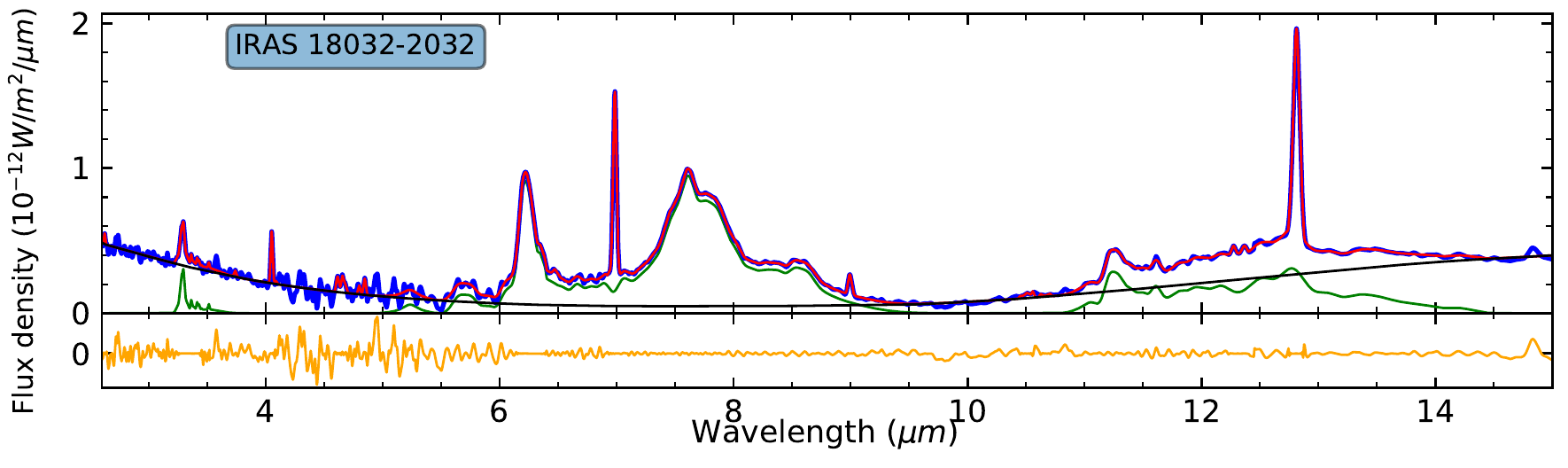}      
\includegraphics{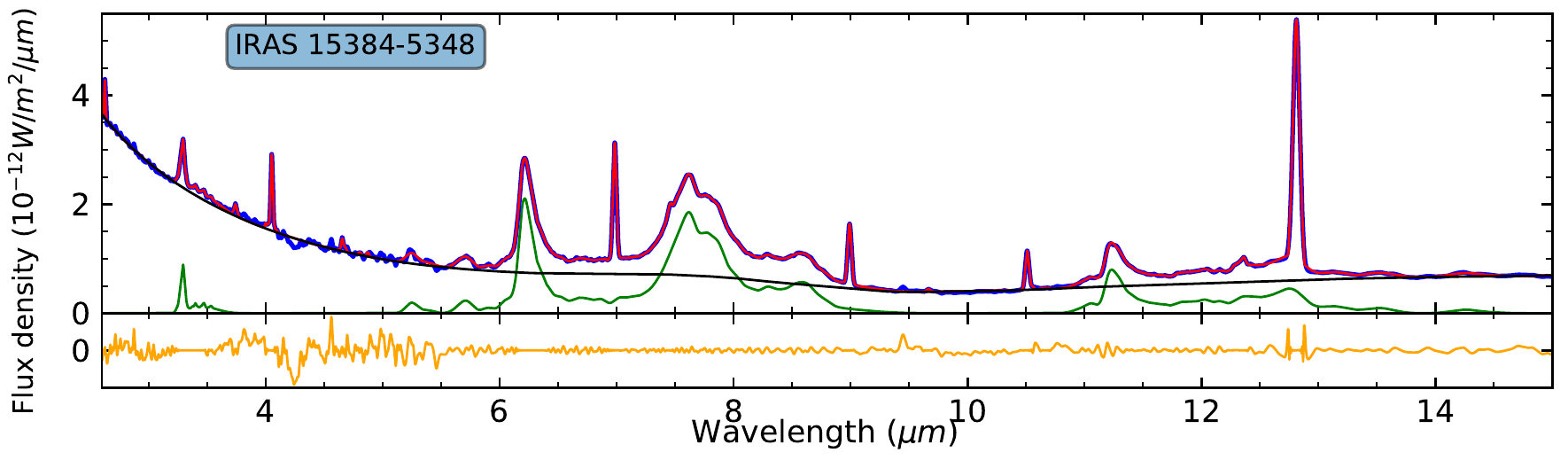}      
\includegraphics{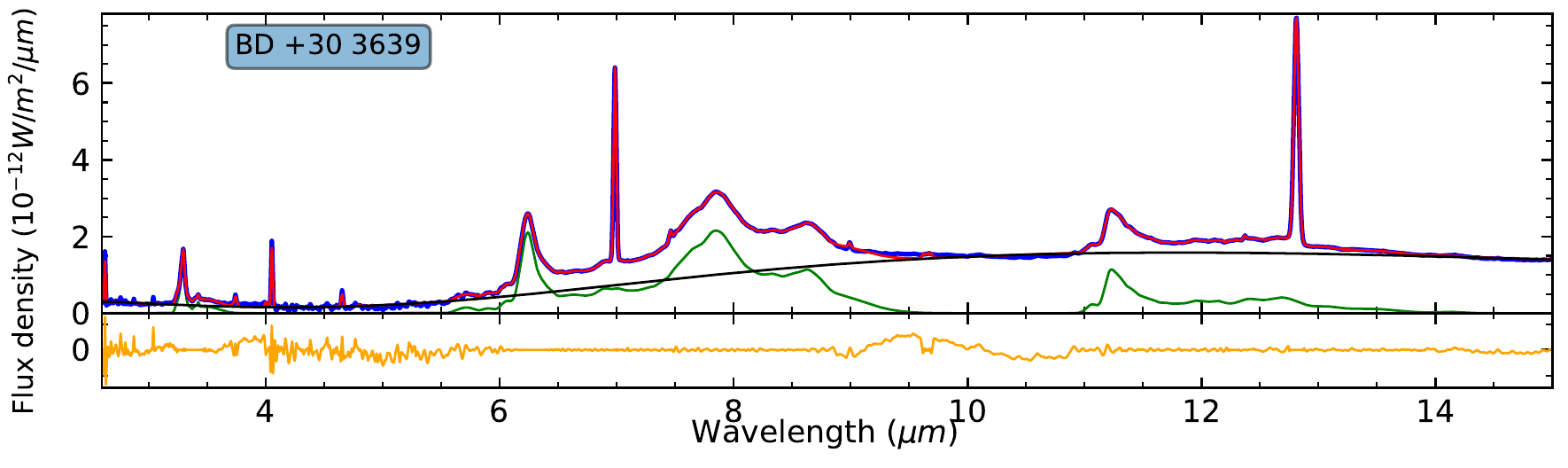}      
\includegraphics{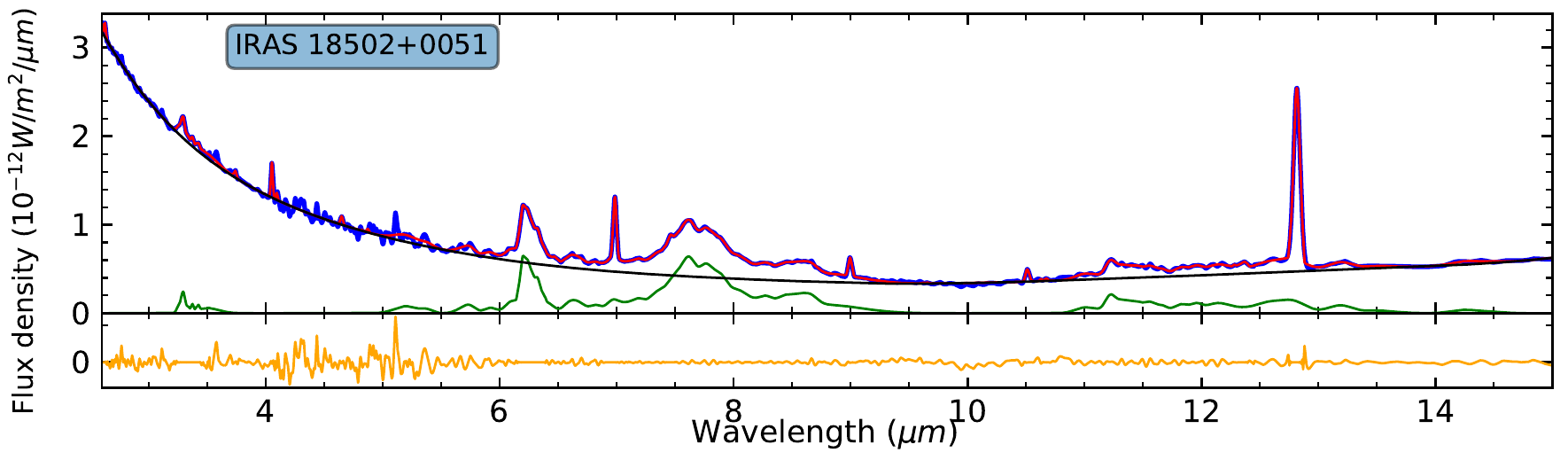}      
\caption{Continued.}
\end{figure*}

\begin{figure*}
\centering
\includegraphics{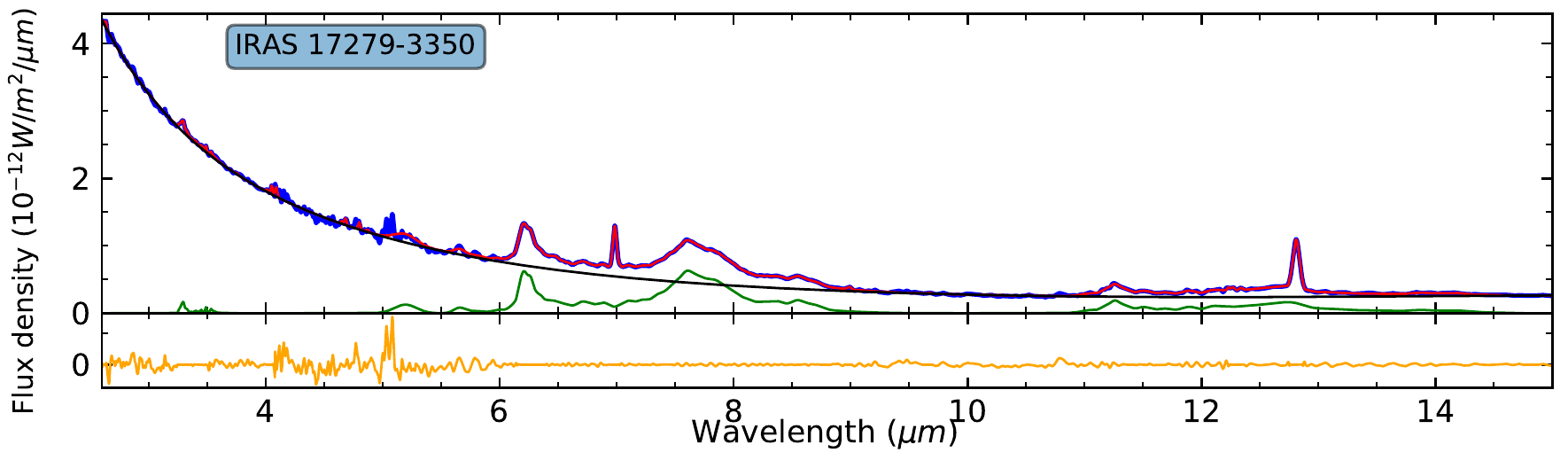}      
\includegraphics{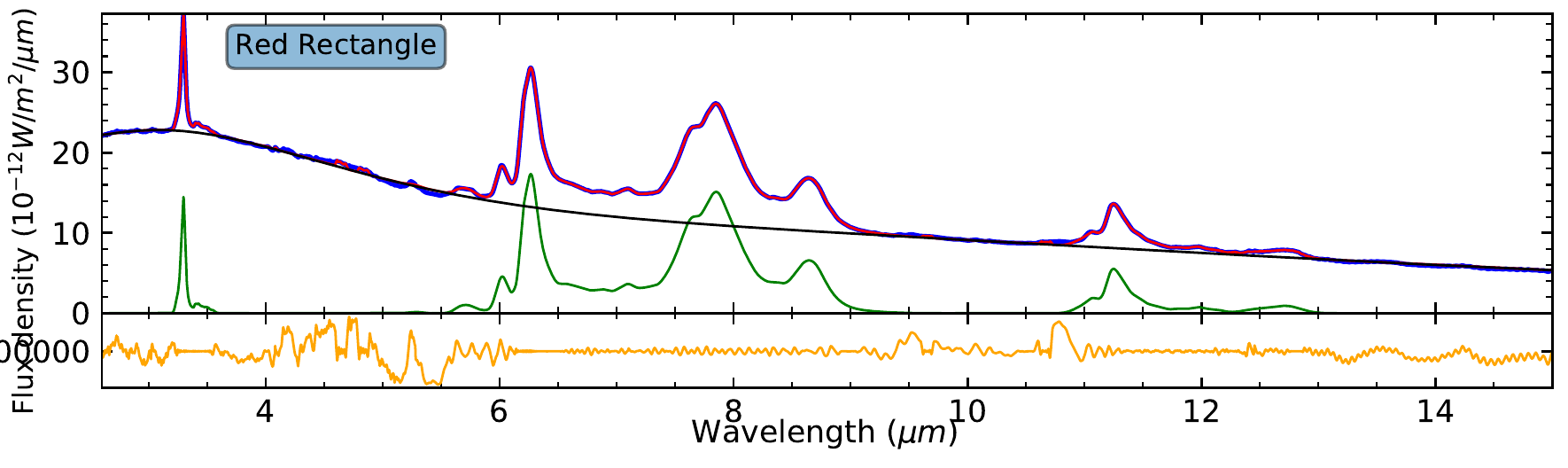}      
\includegraphics{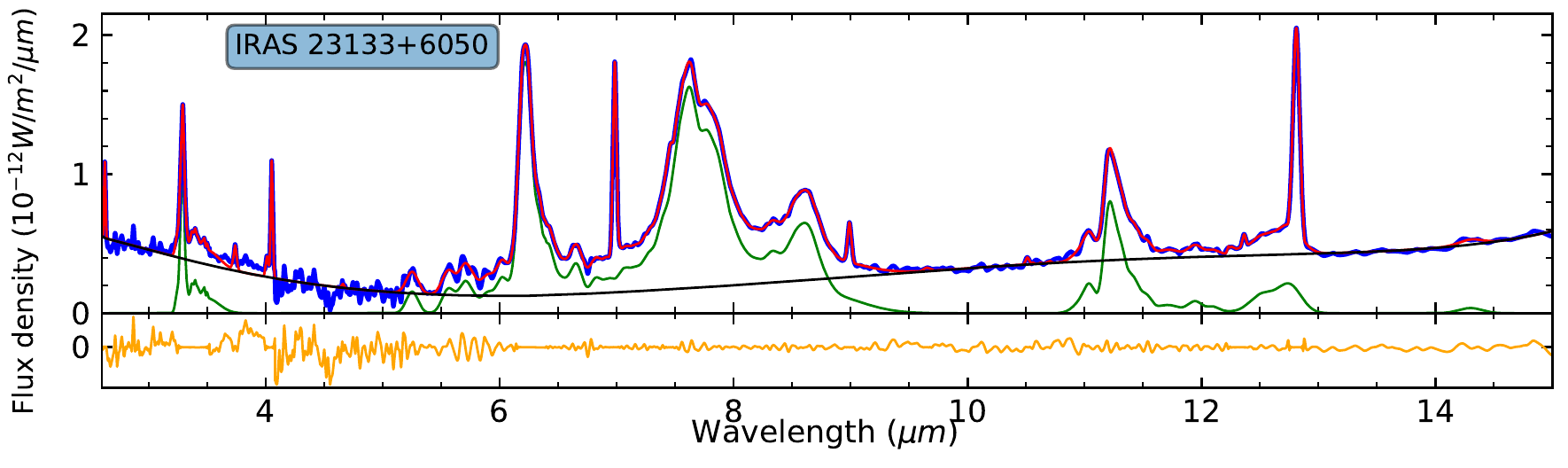}      
\includegraphics{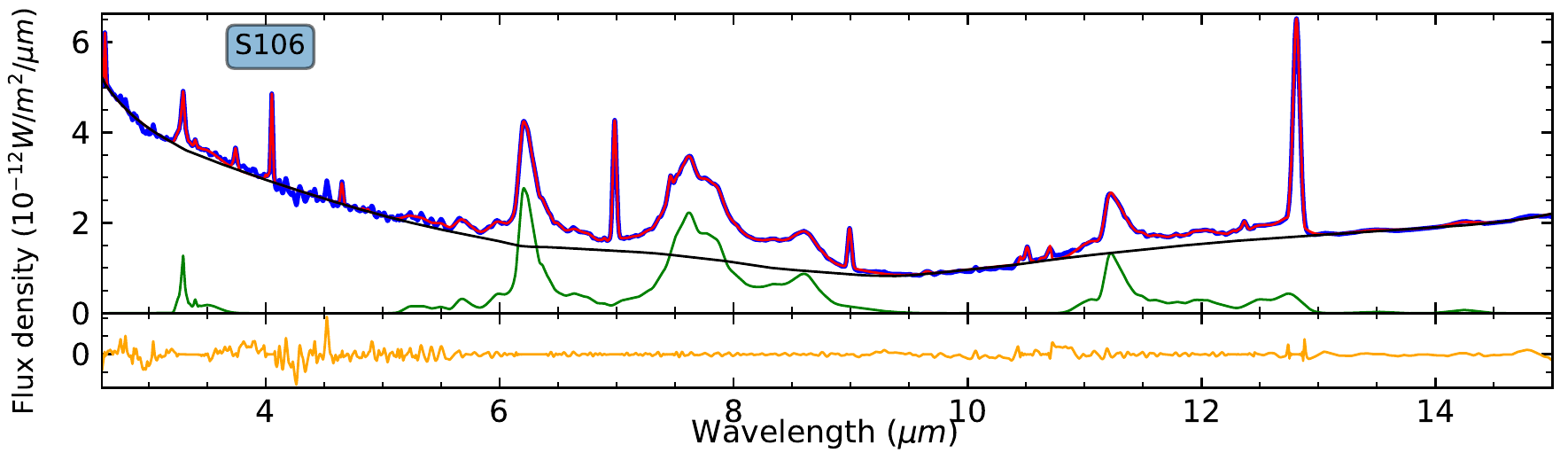}      
\caption{Continued.}
\end{figure*}

\begin{figure*}
\centering
\includegraphics{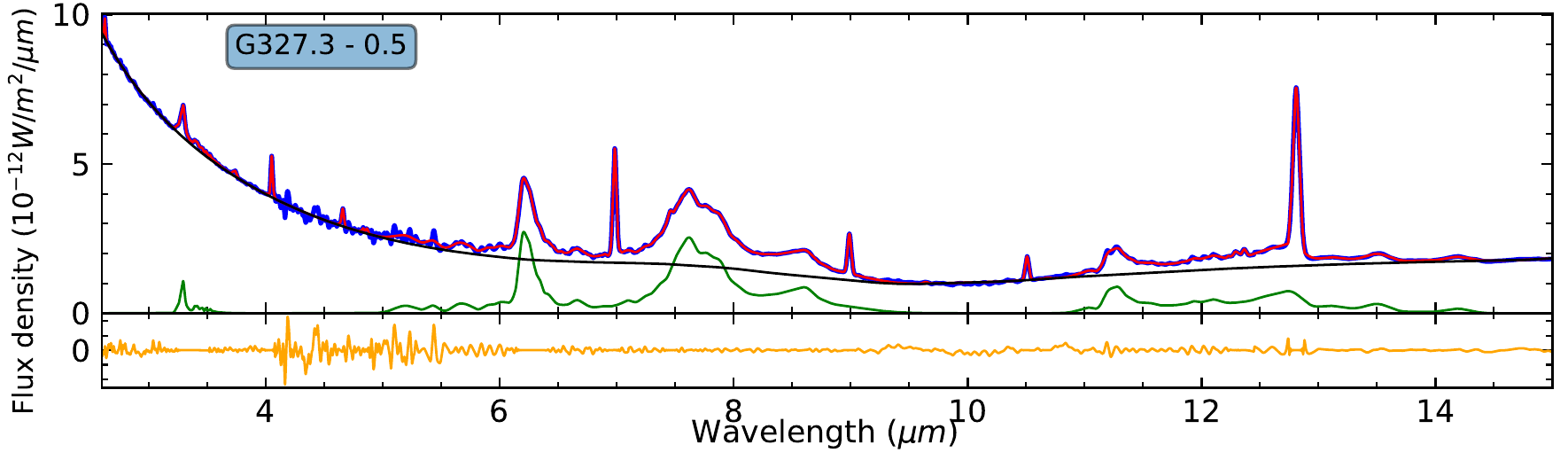}      
\includegraphics{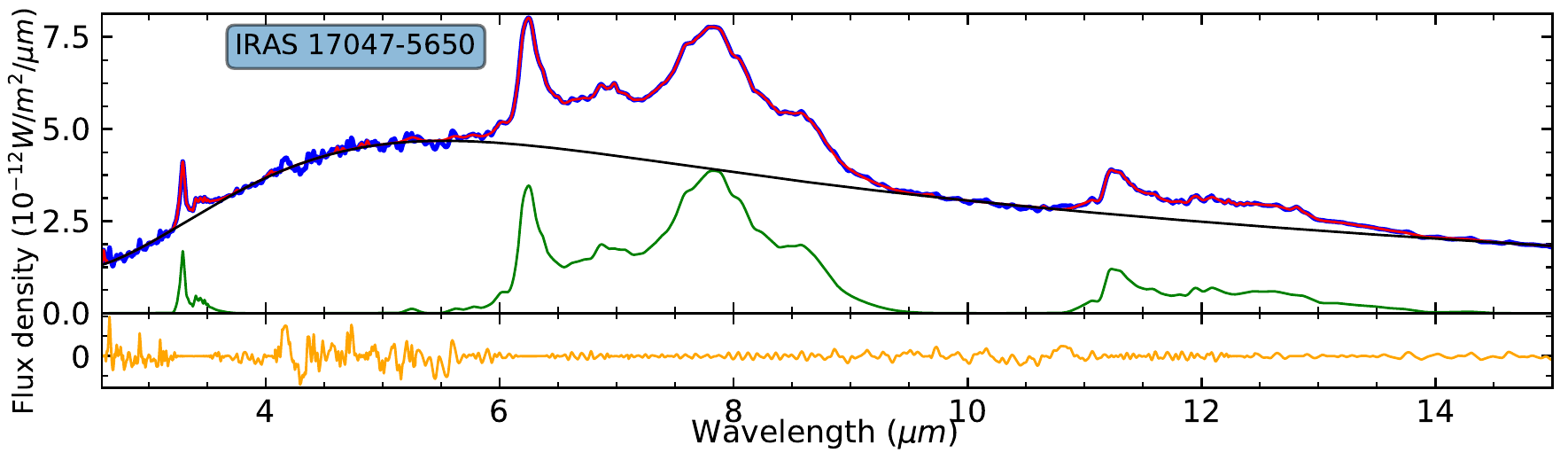}      
\includegraphics{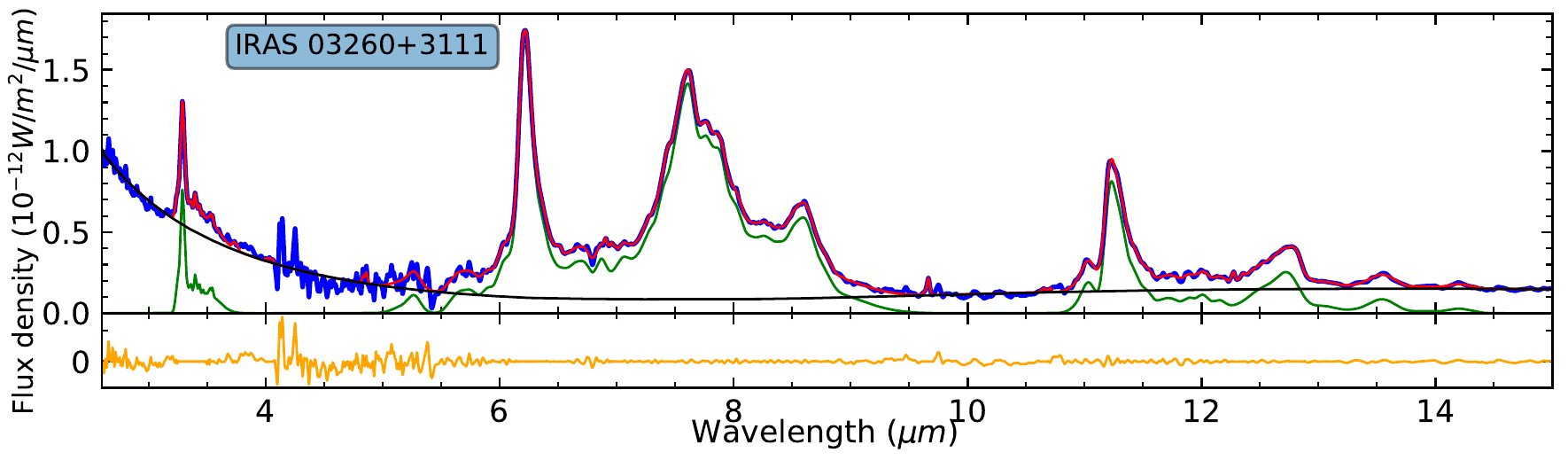}      
\includegraphics{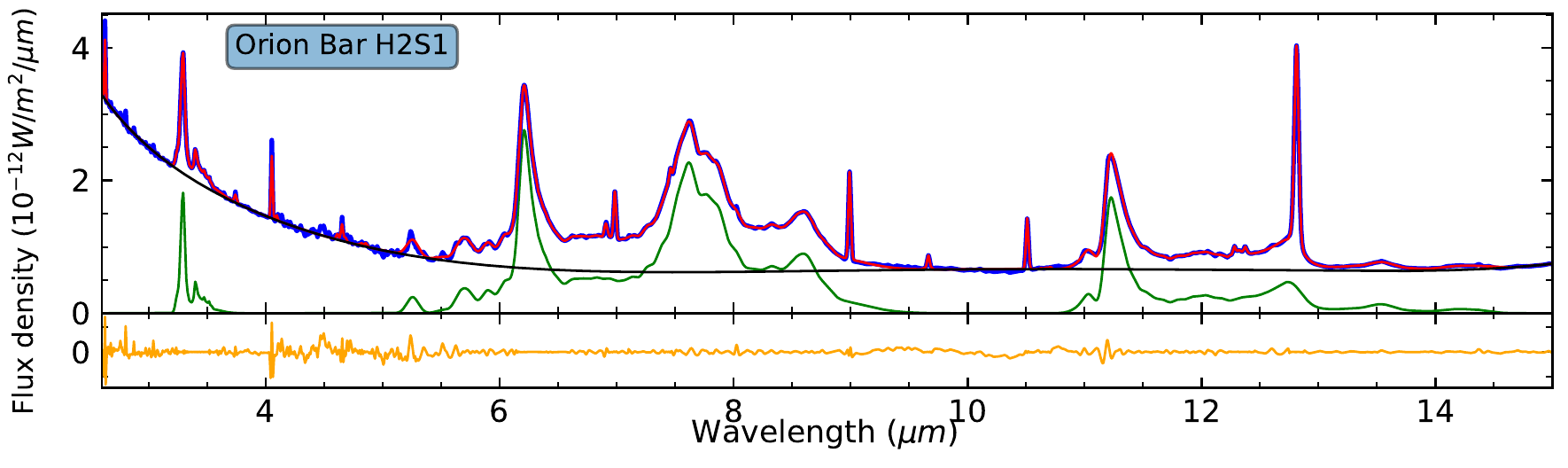}      
\caption{Continued.}
\end{figure*}

\begin{figure*}
\centering
\includegraphics{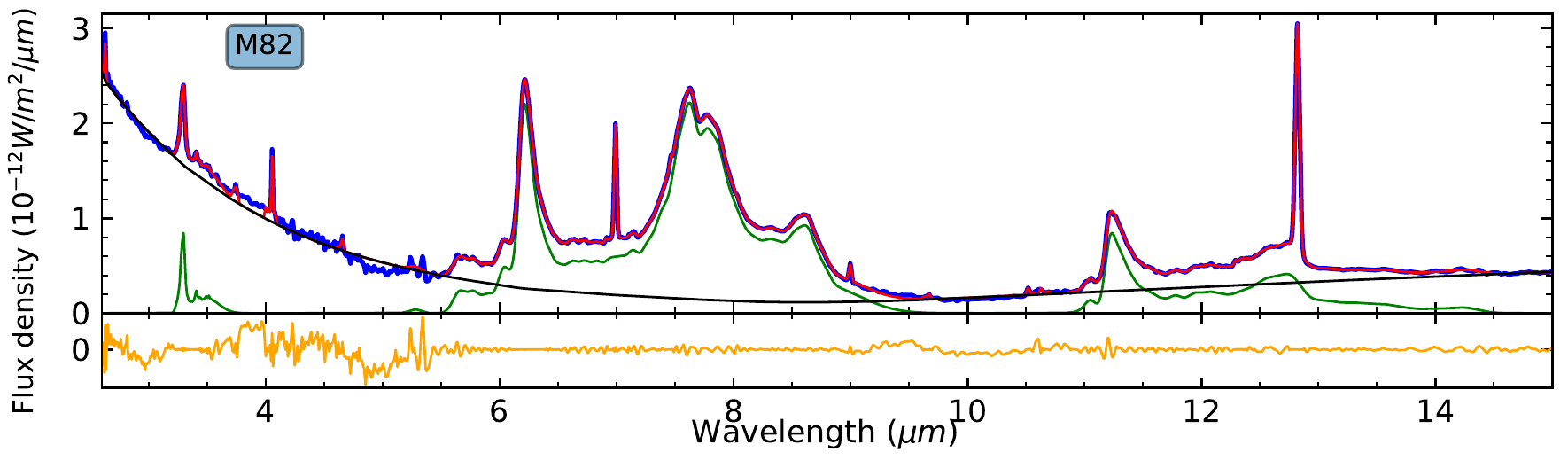}      
\includegraphics{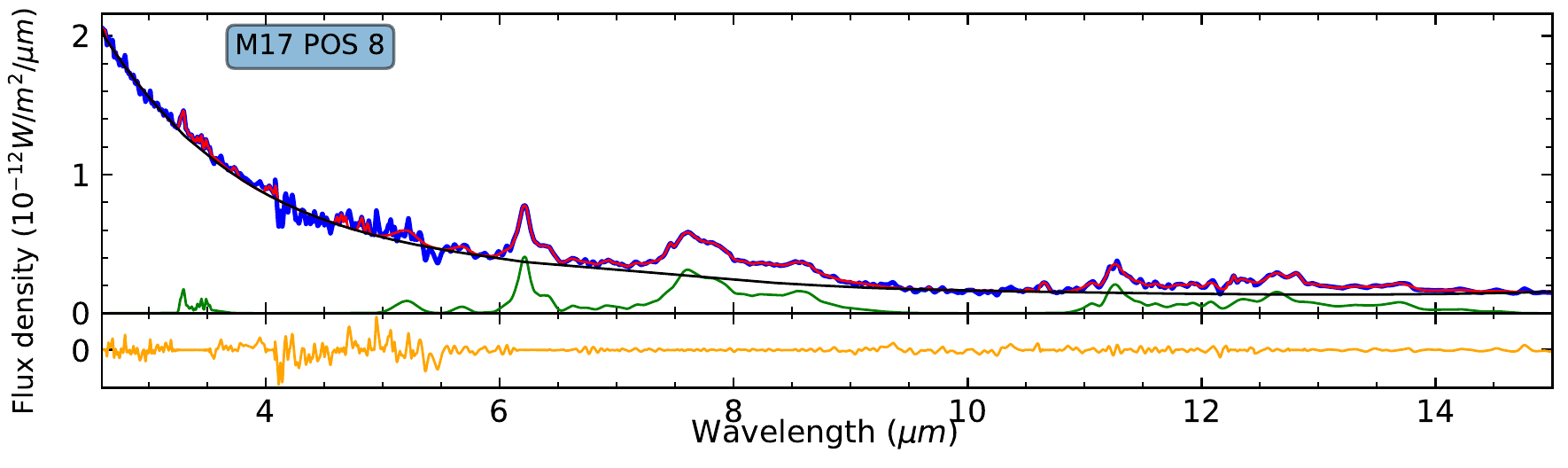}      
\includegraphics{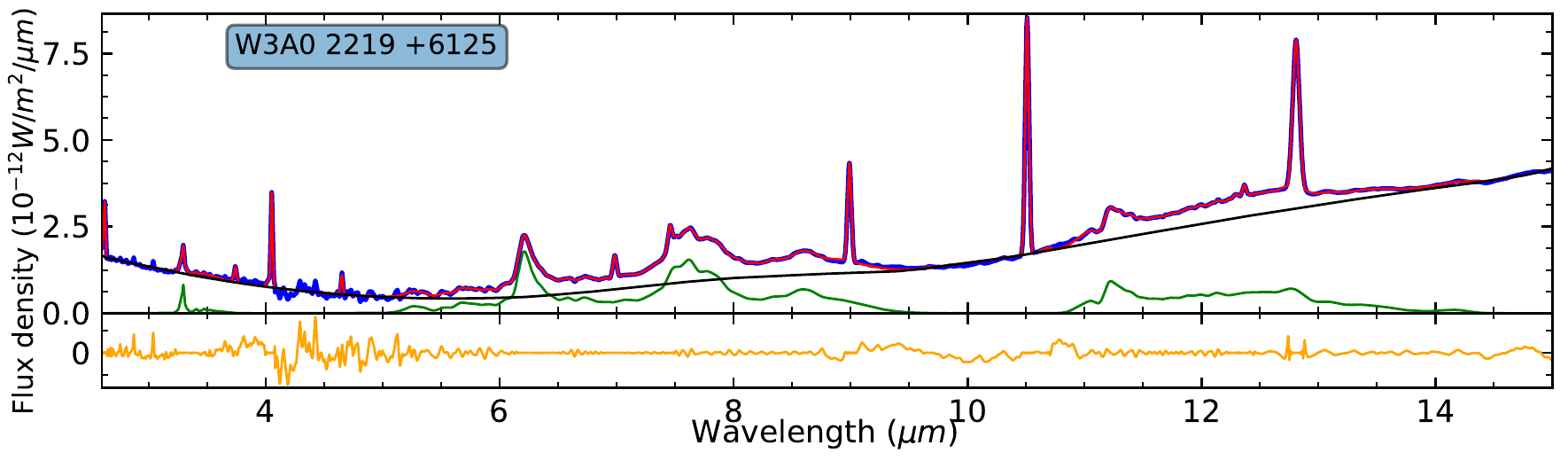}      
\includegraphics{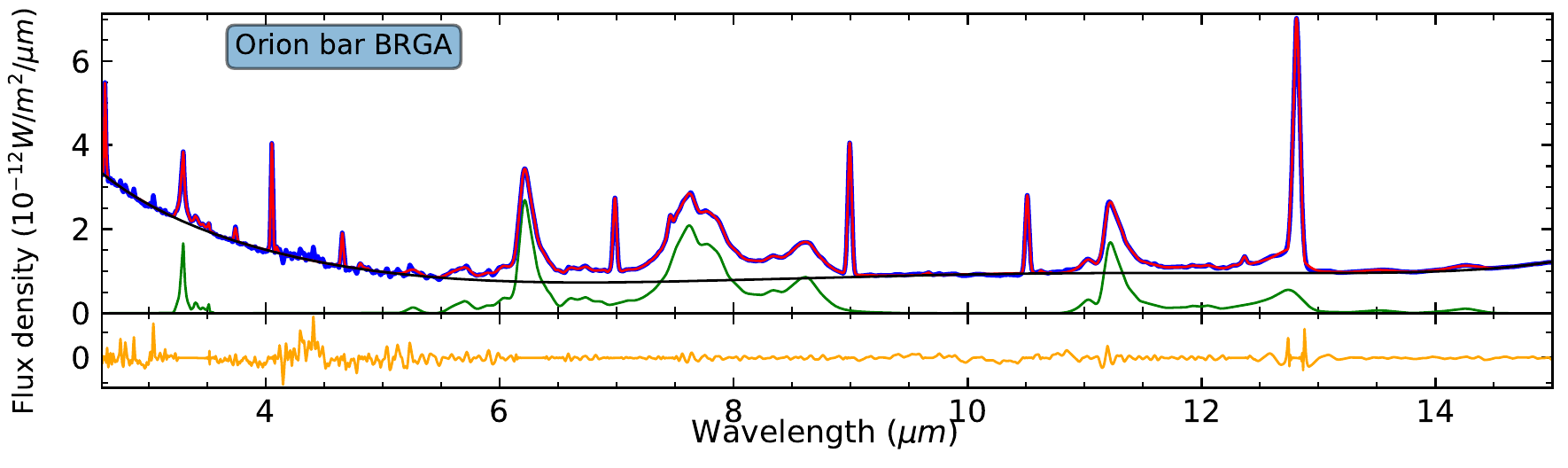}      
\caption{Continued.}
\end{figure*}

\begin{figure*}
\centering
\includegraphics{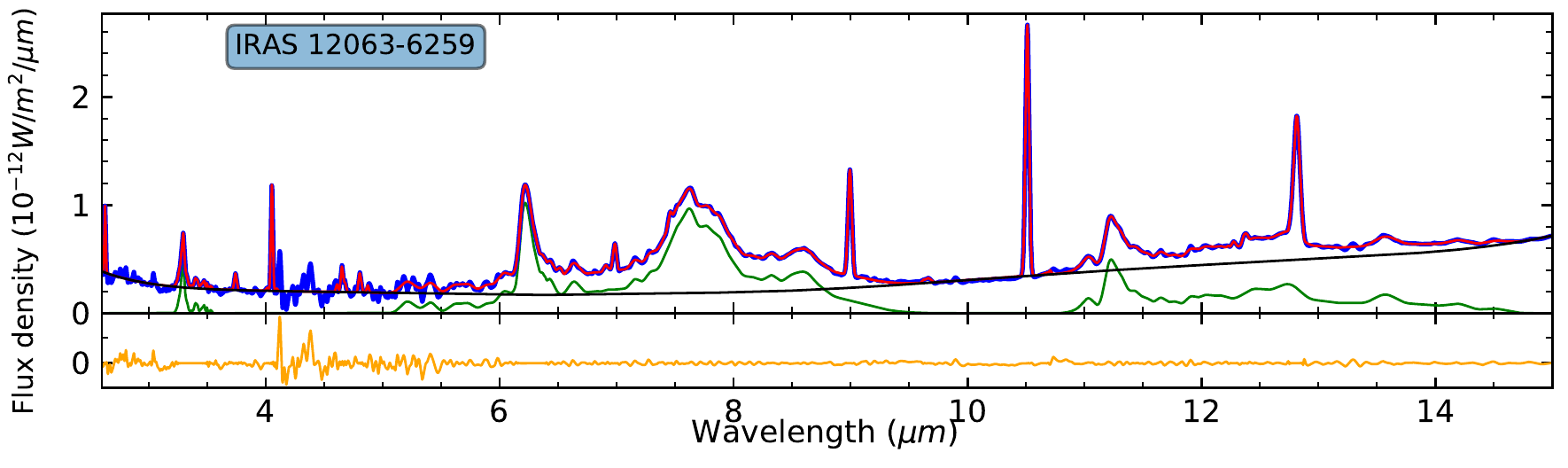}      
\includegraphics{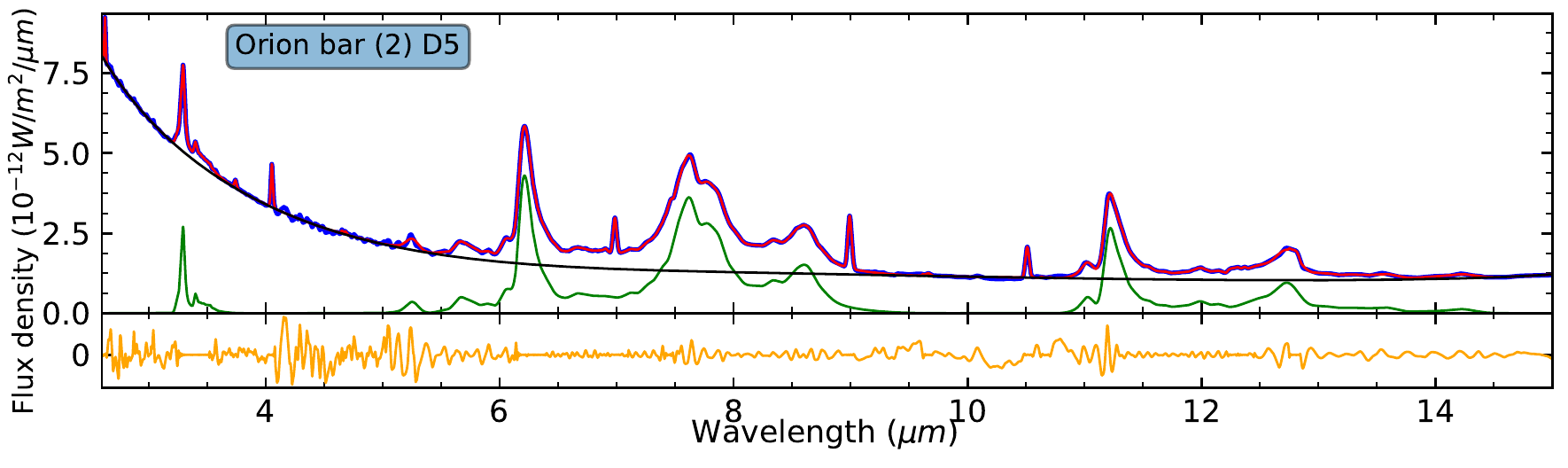}      
\includegraphics{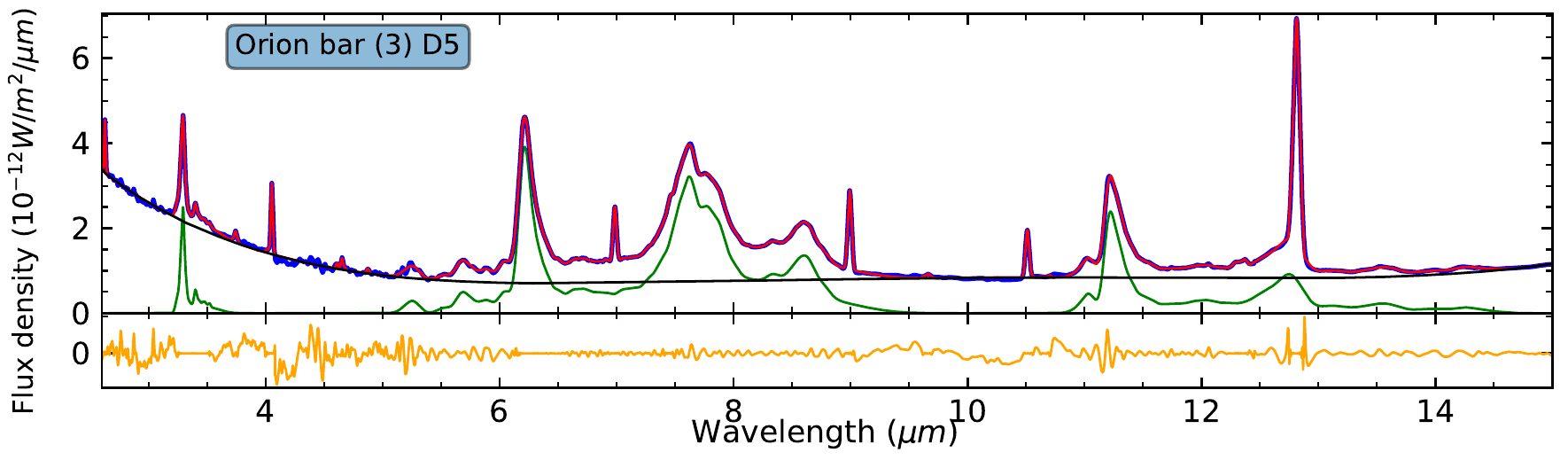}      
\caption{Continued.}
\end{figure*}

\section{Reconstruction of all the observed spectra}

Figures in this section show the reconstruction of each observed spectrum  using the four representative spectra (see Eq.~\ref{eq_model}). The same color-coding is used for all of them. The order of the spectra is the same as in Appendix~\ref{appendix_rec}.

\begin{figure*}
\centering
\includegraphics{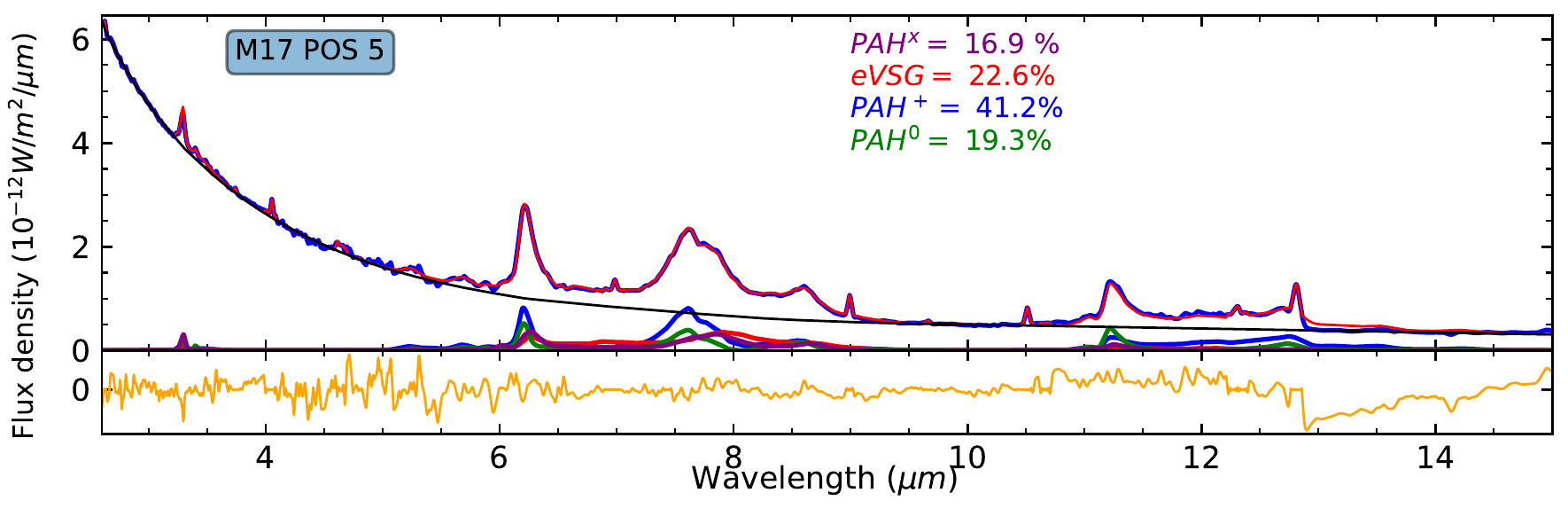}      
\includegraphics{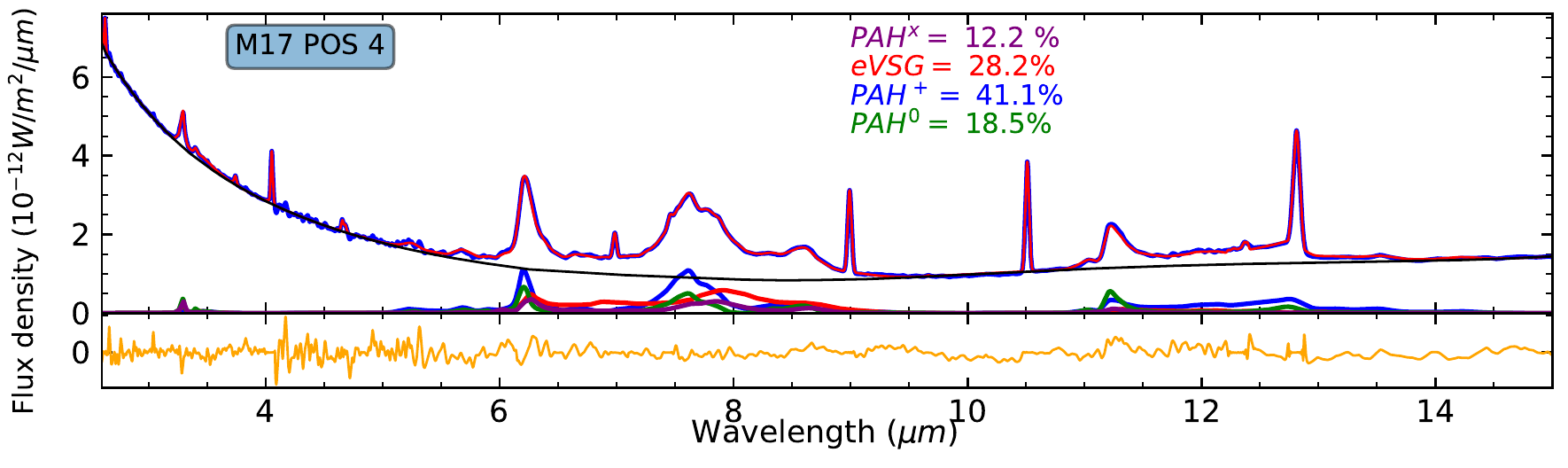}      
\includegraphics{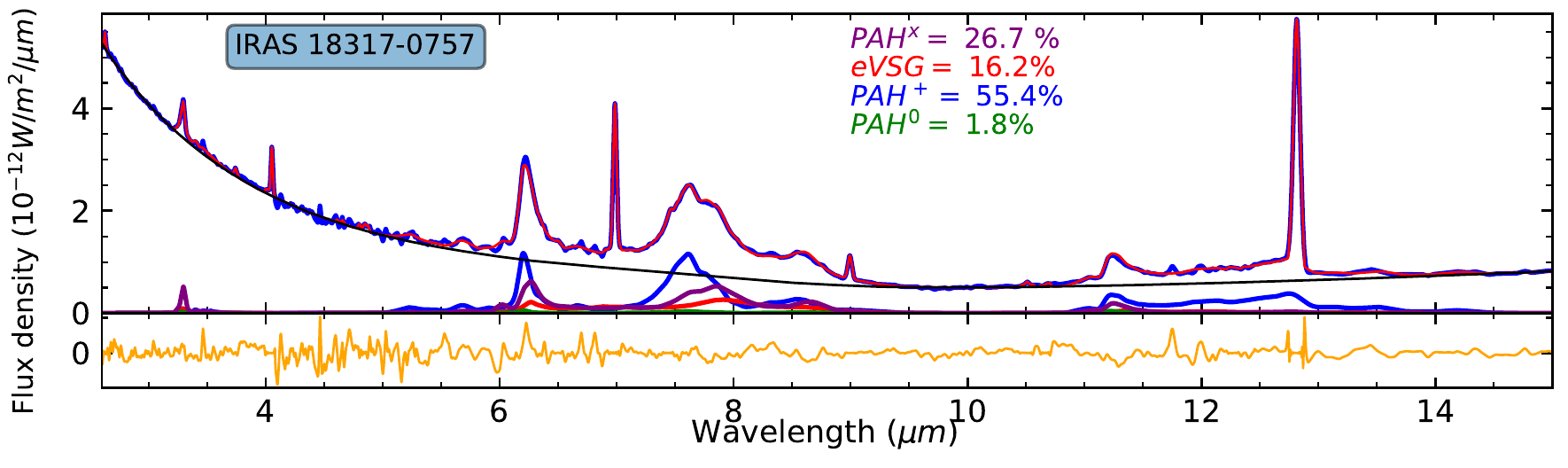}      
\includegraphics{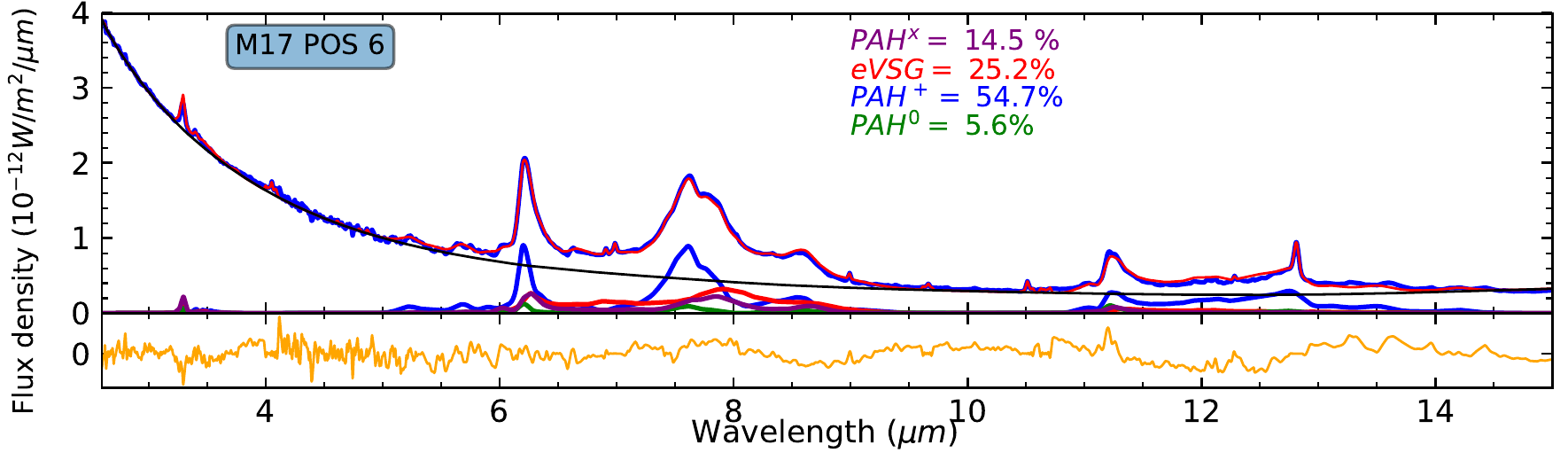}      
\caption{Reconstruction (in red, thin line) of the observed data (in blue), using the reconstruction method described in Sect. \ref{subsect_reconstruction} based on the four representative spectra (Fig.~\ref{fig_templates}). The relative contribution of each of these spectra is indicated. The black line corresponds to the continuum and orange to the residual of the reconstruction. The name of the corresponding astrophysical object is given in the blue box inside the plot.}
\end{figure*}
  
\begin{figure*}
\centering
\includegraphics{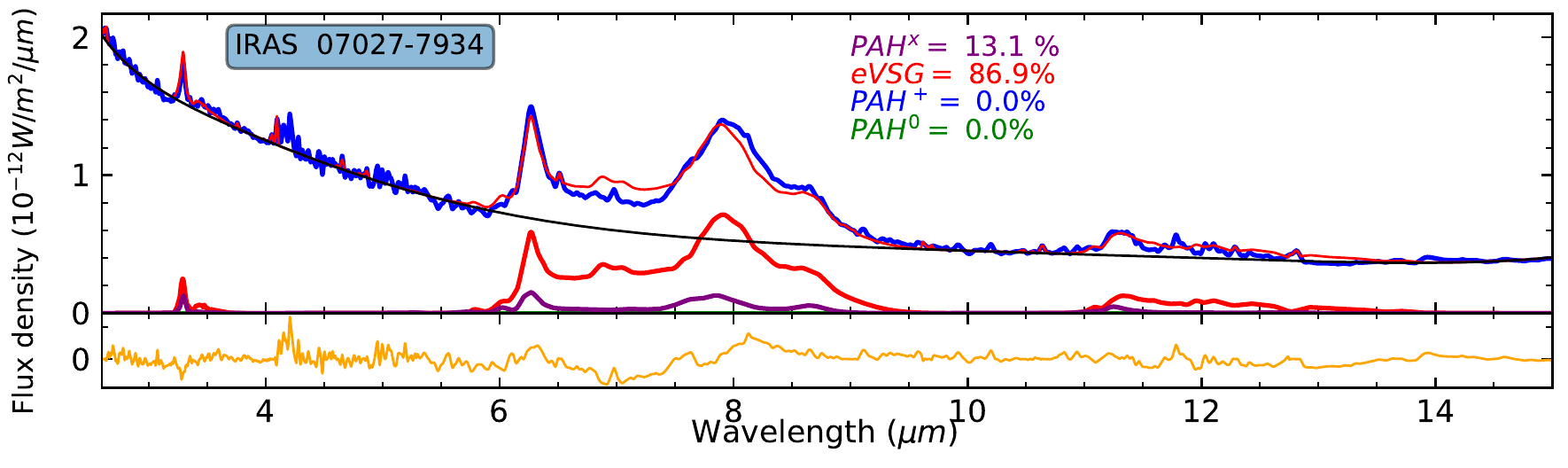}      
\includegraphics{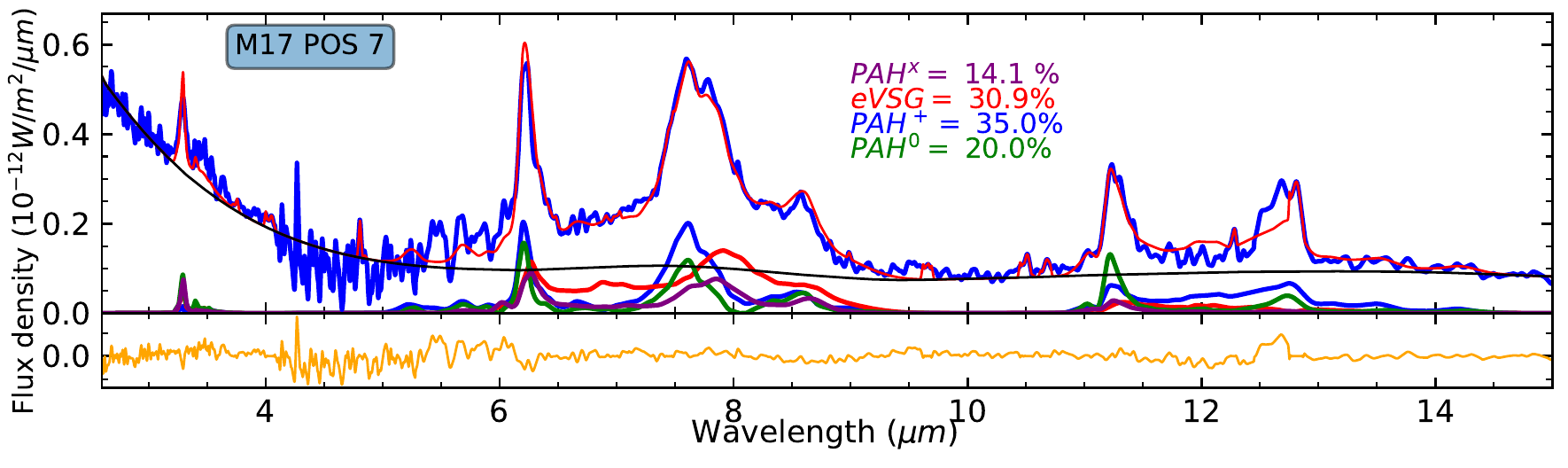}      
\includegraphics{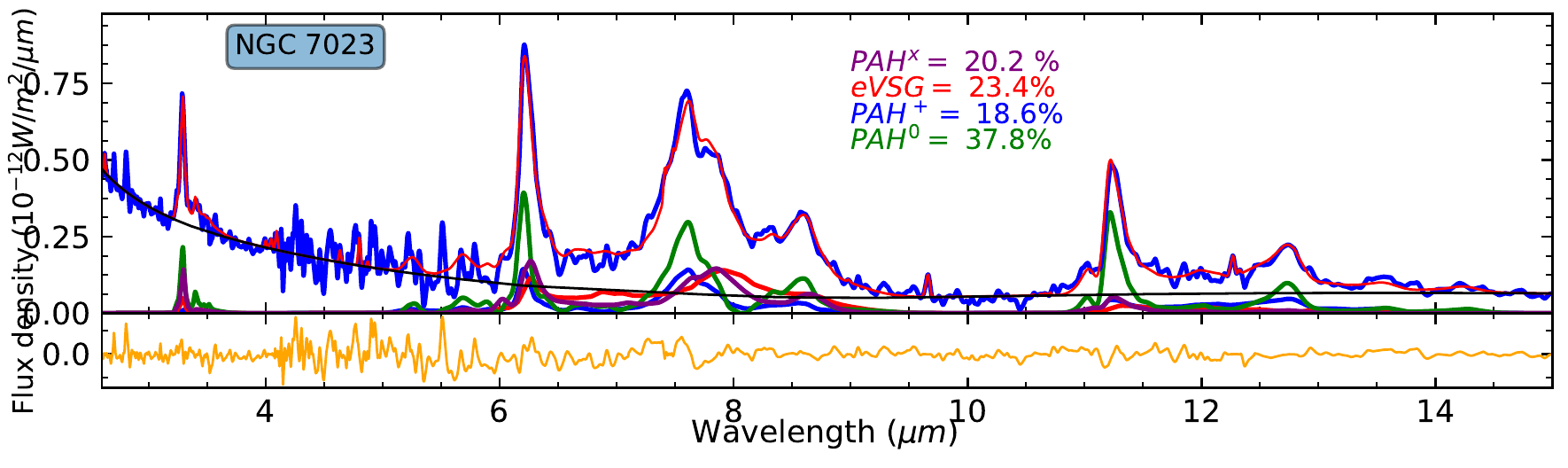}      
\includegraphics{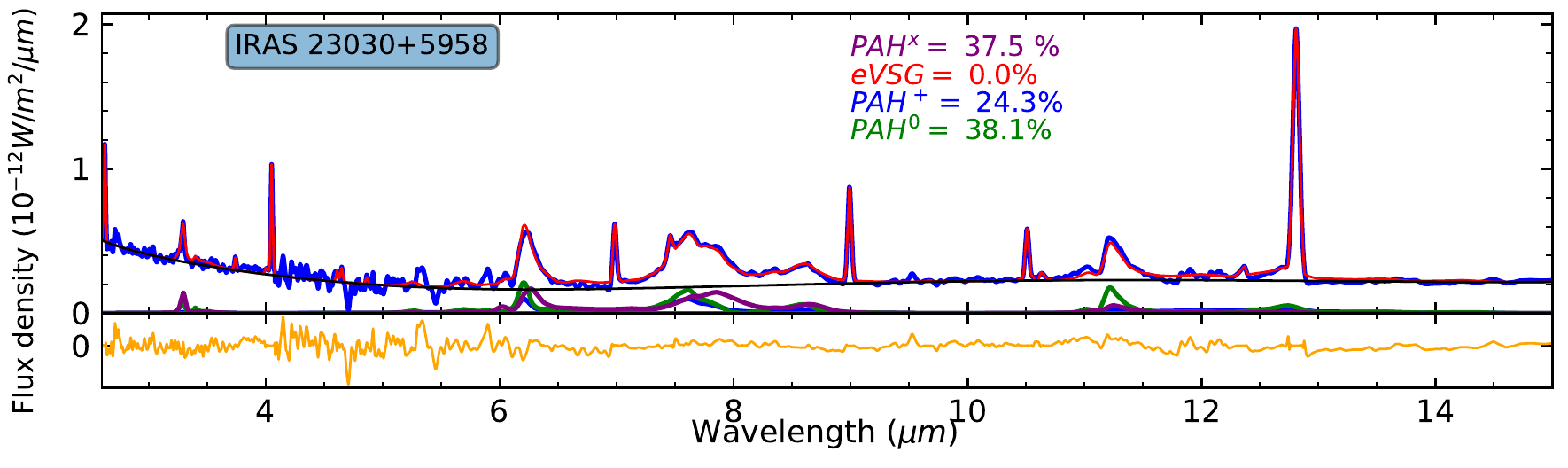}    
\caption{Continued.}
\end{figure*}

\begin{figure*}
\centering
\includegraphics{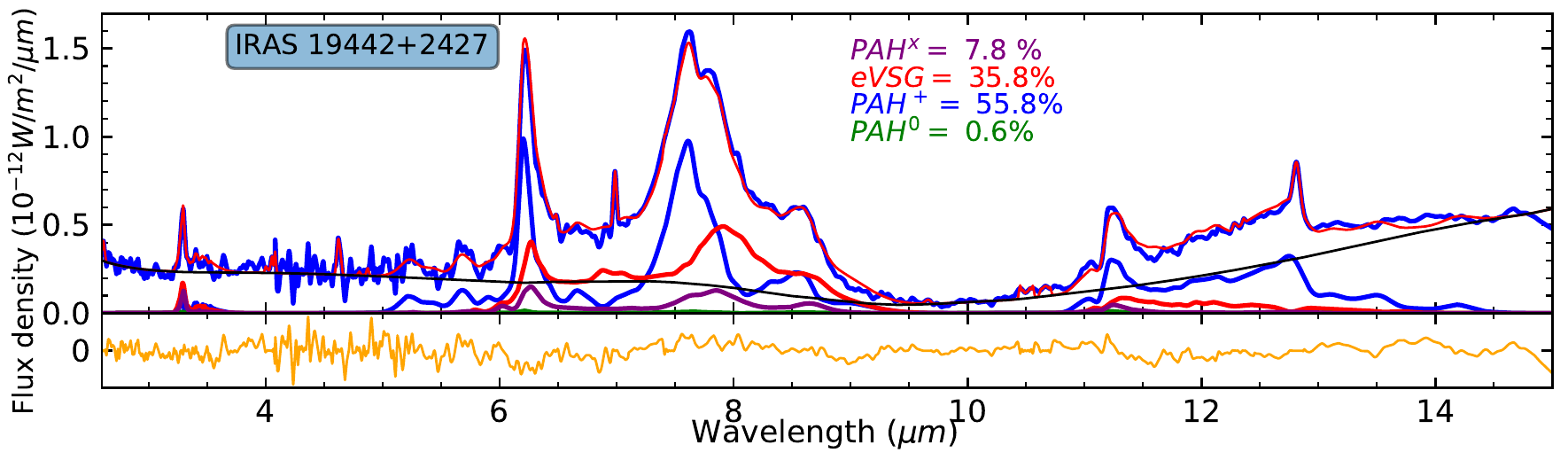}      
\includegraphics{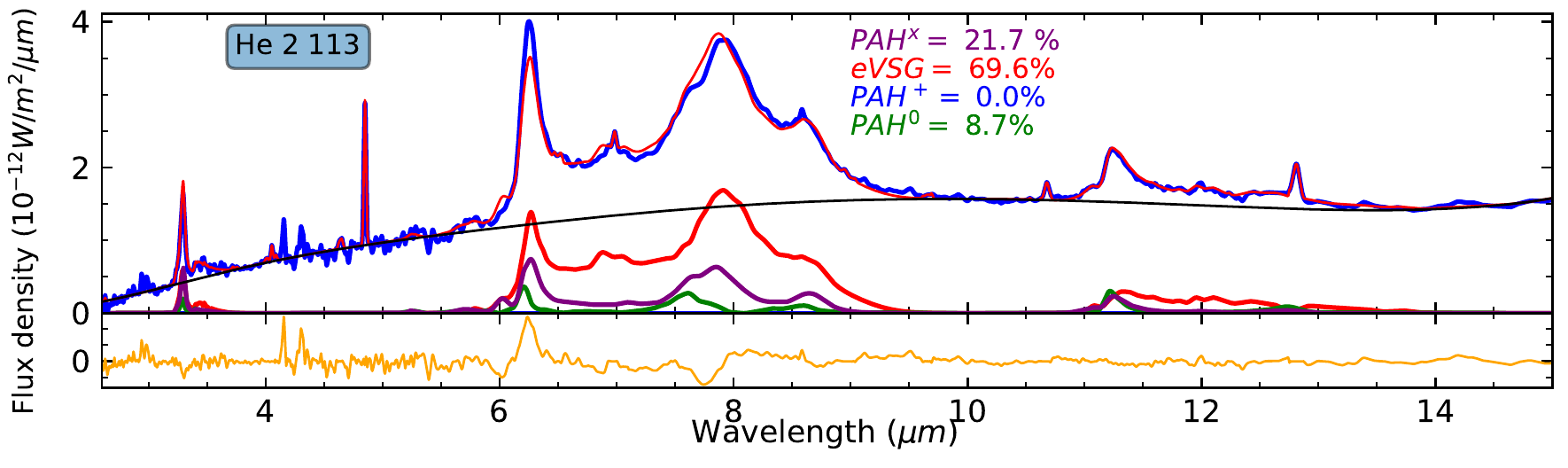}      
\includegraphics{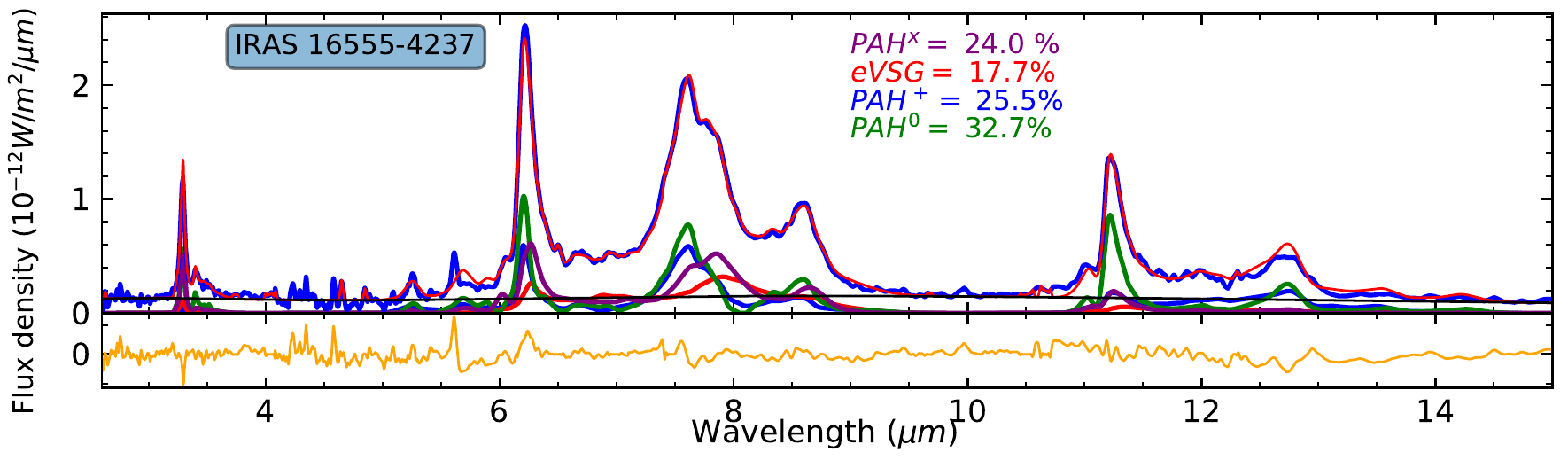}      
\includegraphics{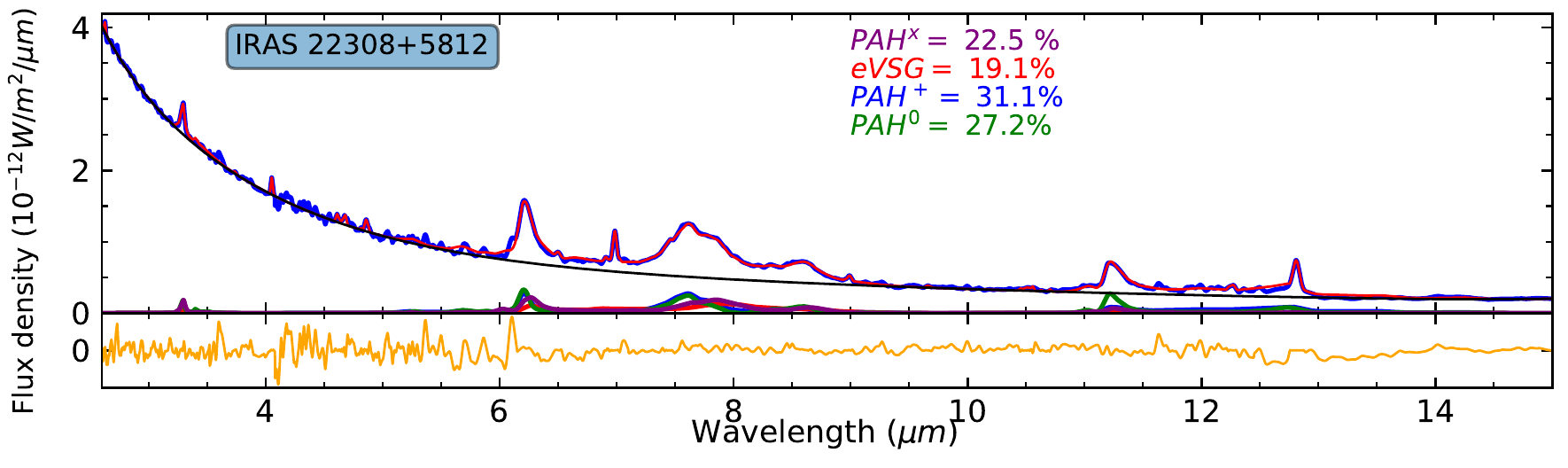} 
\caption{Continued.}
\end{figure*}

\begin{figure*}
\centering
\includegraphics{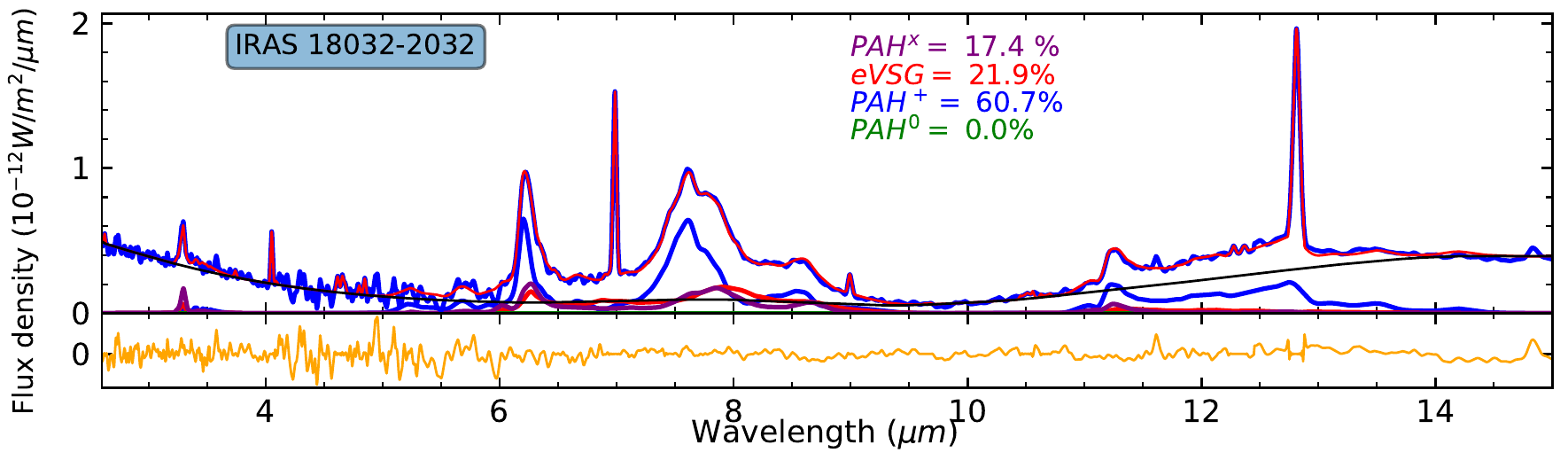}      
\includegraphics{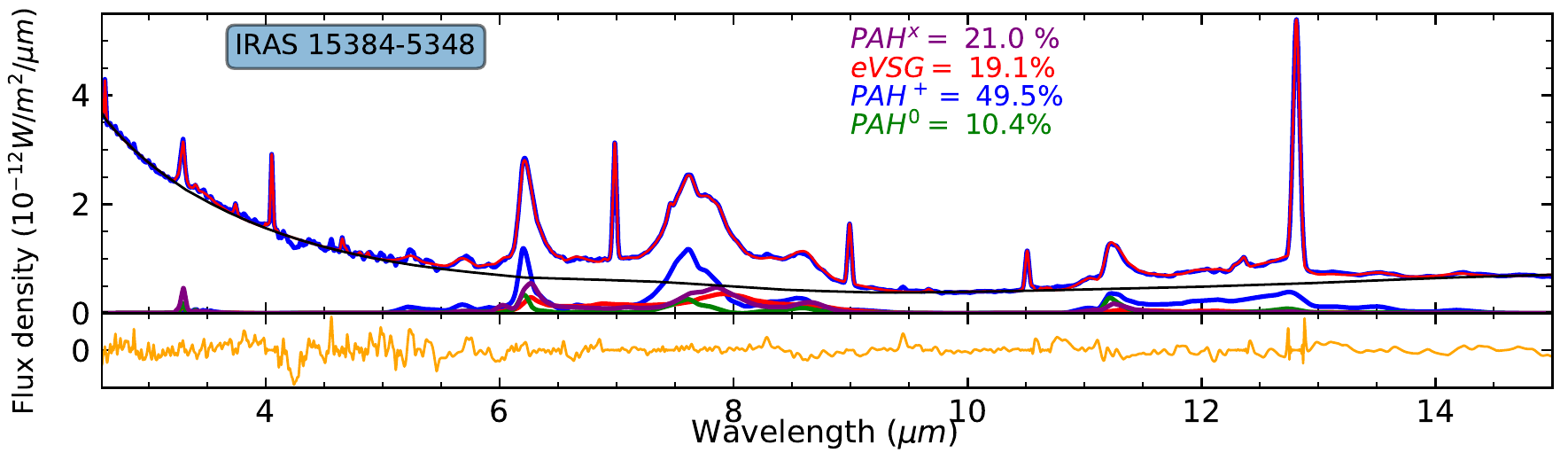}      
\includegraphics{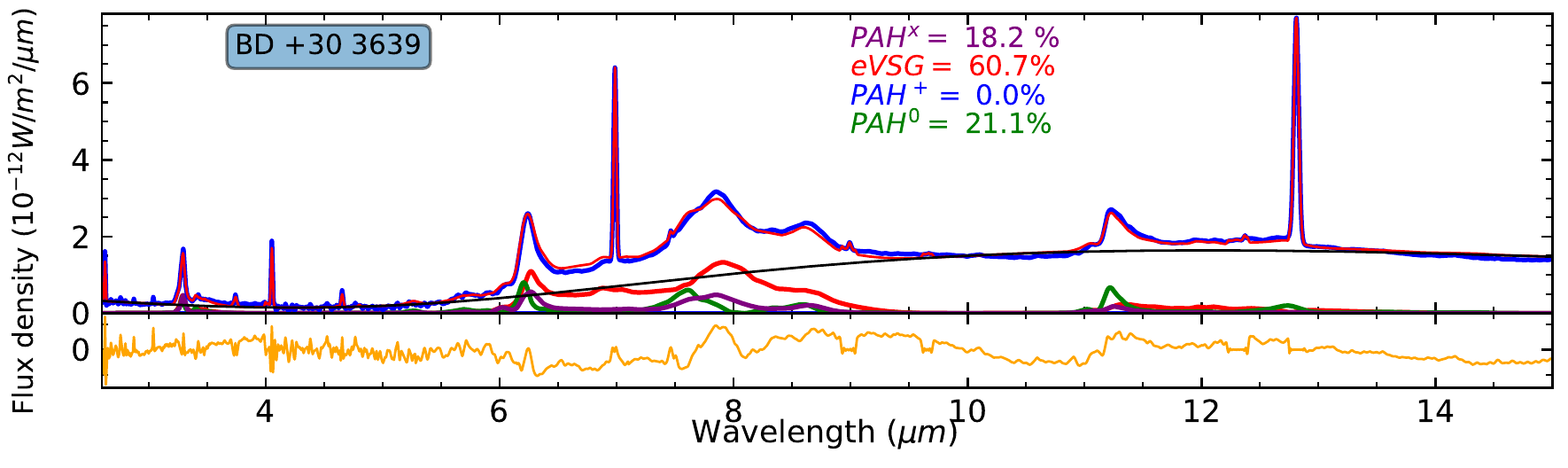}      
\includegraphics{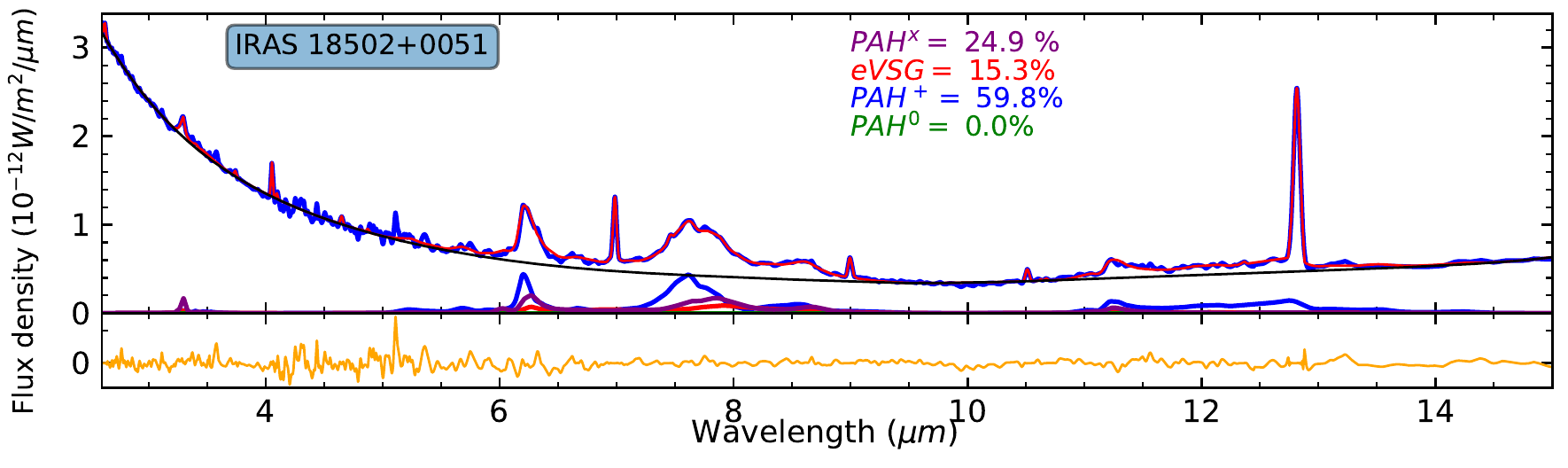}      
\caption{Continued.}
\end{figure*}

\begin{figure*}
\centering
\includegraphics{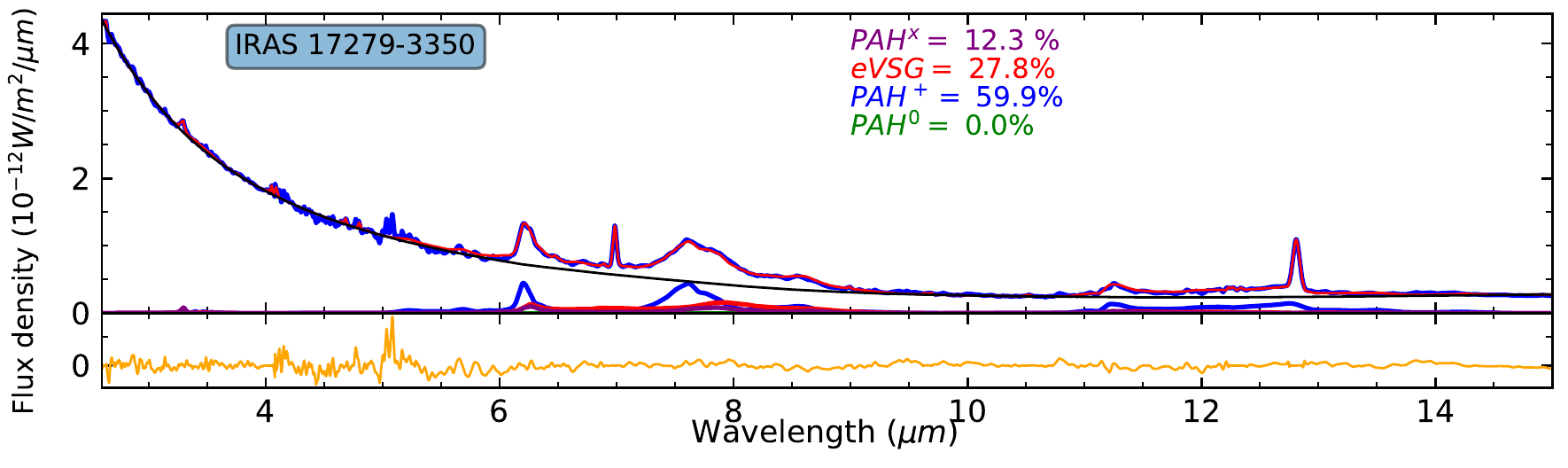}      
\includegraphics{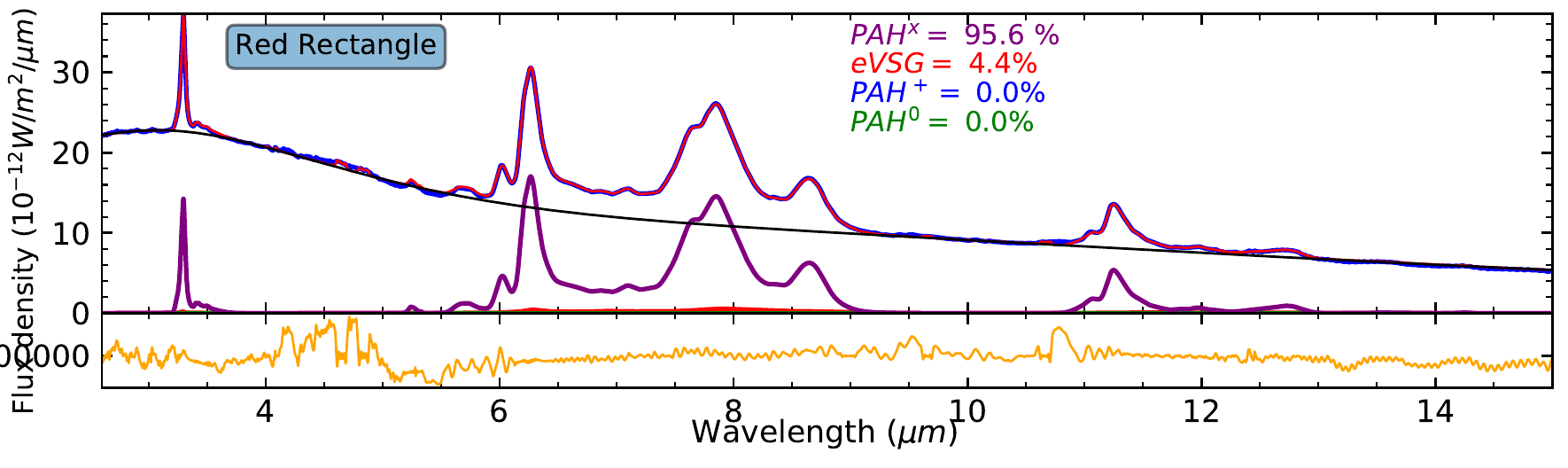}      
\includegraphics{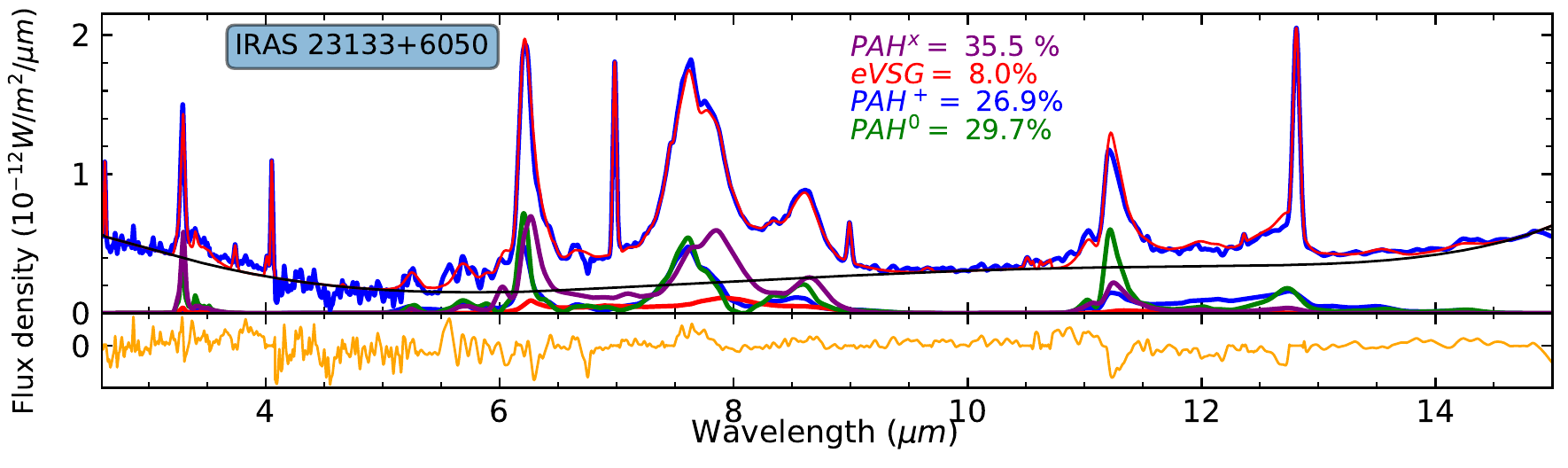}      
\includegraphics{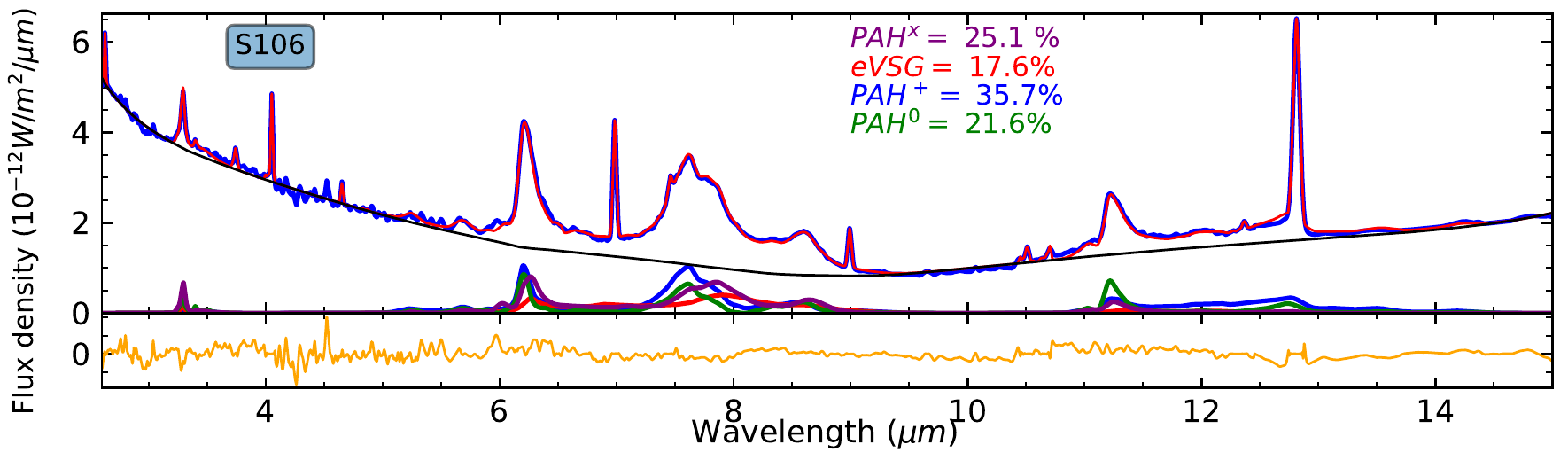}   
\caption{Continued.}
\end{figure*}

\begin{figure*}
\centering
\includegraphics{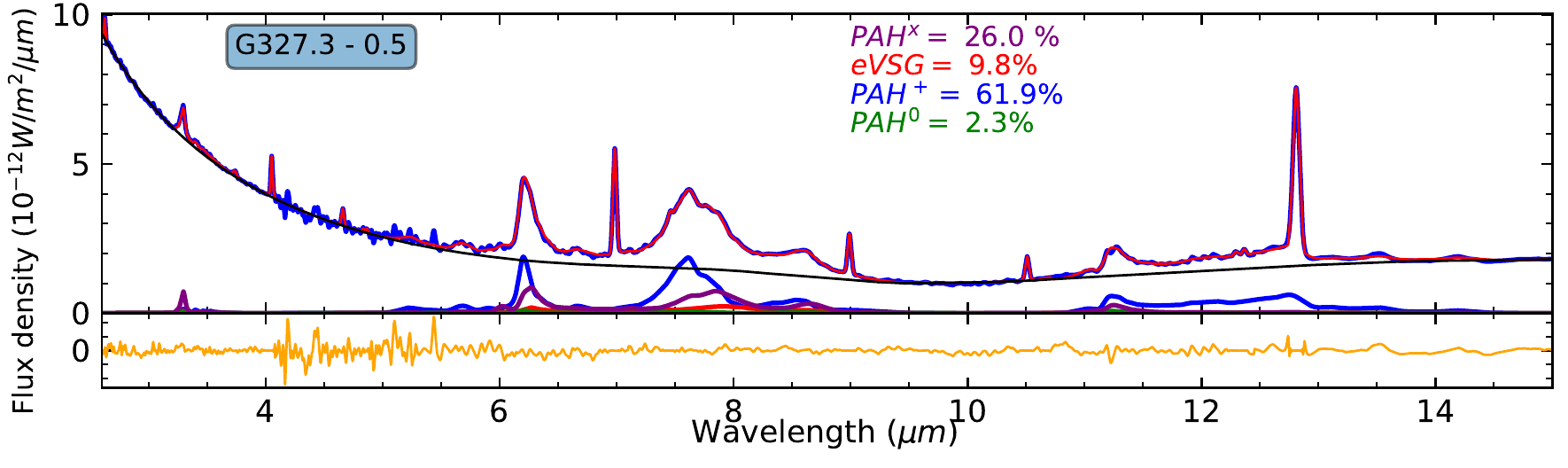}      
\includegraphics{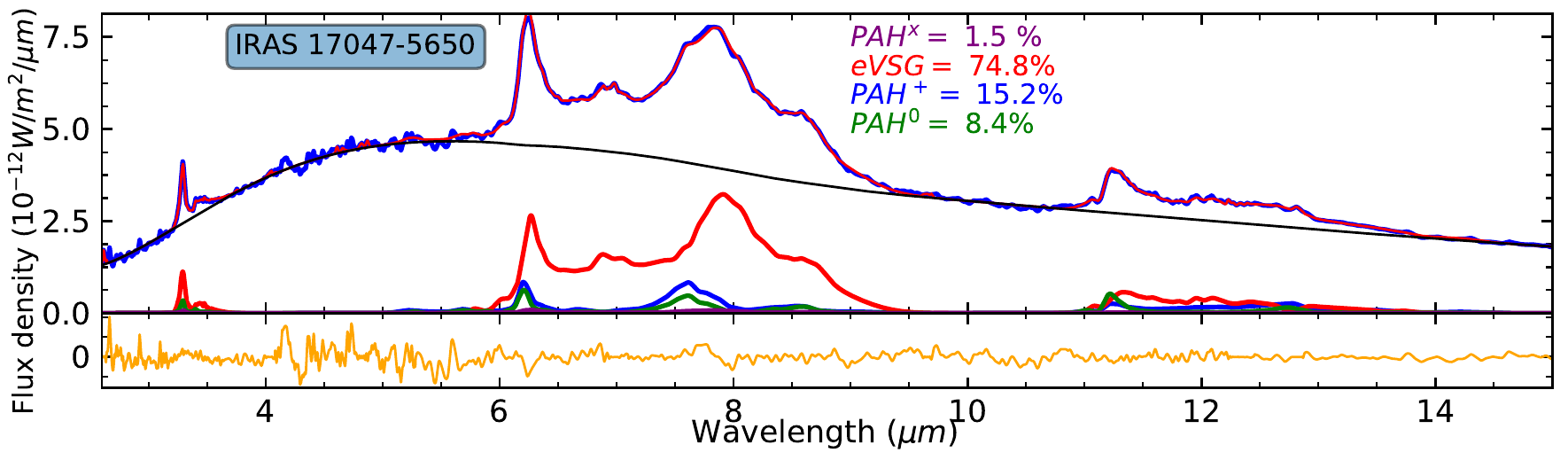}      
\includegraphics{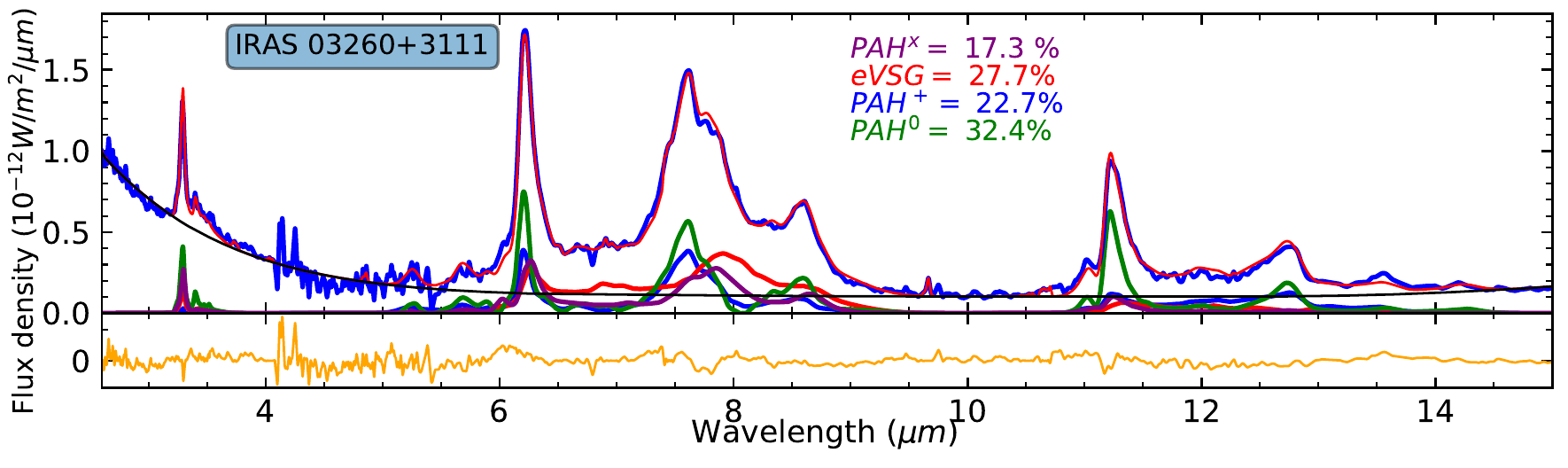}      
\includegraphics{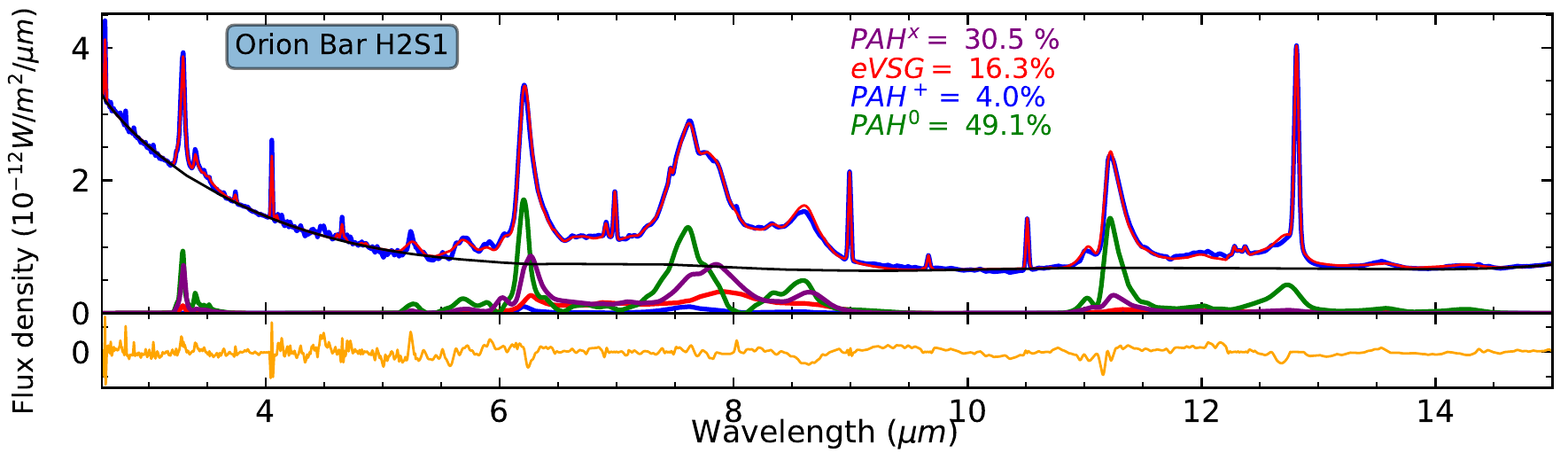}   
\caption{Continued.}
\end{figure*}

\begin{figure*}
\centering
\includegraphics{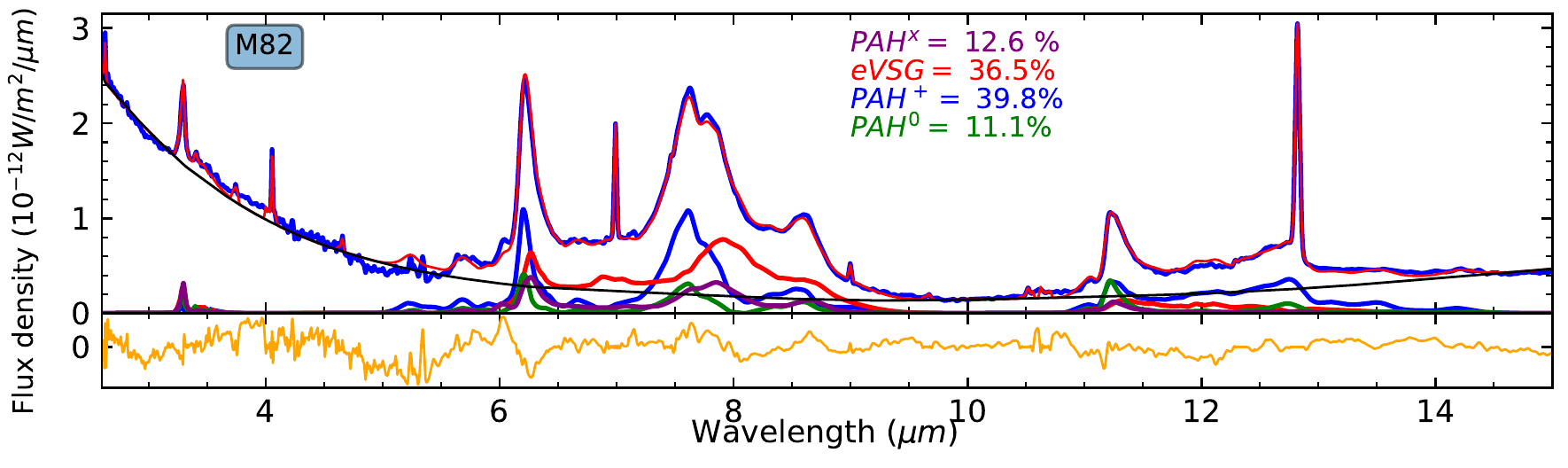}      
\includegraphics{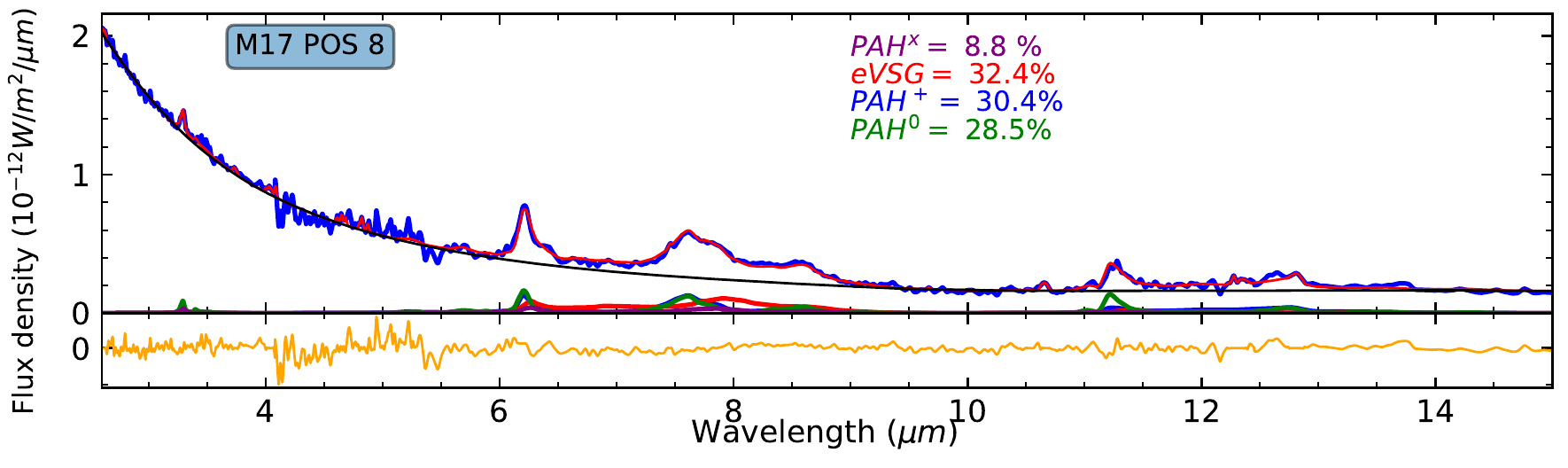}      
\includegraphics{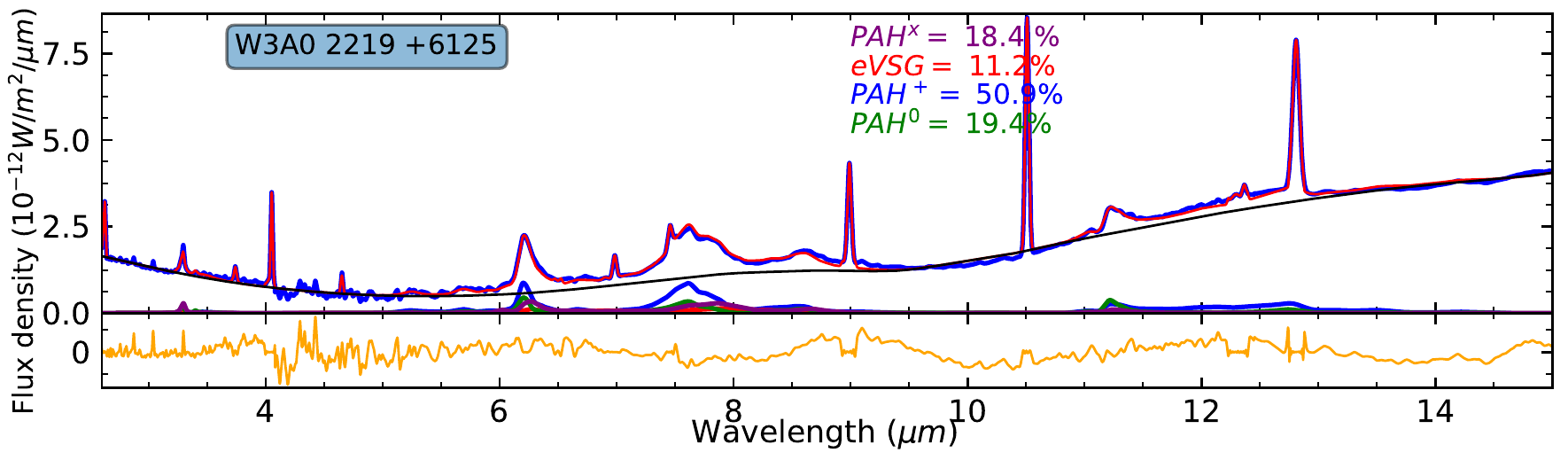}      
\includegraphics{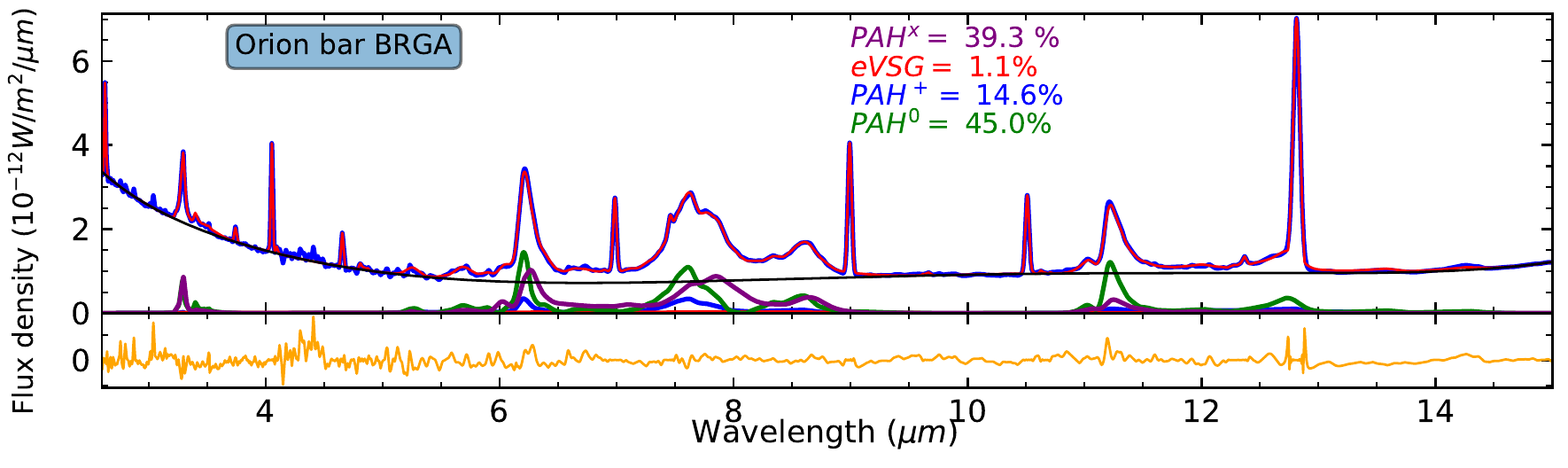}      
\caption{Continued.}
\end{figure*}

\begin{figure*}
\centering
\includegraphics{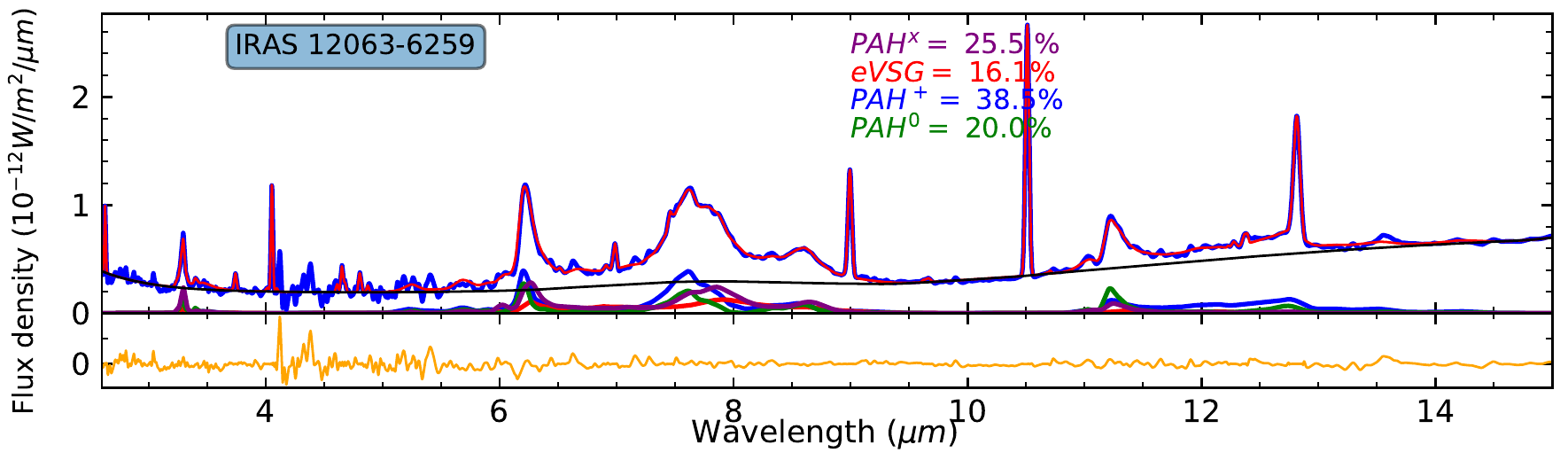}      
\includegraphics{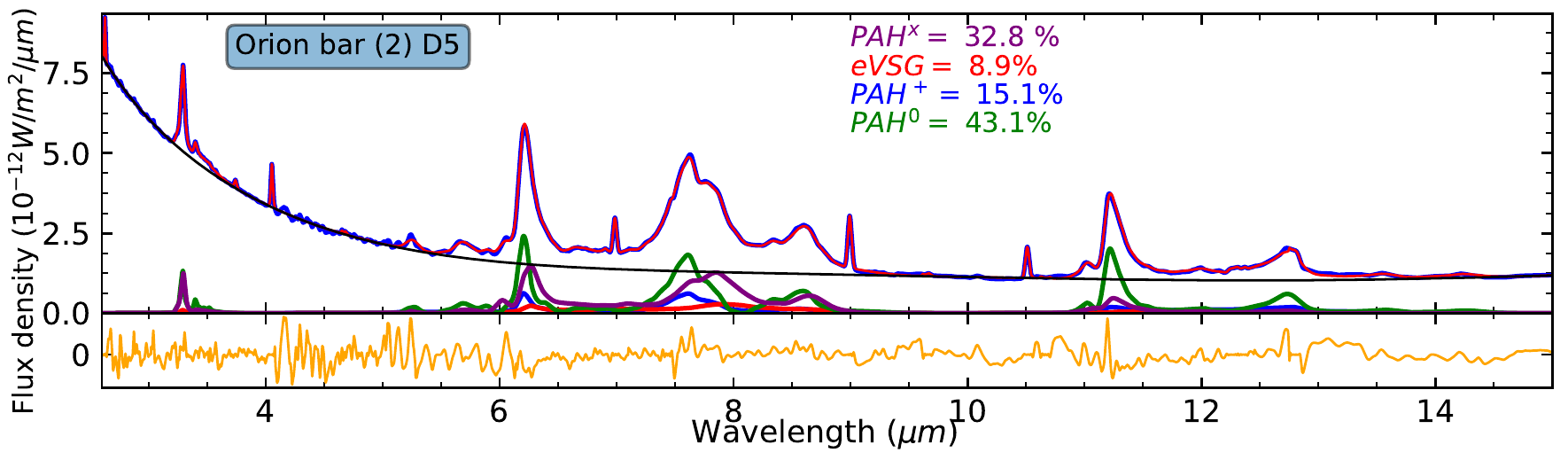}      
\includegraphics{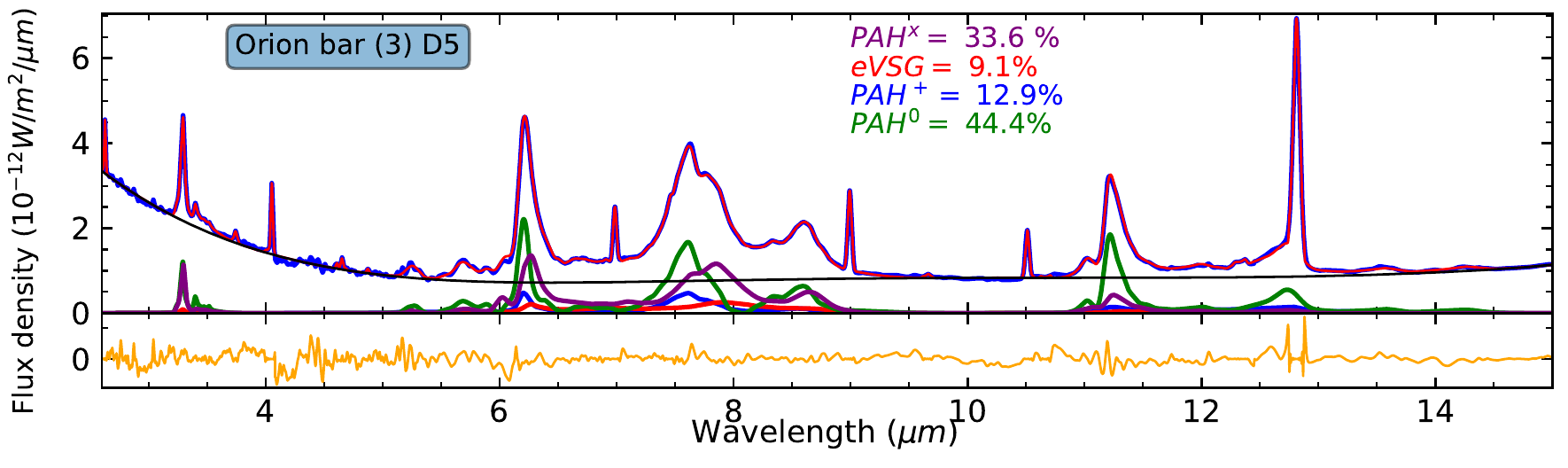}      
\caption{Continued.}
\end{figure*}

\end{appendix}
\end{document}